%% file: main_rmp.tex
\documentclass[rmp, aps,reprint,amsmath,amssymb,graphicx,superscriptaddress]{revtex4-2}

\usepackage{tabularx}
\usepackage{booktabs}
\usepackage{array}

\usepackage[braket,qm]{qcircuit}
\usepackage{mathtools}
\usepackage{hyperref}
\usepackage{cleveref}
\usepackage{xcolor}

\usepackage{amsmath,amsthm,mathrsfs,dsfont}
\usepackage{bm}
\usepackage{amsfonts}
\usepackage{amssymb}
\usepackage{multirow}
\usepackage[figuresright]{rotating}

\newcommand{\lnorm}[1]{\left\lVert #1 \right\rVert}
\newtheorem{challenge}{Challenge}
\newtheorem{definition}{Definition}

\input{environment}
\input{tikz_georgios}

\newcommand{\OxMaths}{\affiliation{Mathematical Institute, University of Oxford, Woodstock Road, Oxford OX2 6GG, United Kingdom}}

\newcommand{\QMT}{\affiliation{Quantum Motion, 9 Sterling Way, London N7 9HJ, United Kingdom}}

\begin{document}

\title{From Bits to Qubits: The Theory and Practice of Quantum Data Encoding}

 \author{Xiao-Ming Zhang}
 \thanks{These authors contributed equally}
 \affiliation{School of Physics, South China Normal University, Guangzhou 510006, China}

 \author{Arthur G. Rattew}
 \thanks{These authors contributed equally}
\affiliation{Department of Materials, University of Oxford, Parks Road, Oxford OX1 3PH, United Kingdom}
\QMT

 \author{Bujiao Wu}
 \affiliation{International Quantum Academy, Shenzhen 518048, China}

 \author{Georgios Styliaris}
 \affiliation{Max Planck Institute of Quantum Optics, Hans-Kopfermann-Str 1, Garching 85748, Germany}
\affiliation{Munich Center for Quantum Science and Technology (MCQST), Schellingstr 4, 80799 M\"{u}nchen, Germany}

\author{Xiaoming Sun}
\affiliation{State Key Lab of Processors, Institute of Computing Technology, Chinese Academy of Sciences, Beĳing 100190, China}

\author{B\'alint Koczor}
\email{balint.koczor@maths.ox.ac.uk}
\OxMaths
\QMT

 \author{Xiao Yuan}
 \email{xiaoyuan@pku.edu.cn}
\affiliation{Center on Frontiers of Computing Studies, School of Computer Science, Peking University, Beijing 100871, China}


\begin{abstract}
Encoding classical data into quantum systems is a foundational step in the execution of nearly all quantum algorithms, and a critical bottleneck in realizing practical quantum advantage. This review provides a comprehensive account of the concepts, algorithms, and practical considerations associated with quantum data encoding. We trace the development from its early conceptual foundations to recent advances, considering commonly used access models, such as quantum state preparation, unitary synthesis, QRAM and block encoding. We survey the circuit size, depth, space-time tradeoffs, as well as non-Clifford resources required for fault-tolerant implementation. We also discuss the roles of different access models in quantum algorithms. Special attention is given to structured data, such as sparse data, Boolean functions and data represented by tensor networks. This review bridges theory and applications, serving both as a pedagogical guide for newcomers and as a reference for active researchers. We also highlight the pivotal role of quantum data encoding in quantum computing and provide insights into future directions that will enable quantum advantage.
\end{abstract}

\maketitle

\tableofcontents

\input{secs/introduction}

\input{secs/state_prepare}

\input{secs/unitary_synthesis}

\input{secs/QRAM_0809}

\input{secs/sparse}

\input{secs/tensor}

\input{secs/Block_encoding}

\input{secs/practical}

\input{secs/application}

\input{secs/discussion}

\section{Acknowledgments}

We thank Ryan Babbush, Thomas Bromley, Earl Campbell, Ignacio Cirac, Alexander Dalzell, Sam Jaques, Zexian Li, Armands Strikis,  Rahul Trivedi and Kewen Wu for helpful discussions. 

X-.M.Z is supported by NSFC (No.~12405013) and the Guangdong Provincial Quantum Science Strategic Initiative (Grants No.GDZX2503008, No.GDZX2503001).
B.K.~thanks UKRI for the Future Leaders Fellowship
project titled Theory to Enable Practical Quantum Advantage (MR/Y015843/1). B.W.~is supported by NSFC (No.12405014). G.S. is funded by the Deutsche Forschungsgemeinschaft (DFG, German Research Foundation) under Germany’s Excellence Strategy -- EXC-2111 -- 390814868.
X.Y.~is supported by Quantum Science and Technology-National Science and Technology Major Project (2023ZD0300200), 
the National Natural Science Foundation of China Grant (No.~12361161602), 
NSAF (Grant No.~U2330201), 
Beijing Natural Science Foundation Z250004, and 
Beijing Science and Technology Planning Project (Grant No.~Z25110100810000).

%

\end{document}

%% file: environment.tex
\newcommand{\cbra}[1]{\{ #1 \}}
\newcommand{\pbra}[1]{\left( #1 \right)}

\newcommand{\Ccal}{\mathcal{C}}

\usepackage{tikz}
\usetikzlibrary{calc,decorations.pathreplacing}

\definecolor{wbjcolor}{RGB}{0, 0, 139}  

%% file: tikz_georgios.tex
\DeclareMathOperator{\polylog}{polylog}

\usepackage{tikz}
\usetikzlibrary{decorations.pathreplacing,calligraphy,decorations.markings}
\usetikzlibrary{fadings}
\usetikzlibrary{arrows.meta}
\usetikzlibrary{calc}

\definecolor{tensorcolor}{rgb}{0.65,0.77,0.95}
\definecolor{btensorcolor}{rgb}{0.85,0.72,0.89}
\definecolor{whitetensorcolor}{rgb}{0.93,0.93,0.93}

\newcommand{\gatevar}[2]{
    \begin{scope}[shift={(#1)}]
        \draw[ thick, fill=whitetensorcolor, rounded corners=1pt] (-\doubledx/2-0.2,0.27) rectangle (\doubledx/2+0.2,-.27); 
	    \draw (0,0) node {\scriptsize #2};
    \end{scope}
        }

\newcommand{\MPSTensor}[5]{
	\begin{scope}[shift={(#1)}]
    \ifnum#5=0
		\draw[thick] (-#2,0) -- (#2,0);
		\draw[thick] (0,#2) -- (0,0);
    \fi
    \ifnum#5=-1
		\draw[thick] (0,0) -- (#2,0);
		\draw[thick] (0,#2) -- (0,0);
    \fi
    \ifnum#5=1
		\draw[thick] (-#2,0) -- (0,0);
		\draw[thick] (0,#2) -- (0,0);
    \fi
        \draw[ thick, fill=tensorcolor, rounded corners=1pt] (-#3,-#3) rectangle (#3,#3);
		\draw (0,0) node {\scriptsize #4};
	\end{scope}
}

\newcommand{\GTensor}[5]{
	\begin{scope}[shift={(#1)}]
    \ifnum#5=0
		\draw[thick] (-#2,0) -- (#2,0);
		\draw[thick] (0,#2) -- (0,0);
    \fi
    \ifnum#5=-1
		\draw[thick] (0,0) -- (#2,0);
		\draw[thick] (0,#2) -- (0,0);
    \fi
    \ifnum#5=1
		\draw[thick] (-#2,0) -- (0,0);
		\draw[thick] (0,#2) -- (0,0);
    \fi
        \draw[ thick, fill=tensorcolor, rounded corners=1pt] (-#3,-#3) rectangle (#3,#3);
		\draw (0,0) node {\scriptsize #4};
	\end{scope}
}

\newcommand{\GDTensor}[5]{
	\begin{scope}[shift={(#1)}]
    \ifnum#5=0
		\draw[thick] (-#2,0) -- (#2,0);
		\draw[thick] (0,#2) -- (0,-#2);
    \fi
    \ifnum#5=-1
		\draw[thick] (0,0) -- (#2,0);
		\draw[thick] (0,#2) -- (0,-#2);
    \fi
    \ifnum#5=1
		\draw[thick] (-#2,0) -- (0,0);
		\draw[thick] (0,#2) -- (0,-#2);
    \fi
        \draw[ thick, fill=tensorcolor, rounded corners=1pt] (-#3,-#3) rectangle (#3,#3);
    \def\dx{#3/3};
	\draw (0,0) node {\scriptsize #4};
	\end{scope}
}

\newcommand{\ATensor}[3]{
    \GTensor{#1}{1}{.5}{#2}{#3};
}

\newcommand{\ETensor}[2]{
	\begin{scope}[shift={(#1)}]
        \GTensor{(0,.8)}{0}{.5}{}{#2};
        \GTensor{(0,-.8)}{0}{.5}{}{#2};
    \def\dx{.75};
	\end{scope}
}

\newcommand{\bTensor}[2]{
	\begin{scope}[shift={(#1)}]
	    \draw [thick] (-1,0) to  (1,0);
		\filldraw[color=black, fill=btensorcolor, thick] (0,0) circle (\stradius);
	\draw (0,0) node {#2};
	\end{scope}
}

\newcommand{\BTensor}[4]{
	\begin{scope}[shift={(#1)}]
        \ifnum#2=-1
            \bTensor{(0,0.9)}{#4};
            \bTensor{(0,-0.9)}{#3};
            \draw [thick] (-1,0.9) to  [bend left=45] (-1.3,1.2);
            \draw [thick] (-1.3,1.2) to  [bend left=45] (-1,1.5);
            \draw [thick] (-1, 1.5) to (-0.6, 1.5);
            \draw [thick] (-1,-0.9) to  [bend right=45] (-1.3,-1.2);
            \draw [thick] (-1.3,-1.2) to  [bend right=45] (-1,-1.5);
            \draw [thick] (-1, -1.5) to (-0.6, -1.5);
        \fi
        \ifnum#2=1
            \draw [thick] (-0.8,0.9) to  [bend right=45] (-0.5,1.2);
            \draw [thick] (-0.5,1.2) to  [bend right=45] (-0.8,1.5);
            \draw [thick] (-0.8, 1.5) to (-1.0, 1.5);
                        \draw [thick] (-0.8,-0.9) to  [bend left=45] (-0.5,-1.2);
            \draw [thick] (-0.5,-1.2) to  [bend left=45] (-0.8,-1.5);
            \draw [thick] (-0.8, -1.5) to (-1.0, -1.5);

        \fi
	\end{scope}
}

\newcommand{\SingleDots}[2]{
	\begin{scope}[shift={(#1)}]
      \draw [thick,  dash pattern=on 1pt off 1.4pt] (-#2*0.6,0) to (#2*0.6,0);
	\end{scope}
}

\newcommand{\IdentityTensor}[3]{
	\begin{scope}[shift={(#1)}]
      \draw [thick] (0,-1) to (0,1);
    \ifnum#3=1
        \draw [thick] (0,-1) to (0,1);
	    \filldraw[color=black, fill=whitetensorcolor, thick] (0,0) circle (\stradius);
	    \draw (0,0) node {#2};
    \fi
	\end{scope}
}

\newcommand{\DoubleIdentityTensor}[3]{
	\begin{scope}[shift={(#1)}]
        \draw [thick] (0,-1.8) to (0,1.8);
    \ifnum#3=1
        \draw [thick] (0,-1.8) to (0,1.8);
	    \filldraw[color=black, fill=whitetensorcolor, thick] (0,0) circle (\stradius);
	    \draw (0,0) node {#2};
    \fi
	\end{scope}
}

\newcommand{\SideIdentityTensor}[4]{
	\begin{scope}[shift={(#1)}]
    \ifnum#4=-1
	   \draw [thick] (\doubledx-1,0.8) to  [bend right=90] (\doubledx-1,-0.8);
    \fi
    \ifnum#4=-2
	   \draw [thick] (\doubledx-1,0.8) to  [bend right=90] (\doubledx-1,-0.8);
      \draw [thick] (\doubledx-1,0.8) -- (\doubledx-0.5,0.8);
      \draw [thick] (\doubledx-1,-0.8) -- (\doubledx-0.5,-0.8);
    \fi
    \ifnum#4=-3
	   \draw [thick] (\doubledx-1,0.8) to  [bend right=90] (\doubledx-1,-0.8);
      \draw [thick] (\doubledx-1,0.8) -- (\doubledx-0.5,0.8);
      \draw [thick] (\doubledx-1,-0.8) -- (\doubledx-0.5,-0.8);
	\filldraw[color=black, fill=whitetensorcolor, thick] (\doubledx-1.4,0) circle (#3);
	\draw (\doubledx-1.4,0) node {#2};
    \fi
    \ifnum#4=1
	   \draw [thick] (-\doubledx+1,0.8) to  [bend left=90] (-\doubledx+1,-0.8);
    \fi
    \ifnum#4=2
	   \draw [thick] (-\doubledx+1,0.8) to  [bend left=90] (-\doubledx+1,-0.8);
      \draw [thick] (-\doubledx+1,0.8) -- (-\doubledx+0.5,0.8);
      \draw [thick] (-\doubledx+1,-0.8) -- (-\doubledx+0.5,-0.8);
    \fi
    \ifnum#4=3
	   \draw [thick] (-\doubledx+1,0.8) to  [bend left=90] (-\doubledx+1,-0.8);
      \draw [thick] (-\doubledx+1,0.8) -- (-\doubledx+0.5,0.8);
      \draw [thick] (-\doubledx+1,-0.8) -- (-\doubledx+0.5,-0.8);
	\filldraw[color=black, fill=whitetensorcolor, thick] (-\doubledx+1.4,0) circle (#3);
	\draw (-\doubledx+1.4,0) node {#2};
    \fi
\end{scope}
}

\newcommand{\myarrow}[2]{
	\begin{scope}[shift={(#1)}]
    \ifnum#2=1
\draw[-{Stealth[length=1mm, width=2.3mm]}] (0,0.0) -- (0,0.03);
    \fi
    \ifnum#2=2
\draw[-{Stealth[length=1mm, width=2.3mm]}] (0,0) -- (0,-0.03);
    \fi
    \ifnum#2=3
\draw[-{Stealth[length=1mm, width=2.3mm]}] (0,0) -- (0.03,0);
    \fi
    \ifnum#2=4
\draw[-{Stealth[length=1mm, width=2.3mm]}] (0,0) -- (-0.03,0);
    \fi
\end{scope}
}

\newcommand{\SideIdentityTensorRT}[4]{
	\begin{scope}[shift={(#1)}]
    \ifnum#4=-1
	   \draw [very thick] (\doubledx-1,0.8) to  [bend right=90] (\doubledx-1,-0.8);
    \fi
    \ifnum#4=-2
	   \draw [very thick] (\doubledx-1,0.8) to  [bend right=90] (\doubledx-1,-0.8);
      \draw [very thick] (\doubledx-1,0.8) -- (\doubledx-0.5,0.8);
      \draw [very thick] (\doubledx-1,-0.8) -- (\doubledx-0.5,-0.8);
    \fi
    \ifnum#4=-3
	   \draw [very thick] (\doubledx-1,0.9) to  [bend right=90] (\doubledx-1,-0.9);
      \draw [very thick] (\doubledx-1,0.9) -- (\doubledx-0.5,0.9);
      \draw [very thick] (\doubledx-1,-0.9) -- (\doubledx-0.5,-0.9);
	\filldraw[color=black, fill=whitetensorcolor, thick] (\doubledx-1.4,0) circle (#3);
	\draw (\doubledx-1.4,0) node {#2};
    \fi
    \ifnum#4=1
	   \draw [very thick] (-\doubledx+1,0.8) to  [bend left=90] (-\doubledx+1,-0.8);
    \fi
    \ifnum#4=2
	   \draw [very thick] (-\doubledx+1,0.8) to  [bend left=90] (-\doubledx+1,-0.8);
      \draw [very thick] (-\doubledx+1,0.8) -- (-\doubledx+0.5,0.8);
      \draw [very thick] (-\doubledx+1,-0.8) -- (-\doubledx+0.5,-0.8);
    \fi
    \ifnum#4=3
	   \draw [very thick] (-\doubledx+1,0.9) to  [bend left=90] (-\doubledx+1,-0.9);
      \draw [very thick] (-\doubledx+1,0.9) -- (-\doubledx+0.5,0.9);
      \draw [very thick] (-\doubledx+1,-0.9) -- (-\doubledx+0.5,-0.9);
	\filldraw[color=black, fill=whitetensorcolor, thick] (-\doubledx+1.4,0) circle (#3);
	\draw (-\doubledx+1.4,0) node {#2};
    \fi
\end{scope}
}

\newcommand\doubledx{1.6}
\newcommand\singledx{1.8}

\newcommand\stradius{0.5}

%% file: secs/introduction.tex
\section{Introduction\label{sec:intro}}

\begin{figure*}[t]
    \centering
          \includegraphics[width=1.8\columnwidth]{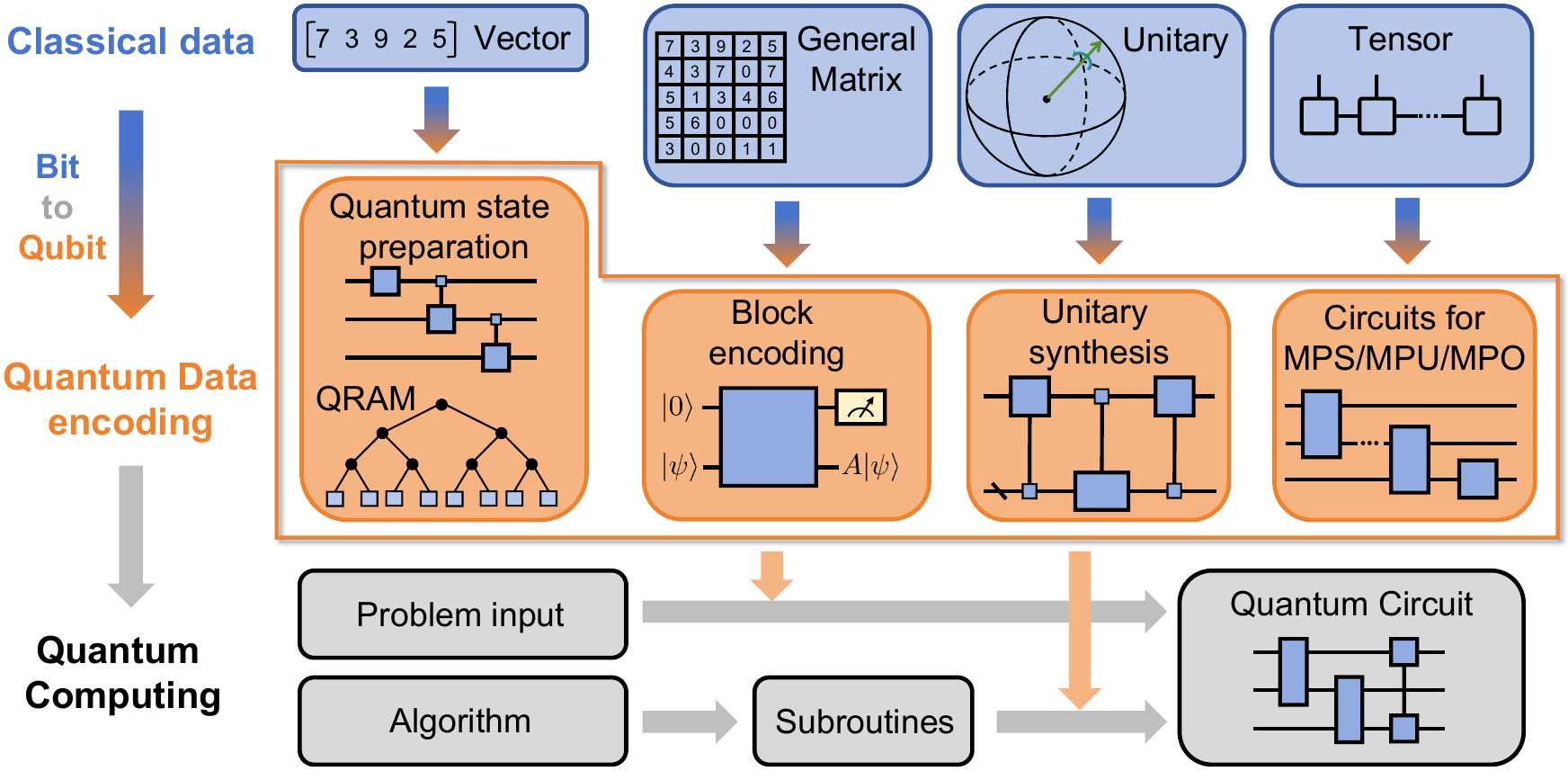}
       \caption{From bit (classical data) to qubit (quantum data). Quantum algorithm and corresponding problem inputs are compiled to concrete quantum circuits. Quantum state preparation: for classical vector $\bm{v}$, the task is to prepare a quantum state $|\psi\rangle=\|\bm{v}\|^{-1}\sum_{j}v_j|j\rangle$. QRAM: For binary vector $\bm{v}$, the task is to construct a unitary transformation that $|j\rangle|0\rangle\rightarrow|j\rangle|v_j\rangle$. Block-encoding: for matrix $A$, the task is to construct a unitary $U_A$ such that 
  $(\bra{0}^{\otimes a}\otimes I_n)U_A(\ket{0}^{\otimes a}\otimes I_n)=A/\alpha$. 
  } \label{fig:sketch}
\end{figure*}

Quantum computing represents a fundamentally new paradigm for information processing. A quantum computer encodes information into quantum bits (qubits), which evolve under unitary transformations implemented by elementary quantum gates. Leveraging the quantum resources like coherence and entanglement, it performs computations that are believed to outperform classical machines. Over the past decades, substantial efforts have been made on developing quantum algorithms for addressing simulation~\cite{georgescu2014quantum,mcardle2020quantum}, optimization~\cite{biamonte2017quantum,blekos2024review,abbas2024challenges}, algebra~\cite{childs2010quantum}, and cryptographic~\cite{portmann2022security} tasks that are believed to be intractable on classical computers. 

Yet, problems and quantum algorithms are written classically. Their practical execution on quantum hardware thus requires translating the program and problem description into sequences of elementary quantum operations. This encoding process, which bridges classical inputs (e.g., vectors, matrices) to quantum representations (e.g., quantum states, block-encodings), is a preliminary step of quantum computing (Fig.~\ref{fig:sketch}). Although it is commonly treated as a \textit{black box} in algorithm design, its implementation complexity plays a decisive role in the overall efficiency, feasibility, and quantum advantage of these algorithms.

The problem of encoding classical data into quantum form has been studied since the early days of quantum computation. Foundational universality results established that any $n$-qubit unitary can be decomposed into elementary three-qubit gates~\cite{deutsch1989quantum} or two-qubit gates~\cite{divincenzo1995two}, with sufficiently long gate sequences.~\citet{barenco1995elementary} further showed that polynomial circuit size with respect to the matrix dimension is enough for arbitrary unitaries. For encoding vectors, pioneer works showed that preparing an arbitrary quantum state can be realized by a sequence of unitary controlled rotations~\citet{Long.01,kaye2001quantum,Grover.02}. Besides,~\citet{Giovannetti.08} introduced an alternative paradigm, quantum random access memory, which coherently encodes classical data $D_x$ into a quantum register via the transformation $|j\rangle|0\rangle\rightarrow|j\rangle|x_j\rangle$~\cite{Giovannetti.08}.

Recently, the field of quantum data encoding has progressed along multiple directions. Improvements have been made in circuit complexity, yielding better circuit depth, size, and fault-tolerant resources (e.g.,~CNOT count, T-count). The scope has been expanded from basic state preparation and unitary synthesis to the encoding of more general classical data, including non-unitary matrices, tensors and functions. Furthermore, simplification has been considered for structured data. Leveraging properties like sparsity, Boolean, tensor network representations and low-rank, more efficient encodings can be made compared to those for unstructured data. Complementing these theoretical advances, classical optimization techniques, including machine learning and randomized compilation have also become powerful tools for encoding practical data. 

This review aims to provide a systematic survey of the progress about the encoding process in quantum computing. For nonspecialists, we aim to make the assumptions and resource costs
behind commonly used access models explicit, so that these parts can be incorporated correctly into end-to-end algorithm analyses.
 In particular, a summary of the circuit complexities for some typical data loading tasks is provided in Table~\ref{table}. For the specialist, we highlight the latest advances in encoding techniques, clarify fundamental limits, and outline some key open questions in the area. 

We restrict our discussion to the encoding of classical data into quantum circuit or states. We do not address quantum algorithms, for which readers are referred to other surveys, e.g.~\cite{childs2010quantum,dalzell2023quantum}. 
Moreover, our scope is limited to the standard qubit-based circuit model, while frameworks based on qudits~\cite{kiktenko2025colloquium} or continuous variables~\cite{weedbrook2012gaussian} are not considered. Lastly, we assume that the classical data are deterministic and fully known in advance, thus excluding cases where data are sampled from probability distributions~\cite{zhao2026exponential}.

The organization of the remaining sections is as follows. We introduce quantum state preparation in Sec.~\ref{sec:qsp} and unitary synthesis in Sec.~\ref{sec:unitary}, corresponding to the encoding of vector and unitary, respectively. In Sec.~\ref{sec:QRAM}, we introduce QRAM within the taxonomy of classical versus quantum data, and further distinguish between active and passive implementations. Sec.~\ref{sec:qsp}~\ref{sec:unitary} and~\ref{sec:QRAM} focus on data without structures. In Sec.~\ref{sec:sparse}, we further introduce data encoding techniques for structured data, including sparse data, isometry, continuous functions, and Boolean data.  Sec.~\ref{sec:tensor} focus on classical data in the form of tensor networks, covering the preparation of tensor network states and the synthesis of matrix product operators. Sec.~\ref{sec:block_encoding} introduces block-encoding, which generalizes the matrix encoding from unitary to arbitrary nonunitary matrices. In Sec.~\ref{sec:practical}, we introduce some classical technique that are useful for optimizing quantum data encoding, including randomization and machine learning. We provide further discussions in Sec.~\ref{sec:application}, which includes an end-to-end quantum algorithm example for Hamiltonian simulation, gathering multiple data encoding techniques into one protocol. We also provide further discussion about the relation between classical data structures and quantum advantage.

\subsection{Definitions}

\subsubsection{Quantum gates}
We begin with the definition of some typical elementary quantum gates. Single qubit Pauli $X, Y$ and $Z$ operations are defined as 
\begin{align}
X=
\begin{pmatrix}
0&1\\
1&0
\end{pmatrix},\quad Y=
\begin{pmatrix}
0&-i\\
i&0
\end{pmatrix},\quad
Z=
\begin{pmatrix}
1&0\\
0&-1
\end{pmatrix},
\end{align}
and arbitrary rotations are defined as
\begin{align}
R_{\bm{\hat n}}(\phi)=e^{-i(\bm{\sigma}\cdot\bm{\hat{n}})\phi/2}
\end{align} for some $\bm{\sigma}=(\sigma_x,\sigma_y,\sigma_z)$ and unit vector $\bm{\hat{n}}$ as the rotation axis. For example, the rotations along $X$, $Y$ and $Z$ axes are:
\begin{subequations}
\begin{align}
R_x(\theta)&=e^{-iX\theta/2}=
\begin{pmatrix}
\cos(\theta/2)&-i\sin(\theta/2)\\
-i\sin(\theta/2)&\cos(\theta/2)
\end{pmatrix}
\\
R_y(\theta)&=e^{-iY\theta/2}=
\begin{pmatrix}
\cos(\theta/2)&-\sin(\theta/2)\\
\sin(\theta/2)&\cos(\theta/2)
\end{pmatrix}
\\
R_z(\theta)&=e^{-iZ\theta/2}=
\begin{pmatrix}
e^{-i\theta/2}&0\\
0&e^{i\theta/2}
\end{pmatrix}
\end{align}
\end{subequations}
The single-qubit identity, Hadamard, and $T$ gates are
\begin{align}
I=
\begin{pmatrix}
1&0\\
0&1
\end{pmatrix},~ H=
\frac{1}{\sqrt{2}}\begin{pmatrix}
1&1\\
1&-1
\end{pmatrix},~
T=
\begin{pmatrix}
1&0\\
0&e^{i\pi/4}
\end{pmatrix},
\end{align}
respectively.

Typical two-qubit gates include CNOT and CZ gates 
\begin{subequations}
\begin{align}
\text{CNOT}=|0\rangle\langle0|\otimes I+|1\rangle\langle 1|\otimes X,\\
\text{CZ}=|0\rangle\langle0|\otimes I+|1\rangle\langle 1|\otimes Z.
\end{align}
\end{subequations}
A general controlled rotation is 
\begin{align}
\text{C-}U=|0\rangle\langle0|\otimes I+|1\rangle\langle 1|\otimes U.
\end{align}
Multi-qubit controlled gates are defined as
\begin{align}
\text{C}^{k}\text{-}U=\left(\sum_{j=0}^{2^k-2}|j\rangle\langle j|\otimes I\right)+|1\cdots1\rangle\langle 1\cdots1|\otimes U.
\end{align}
Another important multi-qubit gate is the uniform controlled rotation (UCR). An $m$-qubit UCR along axis $\bm{\hat{n}}$ with rotation angles $\bm{\theta}=[\theta_{0},\theta_1,\cdots,\theta_{2^m-1}]$ is defined as
\begin{align}\label{eq:Fn}
F_{\bm{\hat{n}}}(\bm{\theta})=\sum_{j=0}^{2^m-1}|j\rangle\langle j|\otimes R_{\bm{\hat{n}}}(\theta_{j}).
\end{align}
As a special case, when $R_{\bm{\hat{n}}}(\theta_{j})\in\{I,X\}$, the operation in Eq.~\eqref{eq:Fn} is equivalent to the one by quantum random access memory (QRAM, Sec.~\ref{sec:QRAM}). For binary classical data $x_j\in\{0,1\}$, the latter aims at performing the transformation $|j\rangle|0\rangle\rightarrow|j\rangle|x_j\rangle$ coherently.

\begin{table*}[htbp]
\centering
\fontsize{7pt}{2pt}\selectfont
\caption{Best known circuit complexities for different tasks. An unlimited number of ancillary qubits and all‑to‑all connectivity are assumed. SP, US and BE refer to state preparation, unitary synthesis and block-encoding respectively. $n$ refers to the number of qubits;  $\varepsilon$ refers to state or unitary accuracy, typically measured as infidelity or operator norm; $k$ refers to the Hamming weight of the target state basis; $S$ refers to the nonzero entries of the target state; $P$ refers to the number of terms in the LCU.}
\begin{tabular}{|c|c|c|c|c|c|c|c|c|c|c|}
\hline
\multicolumn{2}{|c|}{Tasks}  & Section& Circuit size& Circuit size (FT)\footnote{Fault-tolerant decomposition with $\{\text{CNOT},H,T\}$. The scaling is $C\log(C/\varepsilon)$ unless specified, where $C$ is the circuit size.}& Circuit depth &CNOT count\footnote{The scaling is $O(C)$ unless specified.} &T count\footnote{The scaling is $C\log(C/\varepsilon)$ unless specified.} \\ \hline
\multirow{8}{*}{SP} & General &Sec.~\ref{sec:qsp}&$\Theta(2^n)$ &$\Theta\left(\frac{2^n\log(1/\varepsilon)}{n}\right)$&$\Theta(n)$\footnote{reduced to $O(1)$ if measurement, feedback and $\approx O(2^n)$ ancillary qubits are permitted.}&$\sim\frac{11}{12}2^n$&$\Theta(\sqrt{2^n\log(1/\varepsilon)}+\log(1/\varepsilon))$ \\
& Sparse &Sec.~\ref{sec:sparse_sp}&$O\left(\frac{nS}{\log n}+n\right)$ & &$\Theta(\log(nS))$&&  \\ 
& MPS~\footnote{$n$-qubit MPS with constant bond dimension.
} &Sec.~\ref{sec:tensor_MPS}&$\Theta(n)$ & &&&  \\ 
& Dicke &Sec.~\ref{sec:str_other}&$O(kn)$ &$O(kn)$&$O(k\log(n/k))$&&  \\ 
& FHW~\footnote{fixed-Hamming-weight state} &Sec.~\ref{sec:str_other}&$O(\binom{n}{k})$ &&$O(\log\binom{n}{k})$&&  \\[1ex] 
\hline
\multirow{8}{*}{US}& Single qubit &Sec.~\ref{sec:unitary_single}& /&$O(\log(1/\varepsilon))$&$O(\log(1/\varepsilon))$&/&  \\ 
& General &Sec.~\ref{sec:unitary_without},~\ref{sec:unitary_st}&$\Theta(4^n)$ &&$\tilde{O}(2^{n/2})$&$\sim\frac{11}{24}4^n$&$O(2^{4n/3}n^{2/3})$  \\ 
& Sparse &Sec.~\ref{sec:sparse_unitary}&$O(n(S+2^n))$ &&&&  \\ 
& Sparse Boolean &Sec.~\ref{sec:sparse_unitary}&
$O\left(\frac{nS}{\log n}+n\log\min\{S,\log n\}\right)$
&&&&  \\ 
& Boolean &Sec.~\ref{sec:gbu}&$\Theta\left(\frac{2^nn}{\log n}\right)$ &&&&  \\ 
\hline
\multirow{2.5}{*}{BE}&General &Sec.~\ref{sec:bl_general_matrix}&$O(4^n)$ &&$O(2^{n/2})$&$\sim\frac{11}{12}\times4^n$&  \\ 
&LCU~\footnote{Eq.~\eqref{eq:be_lcu} with $U_j$ be $n$-qubit Pauli strings.} &Sec.~\ref{sec:bl_lcu_matrix}&$O(Pn)$ &$O(P(n+\log(1/\varepsilon)))$&$O(\log(Pn))$&&  \\ 
\hline
\multicolumn{2}{|c|}{CNOT/Clifford} &Sec.~\ref{sec:b_CNOT}&$\Theta\left(\frac{n^2}{\log n}\right)$
 &$\Theta\left(\frac{n^2}{\log n}\right)$&$\Theta(\log(n))$&$\Theta\left(\frac{n^2}{\log n}\right)$&0  \\ 
\hline
\multicolumn{2}{|c|}{SAT-oracle\footnote{$m$ clauses, each clause has at most $q$ variables.} }&Sec.~\ref{sec:ecnfg}&$O\left(qm^{1+o(1)}\right)$&&$\tilde{O}(\log(qm))$&&  \\ 
\hline
\multicolumn{2}{|c|}{Isometry }&Sec.~\ref{sec:sparse_iso}&$O(NM)$&&$O(N\log M)$&&  \\ 
\hline
\multicolumn{2}{|c|}{UCR\footnote{See definition in Eq.~\eqref{eq:Fn}}}&Sec.~\ref{sec:qsp_ucr}&$O(2^n)$&$O(2^n\log(1/\varepsilon))$ &$\Theta(n)$&$O(2^n)$&$O(\sqrt{2^n\log(1/\varepsilon)}+\log(1/\varepsilon))$  \\ 
\hline
\multicolumn{2}{|c|}{$C^n$-$X$/$C^n$-$U$ }&Sec.~\ref{sec:cnxg}&$O(n)$&$O(n+\log(1/\varepsilon))$ &$O(\log n)$&$\sim12n$&$\sim16n$/$O(n\log(1/\varepsilon))$   \\ 
\hline
\end{tabular}
\label{table}
\end{table*}

\subsubsection{Circuit complexity metrics}

Below, we summarize several metrics that are important for quantifying the quantum circuit complexity. 
\begin{itemize}
\item \textbf{Circuit size} refers to the total number of elementary quantum gates of the quantum circuit. Unless specified, the elementary gate set contains CNOT and arbitrary single-qubit gates in SU$(2)$.

\item \textbf{Circuit depth} refers to the number of quantum gate layers of the quantum circuit. It also corresponds to the execution time. If the overheads of long-range interaction, measurements, classical post processing, etc are neglected. In case no ancillary qubit is used, circuit depth differs from the circuit size by at most a factor of $n$, but the difference can be significant when the number of ancillary qubits is large.

\item \textbf{CNOT/T/single-qubit gate count} refers to the total number of CNOT/T/single-qubit gates in the quantum circuit. In particular, $T$ count is with respect to the fault-tolerant decomposition using elementary gate set $\{\text{CNOT}, T, H\}$, which is of special interest as implementing $T$ gates is more costly in fault-tolerant setting. A more detailed discussion about $T$ gate is provided in Sec.~\ref{sec:rm}.

\item \textbf{Number of ancillary qubits} corresponds to the space complexity of the quantum circuit. Usually, for a given task, there is a tradeoff between the number of ancillary qubits (space) and the circuit depth (time). 
\end{itemize}

We use $O(\cdot)$, $\Omega(\cdot)$, and $\Theta(\cdot)$ in their standard senses to denote asymptotic upper, lower, and tight bounds, respectively, for describing the scaling of resources with respect to parameters such as $n$ and $\varepsilon$. For example, $f(n)=O(n^2)$ means that there exist constants $c>0$ and $n_0$ such that $f(n)\leq c n^2$ for all $n\geq n_0$. We also use the tilde notation, e.g., $\tilde{O}(\cdot)$, to suppress logarithmic factors.

The above metrics represent distinct optimization objectives, and they are generally not simultaneously minimizable by a single protocol. Thus, the various encoding schemes are usually designed with different optimization goals. In practice, improving one resource metric often comes at the expense of another. For example, reducing circuit depth is usually at the cost of requiring more ancillary qubits. The choice of which metric to prioritize depends on the specific constraints of the quantum hardware.

%% file: secs/state_prepare.tex
\section{Quantum state preparation\label{sec:qsp}}

Quantum state preparation is a standard process that encodes classical data into the wavefunction of a quantum system. Given a classical description of the normalized complex vector $\bm{\alpha}=[\alpha_0,\cdots,\alpha_{N-1}]$ satisfying $\bm{\alpha} \bm{\alpha}^\dag = 1$, the task is to construct the quantum state
\begin{align}\label{eq:sp}
|\psi\rangle=\sum_{j=0}^{2^n-1}\alpha_j|j\rangle
\end{align}
with a binary $n$-bit computational basis representation of $j$. Here, $n=\lceil \log_2 N \rceil$ and $\alpha_j=0$ for $j \ge N$. A standard state preparation precedure is to applying a sequence of elementary quantum gates on some trivial initial state, such as $|0\rangle^{\otimes n}$.  
In case error $\varepsilon$ and ancillary qubits (initialized at $|0\rangle_{\text{anc}}$) are allowed, our target is to realize unitary $U_{\psi}$ satisfying
\begin{align}
U_{\psi}|0\rangle^{\otimes n}\otimes|0\rangle_{\text{anc}}\longrightarrow|\widetilde{\psi}\rangle\otimes|0\rangle_{\text{anc}}
\end{align}
for some $|\langle\psi|\widetilde{\psi}\rangle|^2\geqslant1-\varepsilon$~\footnote{For randomized algorithms, the prepared state may also be mixed; accordingly, the requirement is modified to demand high fidelity between the prepared state and the target state.
}.

From a theoretical perspective, understanding how a state $|\psi\rangle$ can be optimally prepared using a set of elementary operations reflects our fundamental ability to construct quantum systems. From a practical perspective, quantum state preparation is an important subroutine in quantum algorithms. In quantum machine learning, Eq.~\eqref{eq:sp} is equivalent to the \textit{amplitude encoding} of classical data~\cite{Schuld.19,Schuld.20,Lu.20,Huang.21,Du.21,Tian.23,Liao.24,Liao.24_2}, which is an important step of the end-to-end quantum computing pipeline. In quantum simulation, state preparation and its inverse are critical steps in LCU-based block-encoding~\cite{childs2012hamiltonian} (see also Sec.~\ref{sec:block_encoding}), which is useful for simulating both dynamics~\cite{berry2014exponential,low2019hamiltonian} and ground state properties~\cite{gilyen2019quantumtutorial,lin2020near} of many-body systems. We also provide an end-to-end example of applying state preparation and other data encoding techniques to the problem of Hamiltonian simulation in Sec.~\ref{sec:exp}. 

This section is organized as follows. Sec.~\ref{sec:qsp_without} reviews ancilla-free constructions.
Sec.~\ref{sec:qsp_st} discusses space--time tradeoffs enabled by ancillary
qubits. Sec.~\ref{sec:query} discusses black-box and query-complexity models. Sec.~\ref{sec:sp_ft} considers fault-tolerant implementations.
Sec.~\ref{sec:sp_lb} summarizes lower bounds.   Structured target states are treated separately in Sec.~\ref{sec:sparse}.

\subsection{Preparation without ancilla\label{sec:qsp_without}}

\subsubsection{Uniformly controlled rotation methods}\label{sec:qsp_ucr}
A straightforward way of preparing quantum states is to iteratively increase the state dimension with UCRs as defined in Eq.~\eqref{eq:Fn}~\cite{Long.01,kaye2001quantum,Grover.02}.

\begin{figure}[t]
    \centering
          \includegraphics[width=1\columnwidth]{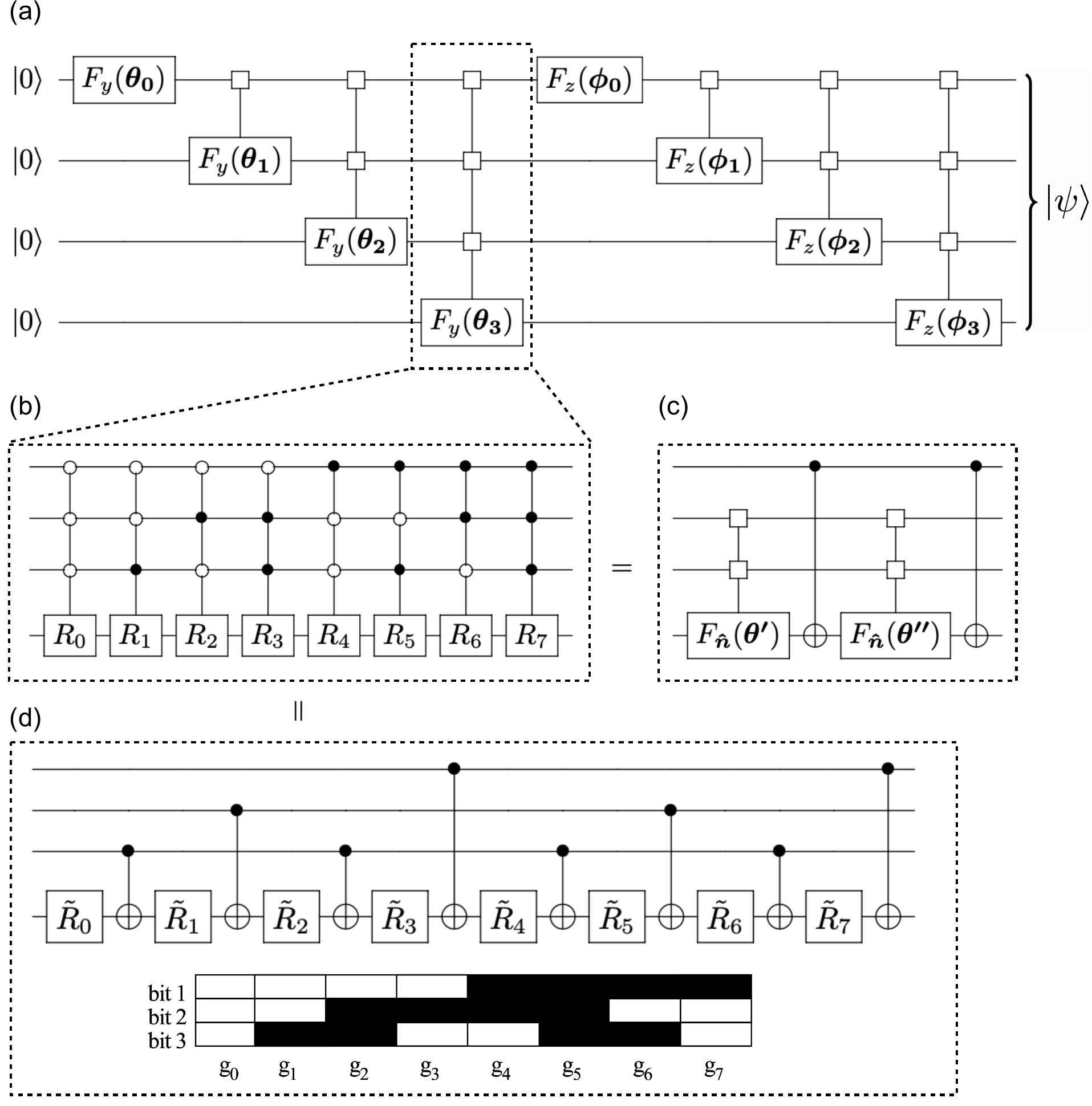}
       \caption{Illustration of quantum state preparation and UCR for $n=4$. (a) Decomposition of quantum state preparation into UCRs. (b) Definition of UCR, where $R_j=R_{y}(\theta_j)$. (c) Recursive decomposition of an $n$-qubit UCR into two $(n-1)$-qubit UCRs and two CNOT gates~\cite{shende2005synthesis}. (d) Gray code based decomposition of UCR that replaces multi-qubit controlled rotations by CNOTs~\cite{mottonen2004quantum}. Lower panel is from~\cite{mottonen2004quantum}, which shows gray code used to define the positions of the control nodes. The black (white) rectangles denote bit values one (zero).
       } \label{fig:sp_ucr}
\end{figure}

For simplicity, we suppose all amplitudes are positive $\alpha_j\geq0$, the process contains $n$ steps, and the intermediate state at the $m$th step is 
\begin{align}
|\psi_{m}\rangle=\sum_{j=0}^{2^m-1}\alpha_{m,j}|j\rangle,
\end{align}
 where the amplitudes are recursively defined as $\alpha_{m,j}=\sqrt{|\alpha_{m+1,2j}|^2+|\alpha_{m+1,2j+1}|^2}$ and $\alpha_{n,j}=\alpha_j$. We then define rotation angles $\bm{\theta}_m=[\theta_{m,0},\theta_{m,1},\cdots,\theta_{m,2^m-1}]$, where 
\begin{align}\label{eq:sp_ang}
\theta_{m,j}\equiv2\text{arctan}(\alpha_{m+1,2j+1}/\alpha_{m+1,2j})
\end{align} 
for $\alpha_{m+1,2j}\neq0$, and $\theta_{m,j}=\pi$ otherwise.  For an $m$-bit basis $|k\rangle$, it can be verified that 
\begin{align}\label{eq:fyt}
F_\textbf{y}(\bm{\theta}_m)|k\rangle|0\rangle=|k\rangle(\cos\frac{\theta_{m,k}}{2}|0\rangle+\sin\frac{\theta_{m,k}}{2}|1\rangle).
\end{align}
 Thus, by introducing an extra qubit and applying UCR to $|\psi_{m}\rangle|0\rangle$, we have 
\begin{align}\label{eq:sp_f}
F_\textbf{y}(\bm{\theta}_m)|\psi_{m}\rangle|0\rangle=|\psi_{m+1}\rangle.
\end{align}
Initializing the quantum state to $|0\rangle^{\otimes n}$, and applying Eq.~\eqref{eq:sp_f} iteratively from $m=0$ to $m=n-1$, we obtain the target state $|\psi\rangle$. For complex quantum states, phase can be added to each amplitude $\alpha_j$ in a similar manner using UCR along z-axis $F_\textbf{z}(\bm{\phi}_m)$, with appropriately chosen rotation angles $\bm{\phi}_m$~\cite{Mottonen.05}. See Fig.~\ref{fig:sp_ucr}(a) for illustration with $n=4$.  

By definition, each UCR can be decomposed into $2^m$ number of $m$-qubit controlled rotations as shown in Fig.~\ref{fig:sp_ucr} (b). Each controlled rotation requires $O(m)$ circuit size~\cite{barenco1995elementary}, so the trivial decomposition results in state preparation circuit size $O(m2^m)$. \citet{Mottonen.05} then developed an improved UCR implementation, in which CNOT gates are set to match the Gray code of each bit string and interleaved with single-qubit rotations with appropriately chosen angles. See  Fig.~\ref{fig:sp_ucr} (d) for illustration. They achieved $O(2^m)$ circuit size and depth for each UCR, so the total circuit size/depth of state preparation is $O(2^n)$. \citet{shende2005synthesis} further introduced a recursive decomposition of UCR, showing that each $m$-qubit UCR can be decomposed into two $(m-1)$-qubit UCRs and two CNOT gates. Because some of the CNOTs can be merged during the recursive decomposition, the circuit sizes of UCR, and thus the total state preparation can be further reduced by a constant factor.  
\citet{sun2023asymptotically} further optimized the circuit depth of $F_\textbf{z}$ gates \footnote{\citet{sun2023asymptotically} considered diagonal matrix, which is equivalent to UCRs along z-axis $F_\textbf{z}$. UCR along different axes $\bm{\hat{n}}$ are interchangeable by two single-qubit rotations.} to $\Theta(2^m/m)$, thus achieving an optimal depth
 $\Theta(2^n/n)$ for quantum state preparation.

 \subsubsection{Schmidt decomposition methods}
Apart from using UCR,~\citet{Plesch.11} developed a method based on Schmidt decomposition. For even $n$, the target state is decomposed as
\begin{align}\label{eq:sp_pb}
|\psi\rangle=\sum_{j=0}^{2^{n/2}-1}\sigma_j|\tau_j\rangle|v_j\rangle,
\end{align}
where $\langle\tau_j|\tau_k\rangle=\langle v_j|v_k\rangle=\delta_{jk}$. 
It can be constructed by one $(n/2)$-qubit state preparation, one layer of CNOT gates, and two $(n/2)$-qubit general unitary synthesis. By applying this decomposition iteratively, they achieve the leading order of CNOT count $\frac{23}{24}2^n$ for even $n$ which is lower than $2^n$ for the method in~\cite{shende2005synthesis}. Based on improved unitary synthesis method~\cite{krol2024beyond}, \citet{li2026reducing} further improved the leading order CNOT count to $\frac{11}{12}2^n$. Yet, there is still a gap to the best-known lower bound $\frac{1}{2}2^n$~\cite{Plesch.11}.

\subsection{Space-time tradeoff \label{sec:qsp_st}}
While the gate count lower bound of preparing an arbitrary quantum state is exponential with respect to $n$, one can still  trade time for space. In other words, circuit depth can be reduced by introducing ancillary qubits. 
As an example, with $O(N)$ ancillary qubits, one can prepare a quantum state $\sum_{j=0}^{N-1}\alpha_j|e_j\rangle$ with $O(n)$ circuit depth, where $|e_j\rangle$ is the unary encoding of $j$ consisting of $N$ qubits~\cite{johri2021nearest}. The unary encoding $e_j$ can then be converted to the binary encoding $j$ with circuit depth $O(n)$~\cite{sun2023asymptotically}, yielding an asymptotically optimal state-preparation circuit depth of $\Theta(n)$.

These low-depth state preparation methods require geometrically non-local gates between qubits. Otherwise the circuit depth is restricted to be $\Omega(\text{poly}(N))$, which can be verified by the lightcone argument. In practice, these non-local operations can be realized by teleportated CNOT gate assisted by preshared entangling photon pairs~\cite{Gottesman.99,Chou.18,Wan.19}. Alternatively, they can also be realized by shuttling for trapped-ion-based or neutral atom-based platforms~\cite{Moses.23,Bluvstein.24}. In surface-code-based fault-tolerant quantum computers~\cite{Fowler2012Surface}, non-local operations can also be realized in constant depth by reshaping the logic qubit patches~\cite{litinski2019game,zhang2022quantum}. 

Below, we introduce some other typical approaches achieving low circuit depth using ancillary qubits. Each approach is grounded in distinct frameworks, offering different advantages.  

\subsubsection{Binary tree approaches}\label{sec:sp_bta}
In ancillary-free state preparation, each UCR is equivalent to a total of $2^m$ controlled rotations with different angles $\theta_{m,j}$. These rotations should be realized in a sequential way due to limited space, even with optimized decomposition methods~\cite{Mottonen.05,sun2023asymptotically}. The main idea of binary tree approaches is to introduce $O(N)$ extra qubits arranged as one or many binary trees. The information of $\theta_{m,j}$ for all $j$ can be effectively injected into the quantum state by a single layer of controlled rotations, which are applied on the same (spatial) layer of the binary tree. In this way, one can prepare an arbitrary quantum state 
\begin{align}
|\Psi\rangle=\sum_{j=0}^{N-1}\alpha_j|0\rangle|\Psi_j\rangle_{\text{anc}}
\end{align} with low circuit depth, where $|\Psi_j\rangle$ is some computational basis of the binary trees. We then perform the computational basis mapping $|0\rangle|\Psi_j\rangle_{\text{anc}}\rightarrow|j\rangle|\Psi_j\rangle_{\text{anc}}$ and uncompute the binary tree $|j\rangle|\Psi_j\rangle_{\text{anc}}\rightarrow|j\rangle|0\cdots0\rangle_{\text{anc}}$. In this way, the target state $|\psi\rangle$ is obtained. 

\begin{figure}[t]
    \centering
          \includegraphics[width=1\columnwidth]{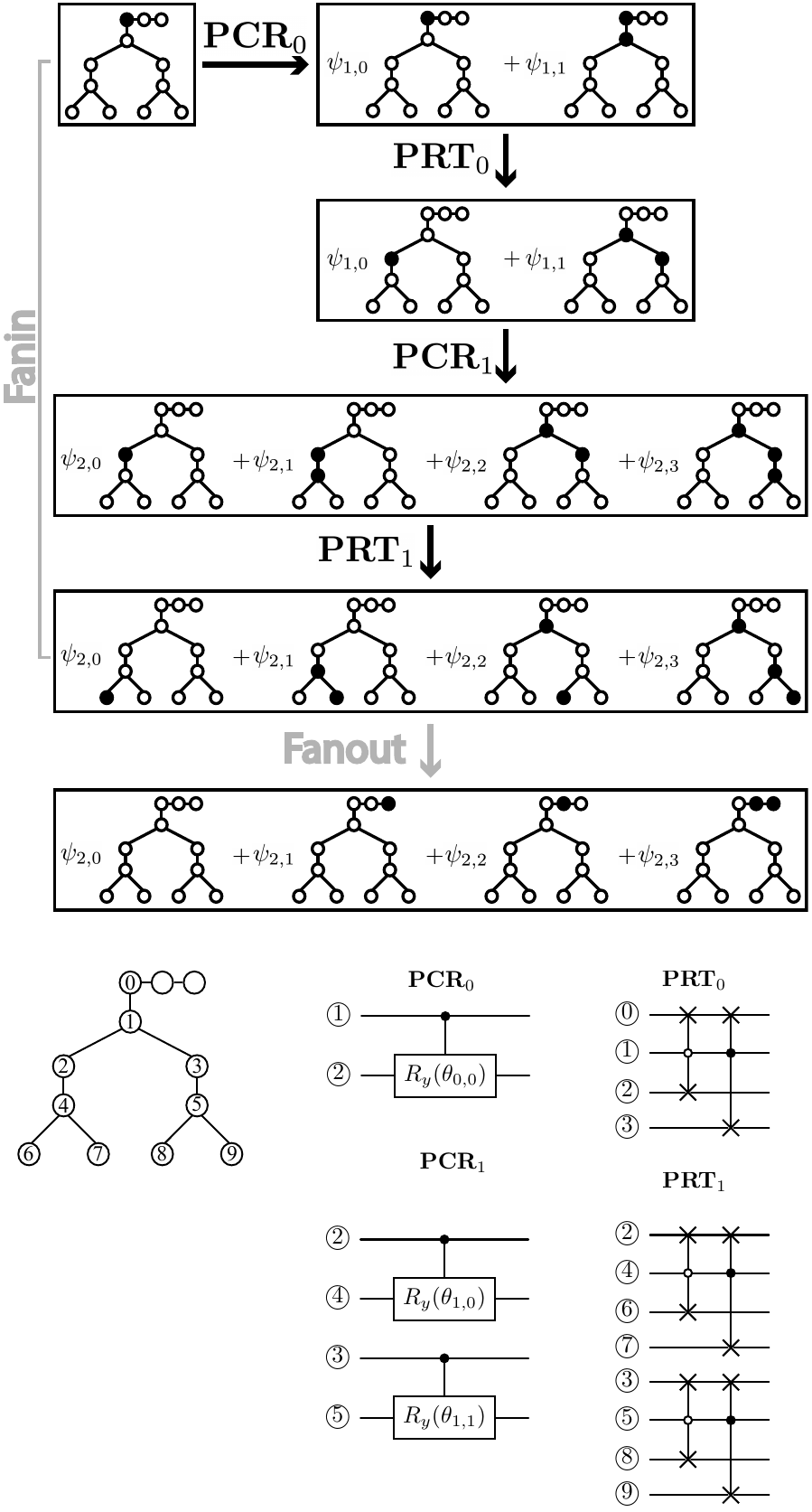}
       \caption{Sketch of the binary-tree approach~\cite{zhang2024practical} for low-depth quantum state preparation for $n=2$. Each circle represents a qubit, two qubit gates are only applied on qubit pairs connected by lines. Hollow circles represent qubits at state $|0\rangle$, and black circle represents qubits at state $|1\rangle$.} \label{fig:1_2}
\end{figure}

\citet{zhang2022quantum} developed a binary-tree-based method with $\Theta(n)$ circuit depth and $O(N)$ ancillary qubits. The qubits are connected as a total of $n$ binary trees.
\citet{zhang2024practical} further simplifies the algorithm by using only single binary tree. A sketch of the algorithm is illustrated in Fig.~\ref{fig:1_2} for $2$-qubit case. The hardware architecture is identical to the qubit-based bucket-brigade QRAM~\cite{Hann.21}, and $|\Psi\rangle_{\text{anc}}$ is also identical to the intermediate state of QRAM. To prepare $|\Psi\rangle_{\text{anc}}$ from all-zero initial state, parallel controlled rotations and parallel routing are implemented alternately. Finally, the mapping $|j\rangle|\Psi\rangle_{\text{anc}}\rightarrow|j\rangle|0\cdots0\rangle_{\text{anc}}$ can be realized by the fanout process of QRAM. Fig.~\ref{fig:1_2} demonstrates the simple 2-qubit-per-node version, which already achieves optimal circuit depth $\Theta(n)$ with single-qubit and CNOT gate. However, its Clifford$+T$ circuit complexity is suboptimal. In the improved 3-qubit-per-node version, the Clifford$+T$ gate count and depth are reduced to $O(N\log(1/\varepsilon))$ and $O(n+\log(1/\varepsilon))$ respectively.

One advantage of the binary tree approaches is that they require fewer non-local operations. In~\cite{zhang2024practical}, instead of the conventional all-to-all connectivity, each qubit only needs to be connected to 3 other qubits, thus minimizing the cost of non-local operations. Another advantage of the binary tree approach is the noise robustness. The bucket-brigade architecture can naturally block the error  propagation between different branches~\cite{arunachalam2015robustness,Hann.21}, so local noise can only affect a small portion of the final state computational bases. In particular, for local depolarizing noise, the infidelity of state preparation can be suppressed to $O(\varepsilon n^2)$, exponentially lower than the scaling $O(\varepsilon N)$ for conventional methods.

The methods above are extreme in which the circuit depth is minimized to $\Theta(n)$ by using $O(N)$ ancillary qubits. In case the number of ancillary qubits is restricted, one can combine the binary tree approaches with ancillary-free approaches in Sec.~\ref{sec:qsp_without}. In this way, one can achieve intermediate circuit depth under an arbitrary number of ancillary qubits~\cite{zhang2024circuit}.  

\subsubsection{Select-swap approaches}\label{sec:sp_ss}
State preparation can be reduced to the construction of some Boolean oracles (see also Sec.~\ref{sec:gra} below). 
Suppose we can construct data-lookup oracle, i.e. a unitary satisfying  
\begin{align}\label{eq:sp_Ox}
O|x\rangle|0\rangle|0\rangle=|x\rangle|a_x\rangle|\text{garb}_x\rangle,
\end{align}
 where $a_x$ is a $d$-digit bitstring and $|\text{garb}_x\rangle$ can be any garbage state in computational basis. Introducing one extra qubit, applying the following controlled rotation conditioned on $|a_x\rangle$,
\[
\Qcircuit @C=.5em @R=0em @!R {
\lstick{|a_{x,0}\rangle}&\qw& \ctrl{4} & \qw& \qw& \qw& \qw& \qw\\
\lstick{|a_{x,1}\rangle}&\qw& \qw & \ctrl{3}& \qw& \qw& \qw& \qw\\
&&&&\cdots \\
\lstick{|a_{x,d}\rangle}&\qw& \qw &\qw&\qw &\qw& \ctrl{1}& \qw\\
\lstick{|0\rangle}&\qw& \gate{R_y\left(\frac{a_{x,0}\pi}{2^1}\right)} &\gate{R_y\left(\frac{a_{x,1}\pi}{2^2}\right)}  &\qw &\qw&\gate{R_y\left(\frac{a_{x,d}\pi}{2^d}\right)}&\qw  
}
\]
we have
 \begin{align}
&|x\rangle|a_x\rangle|\text{garb}_x\rangle|0\rangle\longrightarrow\notag\\
&|x\rangle|a_x\rangle|\text{garb}_x\rangle\left(\cos \frac{a_x\pi}{2^{d+1}}|0\rangle+\sin \frac{a_x\pi}{2^{d+1}}|1\rangle\right).
\end{align}
By applying $O^\dag$, one can reset  $|a_x\rangle|\text{garb}_x\rangle$ to all-zero state. Comparing to the iterative preparation idea in Sec.~\ref{sec:qsp_ucr}, if we set $a_x\pi/2^d$ as a digital approximation of $\theta_{m_j}$, we can well approximate the iteration $|\psi_m\rangle\rightarrow|\psi_{m+1}\rangle$. 

\begin{figure}[t]
    \centering
          \includegraphics[width=1\columnwidth]{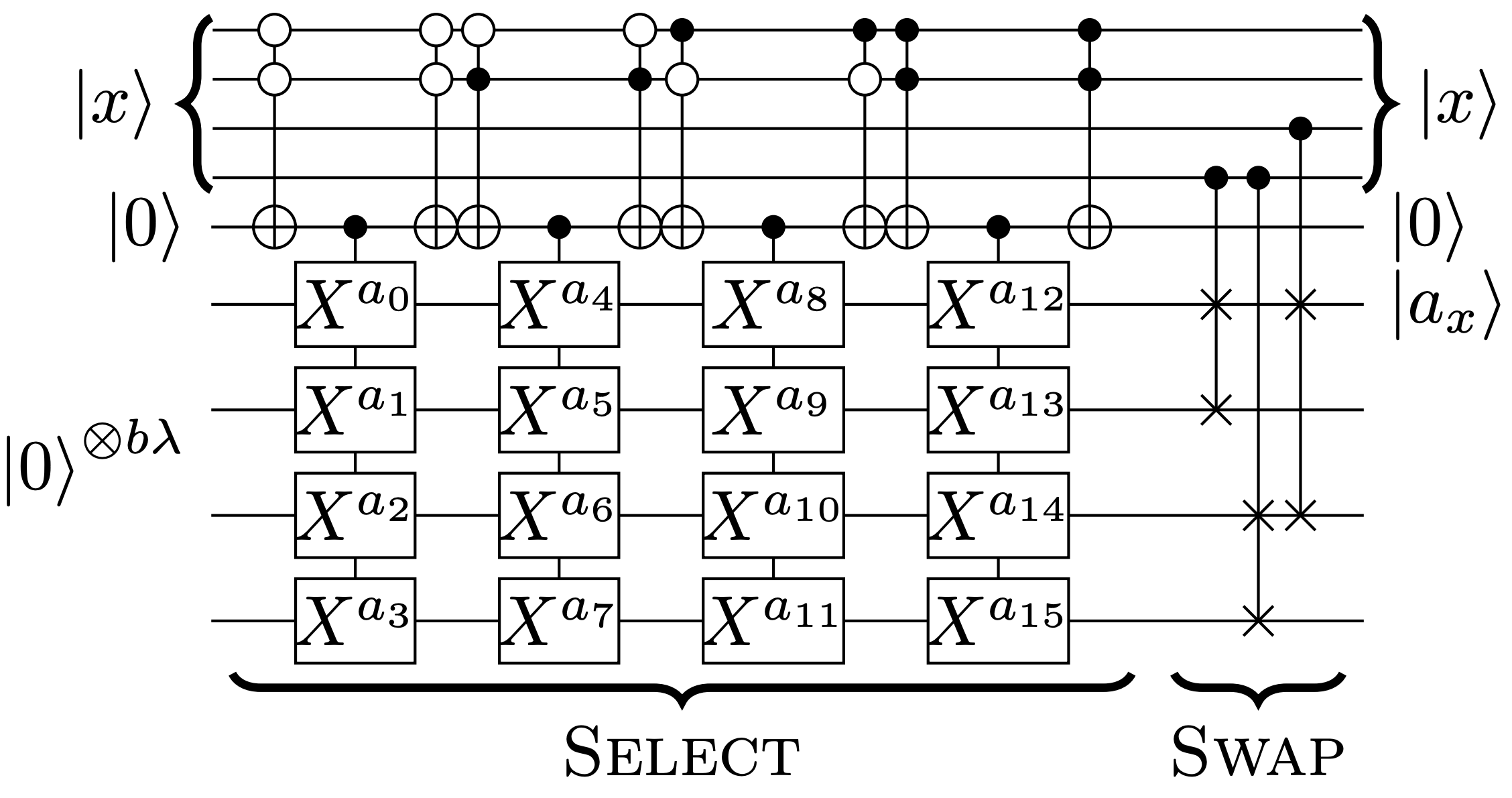}
       \caption{Example of the select-swap approach for $O|x\rangle|0\rangle|0\rangle=|x\rangle|a_x\rangle|\text{garb}_x\rangle$ with $N=16$ and $\lambda=4$. Adapted from~\cite{low2024trading}. $a_{x}=a_{x,0}a_{x,1}\cdots a_{x,b-1}$ is a $b$-digit bitstring, and the corresponding Pauli string $X^{a_{x}}$ is defined as  $X^{a_{x}}=X^{a_{x,0}}\otimes X^{a_{x,1}}\otimes\cdots\otimes X^{a_{x,b-1}}$.} \label{fig:1_ss}
\end{figure}

Select-swap approach enables one to perform a space-time trade-off realization of Eq.~\eqref{eq:sp_Ox}. It is first appeared as a preprint of~\cite{low2024trading} in 2018. In this scheme, bit string $x$ is divided into prefix $x_\text{p}$ and suffix $x_\text{s}$, i.e. $x=x_\text{p}x_\text{s}$. The number of all possible suffixes, denoted as $\lambda\equiv\max(x_\text{s})+1$, controls the circuit depth and ancillary qubit number. The system contains an $n$-qubit control register denoted by $|\cdot\rangle_c$, and a total of $\lambda$ target registers, with the $j$th target register denoted as $|\cdot\rangle_{j}$. For input state $|x_px_s\rangle_{\text{c}}|0\rangle_{0}|0\rangle_{1}\cdots|0\rangle_{\lambda-1}$, the SELECT process creates quantum state $|x_px_s\rangle_{\text{c}}|a_{x_p0}\rangle_{0}|a_{x_p1}\rangle_{1}\cdots|a_{x_p b-1}\rangle_{\lambda-1}$. In the SWAP process, conditioned on the suffix $x_s$ of the control register, one swaps the $0$th target register with the $x_s$th target register. Then, the quantum state becomes $|x_px_s\rangle_{c}|a_{x_px_s}\rangle_0|\text{garb}_{x}\rangle_{1,2,\cdots,b-1}$ as expected. The garbage state $|\text{garb}_{x}\rangle_{1,2,\cdots,b-1}$ for all target registers with label $j\geq1$ is not important for the algorithm. Note that the algorithm works for input in superposition of different $|x_{p}x_s\rangle_c$.

During the SELECT process, for each prefix $x_p$, the preparation of all target registers (for all $x_s\in[0,\lambda-1]$) can be realized in parallel using logarithmic circuit depth. One should just enumerate all prefixes $x_p$ sequentially, so the circuit depth for SELECT is $O(\log(\lambda)N/\lambda)$. Moreover, the SWAP process can be realized by logarithmic circuit depth without ancillae.  Therefore, the circuit depth and ancillary qubit number can be tuned by choosing different $\lambda$. 

In the extreme case, the select-swap approach can achieve circuit depth $\tilde{O}(n^2)$ with $\tilde{O}(N)$ ancillary qubits, where $\tilde{O}$ hides logarithmic factors. Although it is quadratically worse than other approaches in this section, the key advantage of the select-swap approach is that it has much lower $T$ count/depth. In particular, by choosing an intermediate index $\lambda$, one can achieve a nearly optimal $T$ count $O(\sqrt{N n\log(n/\varepsilon)}+\log^2(n/\varepsilon))$, which is quadratically better than other methods. The $T$ count/depth can be further improved by combining with the prerotation~\cite{Clader.22,gui2024spacetime} or linear combination of unitaries~\cite{gosset2026quantum} techniques. See Sec.~\ref{sec:sp_ft} for further discussions.

\subsubsection{Parallel diagonal unitary approach}

As explained in Sec.~\ref{sec:qsp_without}, state preparation can be reduced to the construction of UCR. \citet{sun2023asymptotically} introduces a family of quantum circuit to construct UCR along z-axis, $F_{\bm{z}}$, under different number of ancillary qubits $m$. Note that $F_{\bm{z}}$ is equivalent to diagonal unitaries, and their constructions are also of independent interest. 
The idea is to make phase shifts in Fourier basis. Assisted with Gray code, this can be realized in parallel by careful use of ancillary qubits. Their method is flexible, and for any ancillary qubit numbers $m\in[0,2^n/n\log n]$, optimal circuit depth $\Theta(n+N/(n+m))$ can be achieved. 

\subsubsection{Other approaches}\label{sec:qsp_st_others}
There are other low-depth state preparation approaches using ancillary qubits, e.g.~\cite{zhao2019state,PhysRevResearch3043200,Clader.22,gui2024spacetime,zhang2022quantum,ashhab2022quantum}.  For example, \citet{PhysRevResearch3043200} developed a measurement-based probabilistic approach with $O(n^2)$ circuit depth. The approach is based on the idea in~\cite{araujo2021divide} that preparing quantum states entangled with ancillary qubits. This unwanted entanglement is removed by measurement and post-selections.

\citet{rosenthal2023quantum} showed that arbitrary quantum states can be constructed by $\mathsf{QAC}^{0}_\text{f}$ circuit. Here, $\mathsf{QAC}_\text{f}$ represents quantum circuits with single-qubit gates, generalized Toffoli gates and fanout gates as elementary operations, and $\mathsf{QAC}^0_\text{f}$ is a constant-depth $\mathsf{QAC}_\text{f}$ circuit. Because $\mathsf{QAC}^0_\text{f}$ can be simulated by conventional quantum circuit with $O(n)$ circuit depth and exponential ancillary qubits, this result implies an optimal circuit depth of quantum state preparation. While~\cite{rosenthal2023quantum} uses $O(n2^n)$ ancillary qubits, this space complexity has been further improved by~\cite{Yuan.22}, which also allows a flexible number of ancillary qubits. Specifically, they achieve optimal circuit depth $\Theta(n+N/(n+m))$ for any ancillary qubit number $m$.

In some practical applications, it is acceptable to have a garbage state in an ancillary register, i.e. 
\begin{align}\label{eq:sp_alias}
|\psi'\rangle=\sum_{j=0}^{L-1}\alpha_j|j\rangle|{\rm garb}_{j}\rangle_{\rm{anc}}
\end{align}
for some integer $L$. 
For instance, in LCU-based block-encoding, it suffices to prepare quantum state such that $\langle\psi'|\left(\langle j\rangle\langle j|\otimes I_{\rm{anc}}\right)|\psi' \rangle=w_j$ for some positive target weights $w_j=\sqrt{\alpha_j}$. A commonly used technique for Eq.~\eqref{eq:sp_alias} is the Alias sampling proposed in~\cite{babbush2018encoding}. The idea is to prepare a uniform state $\frac{1}{\sqrt{L}}\sum_{j=0}^{L-1}|j\rangle|{\rm alt}_{j}\rangle|{\rm keep}_{j}\rangle$ with bitstrings $|{\rm alt}_{j}\rangle$ and $|{\rm keep}_{j}\rangle$. Then, conditional swaps are applied between $|j\rangle$ and alternative index $|{\rm alt}_{j}\rangle$ with probability depending on $|{\rm keep}_{j}\rangle$. Both $|{\rm alt}_{j}\rangle$ and $|{\rm keep}_{j}\rangle$ can be loaded from precomputed quantum read only memory (QROM), and the leading order $T$ count of this approach is $4L+O(\log(1/\varepsilon))$.

\subsection{Query complexity \label{sec:query}}
In the query model, except for elementary quantum gates, one is allowed to access extra \textit{oracles} representing Boolean functions. Oracles in quantum computation can be formalized as follows~\cite{Nielsen.02,aaronson2016complexityquantumstatestransformations}. For Boolean function $f:\{0,1\}^{m}\rightarrow\{0,1\}$, we define an $(m+1)$-qubit unitary operation satisfying 
\begin{align}\label{eq:sp_uf}
U_f|x\rangle|b\rangle=|x\rangle|b\oplus f(x)\rangle,
\end{align}
 where $x$ is an $m$-bit string and $b$ is a single bit, and $\oplus$ represents XOR. Note that the input of the quantum oracle is allowed to be in a superposition of different basis. An alternative definition is the $m$-qubit unitary $V_f$,
 \begin{align}\label{eq:sp_vf}
 V_f|x\rangle=(-1)^{f(x)}|x\rangle,
 \end{align}
which is equivalent to $U_f$ by setting $|b\rangle=|-\rangle$. 

The complexity captured by black-box query models can differ substantially from that of explicit, oracle-free computation models, because query models neglect the cost of implementing the oracles, which may not be realizable by polynomial-size quantum circuits. Yet, the query model has its value, and is widely studied in theoretical computer science. An important reason is that query model makes it simpler to provide nontrivial lower bounds and complexity separations~\cite{bernstein1993quantum,watrous2000succinct,aaronson2010bqp,raz2022oracle}. For example, there exists an oracle $A$ such that $\mathsf{BQP}^A\nsubseteq\mathsf{BPP}^A$~\footnote{$\mathsf{BQP}^A$ can be considered as problems efficiently solvable by  quantum circuits $C$ in the form of $C = C_K Q_K \cdots C_1 Q_1 C_0$, where each $C_k$ is a polynomial-size quantum circuit and each $Q_k \in \{A, A^\dagger\}$. The meaning of $\mathsf{BPP}^A$ is similar, with quantum circuit replaced by random Boolean circuit. See~\cite{aaronson2016complexityquantumstatestransformations} for formal definitions.}~\cite{bernstein1993quantum}.

This and other oracle-based separation results give us evidence that quantum computing can provide exponential speedup against the classical one, even though the natural separation between $\mathsf{BQP}$ and $\mathsf{BPP}$ remains an open question.

\subsubsection{Black-box state preparation}\label{sec:sp_bbsp}
In the black-box state preparation framework, we are given the query access to a binary function $f(x)$
\begin{align}\label{eq:sp_bb}
O_{\text{bb}}|x_j\rangle|0\cdots0\rangle=|x_j\rangle|f(x_j)\rangle,
\end{align}
where $f(x_j)$ is a $b$-bit binary representation of the amplitude associated with $x_j$.
The goal is to prepare the state $|\psi_f\rangle
=
\frac{1}{\mathcal N}\sum_{j=0}^{N-1} f(x_j)\,|x_j\rangle,$ where $\mathcal N =\left(\sum_{j=0}^{N-1}|f(x_j)|^2\right)^{1/2},$
whose amplitudes are proportional to the values returned by the oracle. ~\citet{Grover.01} proposed a black-box state-preparation procedure, which can be viewed as a generalization of unstructured quantum search~\cite{grover1996fast,grover1997quantum}, using $O(\sqrt{N}/\mathcal N)$ queries to $O_{\mathrm{bb}}$. The algorithm employs an $n$-qubit output register, a $b$-qubit data register, and a flag qubit. Specifically, one first prepares the output register in the uniform superposition $N^{-1/2}\sum_{j=0}^{N-1}|j\rangle_{\mathrm{out}}$, then queries $O_{\mathrm{bb}}$ to write $\widetilde f(x_j)$ into the data register. A controlled single-qubit rotation subsequently transfers this encoded value to the amplitude of a designated good subspace of the flag qubit. After $O(\sqrt{N}/\mathcal N)$ rounds of amplitude amplification, the data register is uncomputed by applying $O_{\mathrm{bb}}^\dagger$, and measuring the flag projects the output register onto a state proportional to $\sum_j f(x_j)|x_j\rangle$ with a constant success probability. Note that $\mathcal{N}$ is polynomial for certain functions (e.g., continuous functions), as shown in~\cite{Rattew.22}.

Methods above require coherent arithmetic to calculate rotation angles $\theta_{j}$.  To avoid this arithmetic, \citet{sanders2019black} replace the controlled rotation by a COMPARE subroutine $|a\rangle|b\rangle|0\rangle\rightarrow|a\rangle|b\rangle|c\rangle$, where the one-bit state $c=0$ when $a<b$, or $c=1$ otherwise. While the query complexity is the same, the actual gate count is reduced by about two orders of magnitude. Further improvement has been made based on gradient state~\cite{bausch2022fast}, LCU~\cite{wang2021fast}, and quantum singular value transformation~\cite{laneve2023robust}.  
Moreover, \cite{hamoudi2022preparing} improve the black-box preparation of many copies of one quantum state. To prepare $K$ copies of the target state, they improve the query complexity from $O(K\sqrt{N})$ to $\Theta(\sqrt{KN})$. 

Here, $O_{\text{bb}}$ is treated as a black-box, and $O(\sqrt{N})$ is the worst-case performance. In practice, the circuit complexity  depends strongly on the access model, i.e. how $O_{\text{bb}}$ is realized. For some structured data such as continuous and efficiently computable functions, the circuit complexity can be made significantly lower, as will be discussed in Sec.~\ref{sec:sparse}.

\subsubsection{Grover-Rudolph algorithm}\label{sec:gra}
The oracle in Eq.~\eqref{eq:sp_bb} encodes only the information of amplitudes. In contrast, the main idea of Grover-Rudolph algorithm~\cite{Grover.01,aaronson2016complexityquantumstatestransformations} is to query the rotation angles according to relative integral in adjacent intervals, namely $\theta_{m,k}$ in Eq.~\eqref{eq:sp_ang} from Sec.~\ref{sec:qsp_without}. 
At the $m$th step, we introduce a Boolean function $f_{\bm{\theta}_m}(k)=\tilde{\theta}_{m,k}$, where $\tilde{\theta}_{m,k}$ is a $d$-digit approximation of the rotation angle $\theta_{m,k}$. Introducing a $d$-digit ancillary register, one can first query $U_{f_{\bm{\theta}_m}}$, and obtain 
\begin{align}
U_{f_{\bm{\theta}_m}}|k\rangle|0\cdots0\rangle_{\text{anc}}|0\rangle=|k\rangle|\tilde{\theta}_{m,k}\rangle_{\text{anc}}|0\rangle.
\end{align}
Similar to Sec.~\ref{sec:sp_ss}, performing controlled rotations and then applying $U_{f_{\bm{\theta}_m}}$ again gives the transformation $|\psi_{m}\rangle\rightarrow|\psi_{m+1}\rangle$.
By repeating the process above iteratively, quantum state preparation can be realized, with exponentially small errors and $O(n)$ query complexity. Yet, we note again that the total circuit complexity relies of the implementation of the oracle, which can be exponential in the worst case.

\subsubsection{Minimum query}\label{sec:sp_mq}

We have discussed quantum state preparation in the black-box query model. A natural question is then: what is the minimum number of queries required for state preparation, when the form of oracle is arbitrary?

Surprisingly, \citet{Irani2022Quantum} showed that a single query is sufficient to achieve polynomially small error, and two queries are sufficient for exponentially small errors in state preparation. However, the construction in~\cite{Irani2022Quantum} requires an exponential number of extra quantum gates. 

\citet{rosenthal2024efficient} further showed that a constant number of queries and polynomial-size quantum circuit are sufficient to approximate quantum states with arbitrarily small error. The idea is based on the Clifford times phase state (CTPS) and linear combination of unitaries. Let $C\cdot2^{-n/2}\sum_{x\in\{0,1\}^n}\pm|x\rangle$ be the CTPS, where $C$ represents some Clifford unitary. An important property is that any quantum state can be approximated by CTPS with $\Omega(1)$ fidelity~\cite{Irani2022Quantum}. One can recursively define $|\phi_k\rangle$ as a CTPS that is close to $$\frac{|\psi\rangle-\sum_{j=0}^{k-1}c_j|\phi_j\rangle}{\||\psi\rangle-\sum_{j=0}^{k-1}c_j|\phi_j\rangle\|}.$$
By choosing coefficients $c_j$ appropriately, the target state $|\psi\rangle$ can be well approximated by $\sum_{j=0}^{k-1}c_j|\phi_j\rangle$ with exponentially small error.  Furthermore, state $\sum_{j=0}^{k-1}c_j|\phi_j\rangle$ can be constructed in two steps. In the first step, all CTPS $|\phi_j\rangle$ are prepared in parallel. Each $|\phi_j\rangle$ can be constructed with polynomial-size quantum circuit and single query to $V_{f_j}$ for some Boolean function $f_j$. Because $V_{f_1}|x_1\rangle\otimes V_{f_2}|x_2\rangle=V_{f_1\oplus f_2}|x_1x_2\rangle$, querying different oracles $V_{f_j}$ in parallel is equivalent to querying single oracle $V_f$, where $f(x_1\cdots x_k)=\bigoplus_{j=1}^kf_j(x_j)$. Therefore, the first step requires single query to a Boolean function. In the second step, $\sum_{j=0}^{k-1}c_j|\phi_j\rangle$ can be constructed by linear combination of unitaries~\cite{bernstein1993quantum,childs2012hamiltonian}. In this way, with single query (or four queries) and some extra quantum gates, one can non-cleanly (or cleanly) approximate arbitrary quantum state to  accuracy $\varepsilon=1/\text{poly}(n)$ (or $\varepsilon=1/\text{exp}(n)$)\footnote{Here, clean means that all ancillary qubits return to the states $|0\rangle$; non-clean means that ancillary qubits end up with some nontrivial states that are not entangled with $|\psi\rangle$.}. 

Moreover, if all ancillary qubits are required to return to $|0\rangle$ exactly,~\cite{rosenthal2024efficient} showed that ten queries suffices. This result has been improved to two queries~\cite{gosset2026quantum}, whose construction is also useful for oracle-free fault-tolerant state preparation (see Sec.~\ref{sec:sp_ft}).

\subsection{Fault-tolerant preparation \label{sec:sp_ft}}
As discussed above, in fault-tolerant quantum computing, elementary gate sets $\mathcal{G}$ are typically discrete, and all single-qubit rotations should be further decomposed with such a discrete gate set. A commonly used gate set is the two-qubit Clifford gates and the single-qubit $T$ gate, as they are suitable for quantum error correction. With this gate set, circuit size $O(\log(1/\varepsilon_{\text{sig}}))$ is sufficient for achieving accuracy $\varepsilon_{\text{sig}}$. See also Sec.~\ref{sec:u_ds} for detailed discussions. 
To achieve state preparation accuracy $\varepsilon$, the worst-case evaluation indicates that the gate error for each single-qubit rotation should be suppressed to $\varepsilon_{\text{sig}}=O(\varepsilon/N)$, resulting in an extra $O(n)$ depth overhead for every single qubit rotations. Fortunately, the error scaling can be reduced by more careful analysis. 

We begin with the UCR-based method in Sec.~\ref{sec:qsp_without}.  With decomposition in Fig.~\ref{fig:sp_ucr}(b), each single-qubit rotation applies nontrivially only on two computational basis, so the error of each $R_{y}(\theta_{j,k})$ only affect two basis. It can therefore be verified that the total error for UCR is of the same order of $\varepsilon_{\text{sig}}$. Thus, the total Clifford$+T$ count is $O(N\log(n/\varepsilon))$, as opposed to the trivial estimation  $O(N\log(N/\varepsilon))$~\cite{zhang2024circuit}.
One can further optimize the distribution of $\varepsilon_{\text{sig}}$ over different UCRs, i.e.  allow larger $\varepsilon_{\text{sig}}$ for $F_{\bm{z}}(\bm{\phi}_m)$ and $F_{\bm{y}}(\bm{\theta}_m)$ with larger $m$. In this way, the total gate count can be reduced to $O(N\log(1/\varepsilon))$~\cite{zhang2024circuit}. 
Similar techniques are also applicable for other binary tree approaches in Sec.~\ref{sec:sp_bta}. In particular, the Clifford+$T$ count $O(N\log(1/\varepsilon))$ is achievable for different space-time trade-offs~\cite{zhang2024circuit}. The circuit depth $O(n+\log(1/\varepsilon))$ is achievable with $O(N)$ ancillary qubits~\cite{zhang2024circuit,gui2024spacetime}.

When unlimited ancillary qubits are allowed, the best-known Clifford$+T$ count is due to~\cite{rosenthal2024efficient}. Their construction employs an oracle-based method that uses a polynomial-size quantum circuit, and ten queries to an oracle which has $O(n + \log\log(1/\varepsilon))$ input qubits and a single output qubit.
According to the Lupanov's method~\cite{Lupanov.58,jukna2012boolean}, a Boolean function $f: \{0,1\}^{m}\rightarrow \{0,1\}$ can be constructed by $O(2^m/m)$-size Boolean circuit. Because each Boolean circuit can be ideally simulated by quantum Toffoli and NOT gates with constant overhead, this leads to a natural state preparation algorithm with $$O(N\log(1/\varepsilon)/n)$$ Clifford+$T$ count. This result saturates the lower bound in Eq.~\eqref{eq:sp_ganc}.  Yet, the construction in~\cite{rosenthal2024efficient} requires an exponential number of ancillary qubits, so it remains an open question whether the lower bound in Eq.~\eqref{eq:sp_gd1} and Eq.~\eqref{eq:sp_ganc} can be achieved when the number of ancillary qubits is limited. 

The select-swap approach~\cite{low2024trading} in Sec.~\ref{sec:sp_ss} is friendly to $T$ count/depth, as it is free of single-qubit rotations. For parameter $\lambda'\in[\log(n/\varepsilon),N\log(n/\varepsilon)]$, the $T$ count and $T$ depth of state preparation are $O(\lambda' (n+\log (1/\varepsilon))+N/\lambda'\log(n/\varepsilon))$, and $O(N/\lambda'\log(n/\varepsilon)+n\log(\lambda' n/\varepsilon))$ respectively. Specifically, by choosing an intermediate index $\lambda'$, one can achieve a nearly-optimal $T$ count $O(\sqrt{Nn\log(n/\varepsilon)}+\log^2(n/\varepsilon))$.~\citet{gosset2026quantum} further optimized the $T$ count to $$\Theta(\sqrt{N\log(1/\varepsilon)}+\log(1/\varepsilon))$$ by combining the select-swap approach with the LCU technique in~\cite{rosenthal2024efficient}. This result matches the fundamental lower bound in Eq.~\eqref{eq:sp_Tv}.

Regarding the $T$-depth, results in~\cite{low2024trading} scale at least quadratically with $n$. \citet{Clader.22} introduces a prerotation technique that significantly improves the $T$-depth to $\Theta(n+\log(1/\varepsilon))$, which is optimal.  
The idea is to encode all rotation angles in parallel prior to the state preparation. In subsequent process, controlled rotations with certain angles are enacted by swap gates. 
Note that methods in~\cite{Clader.22} require $O(n^2)$ depth of Clifford gates. Subsequent work has further improved the total Clifford+$T$ depth to $\Theta(n+\log(1/\varepsilon))$~\cite{gui2024spacetime}.

\subsection{Lower bound \label{sec:sp_lb}}
Given a specific gate set $\mathcal{G}$, we define $G_{\mathcal{S}}(|\psi\rangle,\varepsilon)$ as the minimum gate count required to prepare $|\psi\rangle$ with quantum gates in $\mathcal{G}$, up to infidelity $\varepsilon$. The lower bound of arbitrary quantum state preparation is defined as the maximal $G_{\mathcal{G}}(|\psi\rangle,\varepsilon)$ over all possible quantum states, i.e.
$G_{\mathcal{G}}(\varepsilon)\equiv\max_{|\psi\rangle}G_{\mathcal{G}}(|\psi\rangle,\varepsilon)$.

We first consider the circuit complexity lower bound when $\mathcal{G}$ is the gate set of arbitrary single- and two-qubit rotations, denoted as $G_{\text{sig}+\text{two}}(\varepsilon)$. To match the exponential degree of freedom of Eq.~\eqref{eq:sp}, we have~\cite{Mottonen.05}
\begin{align}\label{sec:Gst}
G_{\text{sig}+\text{two}}(0)=\Omega(2^n).
\end{align}
Eq.~\eqref{sec:Gst} also implies the lower bound for circuit depth. Specifically, the circuit depth without ancillary qubit is lower bounded by $\Omega(2^n/n)$ for ancillary-free case. Provided an infinite number of ancillae, the lower bound is $\Omega(n)$~\cite{sun2023asymptotically,zhang2022quantum}, which can be obtained by lightcone and counting argument. In the intermediate case, with $m$ ancillary qubits, the circuit depth lower bound is $\Omega(n+2^n/(n+m))$~\cite{sun2023asymptotically}, which has been proven to be tight by~\cite{Yuan.22} (see also Sec.~\ref{sec:qsp_st_others}).

Another practical scenario is when $\mathcal{G}$ is a discrete, finite gate set (as opposed to arbitrary single-qubit gates that can be varied continuously). This scenario includes the fault-tolerant quantum computing framework, in which only a finite number of discrete quantum gates can be realized fault-tolerantly, and single-qubit gates should be further decomposed~\cite{selinger2012efficient,ross2016optimal,kliuchnikov2014asymptotically}. 
In~\cite[Sec.~4.5.4]{Nielsen.02}, the lower bound is obtained via a covering argument on the space of quantum states. Specifically, the set of normalized state vectors in an $N$-dimensional Hilbert space can be identified with the unit sphere $S^{2N-1}\subset \mathbb{R}^{2N}$. For a fixed state, all states within distance $\varepsilon$ form a local $\varepsilon$-patch on this sphere. Since the volume of such a patch scales as $\varepsilon^{2N-1}$, covering the entire sphere requires at least $\Omega(\varepsilon^{-(2N-1)})$ patches. On the other hand, if one has a finite gate set with $g$ gate types, each acting on at most $f$ qubits, then an $m$-gate circuit can generate at most $O((n^f g)^m)$ distinct outputs from the reference state $|0\rangle^{\otimes n}$. Therefore, to approximate all $n$-qubit states to accuracy $\varepsilon$, one must have $O((n^f g)^m)\ge \Omega(\varepsilon^{-(2^{n+1}-1)})$, which implies $m=\Omega(2^n\log(1/\varepsilon)/\log n)$.
Let $G_{\text{discrete}}(\varepsilon)$ be the circuit size lower bound when $\mathcal{G}$ is an arbitrary finite gate set applied at constant number of qubits, and ancillary qubit is not allow,~\citet {Nielsen.02} shows that
\begin{align}\label{eq:sp_gd1}
G_{\text{discrete}}(\varepsilon)=\Omega\left(\frac{N\log(1/\varepsilon)}{\log n}\right)
\end{align}
A similar idea is also applicable when ancillary qubits are allowed.  
For example,~\citet{rosenthal2024efficient} showed that 
\begin{align}\label{eq:sp_ganc}
G^{(\text{anc})}_{\text{discrete}}(\varepsilon)=\Omega\left(\frac{N\log(1/\varepsilon)}{n}\right)
\end{align}
for $1/4\geqslant\varepsilon\geqslant\exp(-\text{poly}(n))$, where the definition of $G^{(\text{anc})}_{\text{discrete}}$ is similar to $G_{\text{discrete}}$ except that ancillary qubits are allowed.~\citet{rosenthal2024efficient} also proves that this lower bound is tight. Slightly worse lower bounds have also been obtained  in~\cite{zhang2024circuit,gui2024spacetime}. 
\citet{gui2024spacetime} also gives a lower bound $\Omega(\frac{N\log(1/\varepsilon)}{n+\log\log(1/\varepsilon)})$ for space-time allocation, i.e. the total time that each individual qubit must be not in state $|0\rangle$.

An important discrete gate set is the two-qubit Clifford$+T$ gates, because it is the elementary gate set for surface code~\cite{Fowler2012Surface}. 
An interesting result is that the $T$ count of quantum state preparation can be significantly lower than the total gate count. Specifically, with totally $q$ qubits (including system and ancillary qubits) and accuracy $\varepsilon$, \citet{low2024trading} gives a $T$ count lower bound
\begin{align}\label{eq:sp_Tvq}
T(\varepsilon,q)=\Omega\left(\frac{N\log(1/\varepsilon)}{q}-q\right).
\end{align}
The idea is as follows. A $q$-qubit Clifford$+T$ circuit with $\Gamma$ number of $T$ gates can always be represented as $C\cdot\prod_{j=1}^\Gamma e^{-i\pi P_j/8}$ for some Clifford gate $C$ and $q$-qubit Pauli string $P_j$~\cite{Gosset2014algorithm}. There are $2^{O(q^2)}$ possible choices of $C$ and $O(4^q)$ possible choices of each $P_j$. So there are at most $O(4^{q\Gamma+O(q^2)})$ unique quantum circuits.  Eq.~\eqref{eq:sp_Tvq} can then be obtained by comparing with the $\Omega(\varepsilon^{2N-1})$ lower bound for unique quantum circuits. 

\citet{low2024trading} also generalizes the lower bound for adaptive Clifford$+T$ circuit where measurement and adaptive feedback are allowed. They show that regardless of the number of ancillary qubits, there is a $T$ count lower bound $T(\varepsilon)=\Omega(\sqrt{N\log(1/\varepsilon)})$, which is quadratically lower than the total gate count in Eq.~\eqref{eq:sp_ganc}. 
\citet{gosset2026quantum} further improved this lower bound to 
\begin{align}\label{eq:sp_Tv}
T(\varepsilon)=\Omega\left(\sqrt{N\log(1/\varepsilon)}+\log(1/\varepsilon)\right)
\end{align}
 and show that Eq.~\eqref{eq:sp_Tv} is tight.

%% file: secs/unitary_synthesis.tex
\section{Universal unitary synthesis\label{sec:unitary}}

Given a classical description of a general $N$-dimensional unitary $U\in\mathrm{SU}(N)$, this section considers the synthesis of quantum circuits that realize $U$. We focus on the $n$-qubit setting, where $N=2^n$. 
Compared with the state-preparation task discussed in the previous section, unitary synthesis is more challenging because it involves specifying and implementing an entire $N\times N$ transformation rather than a single $N$-dimensional state.

This section is organized as follows. In Sec.~\ref{sec:unitary_single}, we begin with the synthesis of single-qubit unitaries, either with continuous (Sec.~\ref{sec:u_cs}) or discrete (Sec.~\ref{sec:u_ds}) elementary gate set. In Sec.~\ref{sec:unitary_without}, we consider the synthesis of general multi-qubit unitary and focus primarily on the case without ancillary qubits. This includes historical efforts in studying the universality decomposition, reduction of the circuit complexity with single-qubit and CNOT gates. In Sec.~\ref{sec:unitary_st},  we consider the space-time tradeoff, i.e.~reducing circuit depth with ancillary qubits. In Sec.~\ref{sec:unitary_query}, we discuss the query complexity of unitary synthesis. In Sec.~\ref{sec:rm}, we further introduce several techniques for resource minimization in fault-tolerant architecture. 

\subsection{Single-qubit unitaries\label{sec:unitary_single}}

We begin with single-qubit unitaries $U\in\mathrm{SU}(2)$, for which two synthesis settings are commonly considered. The first is ideal synthesis using a continuous set of rotations. This setting is relevant to physical qubits, where the available elementary gates depend on the experimental platform. The second is approximate synthesis over a discrete gate set. This setting is typical for error-corrected quantum computers, where only a finite set of elementary gates can be implemented fault-tolerantly.

\subsubsection{Ideal synthesis with continuous set}\label{sec:u_cs}
One of the typical scenarios is that we are allowed to perform rotations only along a pair of orthogonal axes. With Euler decomposition, three layers are always sufficient for decomposing an arbitrary $U\in\text{SU}(2)$. For example, the $z$-$y$-$z$ decomposition is
\begin{align}\label{eq:u_zyz}
U=R_z(\beta)R_y(\gamma)R_z(\lambda)
\end{align}
with appropriately chosen angles $\beta,\gamma,\lambda\in[0,2\pi)$.  
More generally, any pair of \textit{nonparallel} rotation axes is sufficient for decomposing arbitrary single-qubit unitaries,
i.e.~$U=R_{\bm{\hat n}}(\beta_1)R_{\bm{\hat m}}(\gamma_1)R_{\bm{\hat n}}(\beta_2)R_{\bm{\hat m}}(\gamma_2)\cdots$ with rotation axes ${\bm{\hat n}}\neq\pm{\bm{\hat m}}$. 
Specifically, let $q_{\min}(U)$ be the minimal circuit depth for target gate $U$, we define the minimal circuit depth for arbitrary single-qubit gate as $q_{\min}=\max\limits_{U\in\text{SU}(2)}q_{\min}(U)$. Suppose the angle between $\bm{\hat{n}}$ and $\bm{\hat{m}}$ is $\theta_{\rm axes}$. Then we have~\cite{lowenthal1971uniform}  
\begin{align}\label{eq:u_qm}
q_{\min}=\left\lceil\frac{\pi}{\theta_{\rm axes}}\right\rceil+1.
\end{align}
In the context of quantum computing, \citet{divincenzo2000universal} numerically discovered the special case $\theta_{\rm axes}=\pi/3$  of Eq.~\eqref{eq:u_qm}, and~\citet{hanson2007universal} discovered the special case $\theta_{\rm axes}\geq\pi/4$.  
Moreover, for any given single-qubit gate, $q_{\text{min}}(U)$ can be lower than the upper bound in Eq.~\eqref{eq:u_qm}. Two analytical expressions for $q_{\text{min}}(U)$ are given in~\cite{hamada2014minimum,zhang2019minimal}.

A more flexible scenario is when the rotation axes are also continuously changeable, i.e.~elementary operations are $\mathcal{G}_{\theta_{\rm axes}}=\{R_{\bm{\hat{n}}}(\phi)|\bm{\hat{n}}=(\sin\theta,0,\cos\theta),0\leq\theta\leq\theta_{\rm axes}\}$. A typical example is the unitaries generated by Landau-Zener-type Hamiltonian~\cite{landau1932theorie,zener1932non,greilich2009ultrafast,petta2005coherent}.~\cite{shim2013single} shows that  when rotation axes can be chosen freely in a plane, i.e.~$\mathcal{G}_\pi$, two layers are always sufficient.~\cite{zhang2019minimal} discusses  general case for all $0\leq\theta_{\rm axes}\leq\pi$, showing that the minimal number of layers is still given by Eq.~\eqref{eq:u_qm}. Specifically, for $0\leq\theta_{\rm axes}\leq\pi/2$, the optimal decomposition can be achieved with two rotation axes at the boundary of $\mathcal{G}_{\xi}$ (i.e.~with $\theta=0,\theta_{\rm axes}$). In other words, the flexibility of rotation axes will not reduce $q_{\text{min}}(U)$.

\subsubsection{Approximated synthesis with discrete set}\label{sec:u_ds}

We now turn to the second setting: approximate synthesis over a discrete finite gate set. We first consider the general case in which the gate set $\mathcal{G}$ is universal, meaning that any $U\in\mathrm{SU}(2)$ can be approximated to arbitrary accuracy $\varepsilon$ using gates from $\mathcal{G}$. Due to the well-known  Solovay-Kitaev theorem~\cite{kitaev1997quantum,kitaev2002classical,Nielsen.02}, whenever $\mathcal{G}$ is closed under inversion and its generated subgroup is dense in SU$(2)$, circuit depth $O(\log^{c}(1/\varepsilon))$ is always sufficient to achieve accuracy $\varepsilon$. Specifically, Appendix A3 of~\cite{Nielsen.02} achieved constant $c=3.97$, and Theorem 8.3 of~\cite{kitaev2002classical}  achieved $c=3+o(1)$. 

A special gate set that is commonly studied is $\mathcal{G}=\{H,T\}$, as it is suitable  for fault-tolerant quantum computing. 
The universality of $\{H,T\}$ is due to \citet{boykin1999universal}. They construct two rotations $R_{\hat{n}}(\alpha)$ and $R_{\hat{m}}(\alpha)$ using $\{H,T\}$, where $\alpha$ is an irrational rotation angle, and $\hat{n}\neq\pm\hat{m}$ is a pair of nonparallel rotation axes. The irrationality of $\alpha$ ensures that by repeating the rotations, one can approximate $R_{\hat{n}}(\theta)$ for any $\theta\in[0,2\pi)$ with $R_{\hat{n}}(\alpha m)$ and  $m\in\mathbb{Z}^+$. The nonparallel axes ensure that one can construct arbitrary $U\in\text{SU}(2)$, as discussed in Sec.~\ref{sec:u_cs}. 
Due to the Solovay-Kitaev theorem, universality directly implies that $O(\log^{c}(1/\varepsilon))$ circuit depth is sufficient. 

Further efforts have  been made for optimizing the circuit depth of $\{H,T\}$ decomposition~\cite{fowler2011constructing,kliuchnikov2012fast,selinger2012efficient,kliuchnikov2013asymptotically,Ross2015CV,kliuchnikov2015practical,paetznick2014repeat,bocharov2015efficient}.
For example,~\citet{kliuchnikov2012fast} pointed out that due to the property of $H$ and $T$ gates, a unitary has exact synthesis if and only if all its entries are in the ring 
\begin{align}
&\mathbb{Z}[1/\sqrt{2},i]\notag\\
\equiv&\Big\{\frac{a\omega^3+b\omega^2+c\omega+d}{(\sqrt{2})^k}\Big|k\in\mathbb{N},a,b,c,d\in\mathbb{Z},\omega=e^{i\pi/4}\Big\}.\notag
\end{align}
Based on this result, several synthesis algorithms achieving or close to the optimal scaling $O(\log(1/\varepsilon))$ have been proposed~\cite{selinger2012efficient,kliuchnikov2013asymptotically,Ross2015CV,kliuchnikov2015practical}.
In particular,~\cite{Ross2015CV} introduced an efficient synthesis algorithm for $R_z(\theta)$ with $T$ count $\approx 3.02\log_{2} (1/\epsilon )+1.77$, estimated by linear fit~\cite{kliuchnikov2023shorter}.  
This result is close to the information-theoretic bound for ancillary-free synthesis~\cite{selinger2012efficient}.
The idea is to approximate $R_z(\theta)$ with unitary
$\tilde{U}=\begin{pmatrix}u&-t^\dag\\t&u^\dag\end{pmatrix}$
for some $u,t\in\mathbb{Z}[1/\sqrt{2},i]$. This is realized by first finding a suitable candidate for $u\approx e^{i\theta/2}$, and then solving $t$ to ensure that $\tilde{U}$ is a unitary. Together with Eq.~\eqref{eq:u_zyz}, this approach can be used to approximate  universal SU$(2)$. 
Using a similar approach, one can also obtain nearly-optimal decomposition with other fault-tolerant gate sets, such as Pauli$+V$~\cite{blass2015optimal}, Clifford$+V$~\cite{Ross2015CV} or topological braiding with Fibonacci anyons~\cite{kliuchnikov2014asymptotically}. 

In quantum computing, minimizing $T$ gates or other non-Clifford gates plays a central role in the total resource frugal in the fault-tolerant setting. In Sec.~\ref{sec:rm}, we will further introduce other higher-level techniques for reducing the $T$ count for single-qubit rotations, as well as the multi-qubit quantum circuit.

\subsection{Ancilla-free multi-qubit unitary synthesis\label{sec:unitary_without}}
The discussion about the multi-qubit quantum gate synthesis dates back to 
\citet{deutsch1989quantum}, who showed that the following controlled-controlled rotation

\vspace{.1cm}
{\centering

\,\Qcircuit @C=1em @R=.7em {
& \ctrl{1} &\qw\\
& \ctrl{1} & \qw\\
& \gate{iR_x(\pi\alpha)} & \qw
\inputgrouph{1}{3}{1.1em}{D(\alpha)=}{2.2em}
}

}
\vspace{.2cm}

\noindent is universal when $\alpha$ is an irrational number. $D(\alpha)$ can be considered as a quantum generalization of the Toffoli gate, and is also known as the \textit{Deutsch gate}.
The main idea is as follows. By repeating $D(\alpha)$, one can construct $D(\alpha m)$ for all $m\in \mathbb{Z}^{+}$. Due to the irrationality of $\alpha$, one can approximate arbitrary $D(x)$ with $x\in \mathbb{R}$ to any accuracy $\varepsilon>0$. In particular, Toffoli gate corresponds to $D(1)$ and thus can be well-approximated. Combining Toffoli gate with $D(x)$, one can further construct controlled-controlled-rotations along two other orthogonal axes, and thus $\text{C}^2\text{-}U$ for arbitrary $U\in\text{SU}(2)$. Finally, a general $\text{SU}(N)$ can be constructed using $\text{C}^2\text{-}U$ and Toffoli gates.

\citet{divincenzo1995two} further proved the universality of two-qubit gates. Specifically, it is shown that $D(\lambda)$ with infinitesimal rotation angles $\lambda$ can be decomposed into two-qubit gates. 
The universality then follows directly from the Deutsch gate construction  above. \citet{Sleator1995realizable} further introduced the two-qubit decomposition of $D(x)$ for all $x\in\mathbb{R}$, thus simplifying the two-qubit construction.

Universal decompositions above are based on specific quantum gates. \citet{deutsch1995universality} and~\citet{lloyd1995almost} independently showed that \textit{almost all} two-qubit gates are universal. Their idea is as follows. Due to the universality of two-qubit gates, it suffices to show that one can construct arbitrary $\text{SU}(4)$ from a single two-qubit unitary gate.  Given a two-qubit unitary $e^{iA}$, one can construct $e^{iAt}$ for arbitrary $t\in\mathbb{R}$ by repeating $e^{iA}$, if the eigenvalues of $A$ are irrational. This can be satisfied for almost all unitaries. Let $e^{iB}$ be the unitary that only switches two inputs of $A$. 
Then, one can construct unitaries in the form of $e^{iLt}=\cdots e^{iAt_4}e^{iBt_3}e^{iAt_2}e^{iBt_1}$. $L$ can be any Hermitian operator spanned by the iterative commutator $\{A,B,i[A,B],[A,[A,B]],\cdots\}$. With unit probability, this iterative commutation can generate $15$ linearly-independent vectors, thus spans the entire $\text{SU}(4)$ for two-qubit gates. So $e^{iLt}$ is dense in $\text{SU}(4)$.

While the constructions discussed above establish universality, they are generally too complicated to be practical. Subsequent work has therefore focused on reducing the circuit complexity of unitary synthesis. Below, we review several representative synthesis schemes. A standard approach is to decompose multi-qubit unitaries into single-qubit gates in $\mathrm{SU}(2)$ and CNOT gates. The single-qubit gates can then be further compiled into more elementary operations using the methods discussed in Sec.~\ref{sec:unitary_single}, while the CNOT gate serves as a standard entangling two-qubit operation.

\subsubsection{QR decomposition}
\citet{reck1994experimental} showed that unitaries can be universally decomposed into two-level rotations. Specifically, let
\begin{align}
T_{j,k}=\sin\omega(e^{i\phi}|j\rangle\langle j|-|k\rangle\langle k|)
+\cos\omega(e^{i\phi}|j\rangle\langle k|+|k\rangle\langle j|)
\end{align}
be some Givens rotations applied at the subspace $\{|j\rangle,|k\rangle\}$. For any $U\in\text{SU}(N)$, one can always find some $T_{N-1,k}$, such that 

\begin{align}\label{eq:u_r1}
U\prod_{k=0}^{N-2}T_{N-1,k}=\begin{pmatrix}U_{N-1}&\begin{matrix}0\\\vdots\end{matrix}\\\begin{matrix}0&\cdots\end{matrix}&e^{i\alpha}\end{pmatrix},
\end{align}
where $U_{N-1}$ is an $(N-1)$-dimensional unitary.  Applying this idea recursively, we have  
\begin{align}\label{eq:u_r2}
U\prod_{j=1}^{N-1}\prod_{k=0}^{j-1}T_{j,k}=D,
\end{align}
where $D$ is a diagonal unitary. 
Accordingly, one can synthesize arbitrary $U$ by
\begin{align}\label{eq:u_r3}
D\prod_{j=N-1}^{1}\prod_{k=j-1}^{0}T_{j,k}^\dag=U.
\end{align}
See also Section 4.5.2 of~\cite{Nielsen.02} for a concrete example.
This method is  commonly referred to as a QR decomposition, because Eq.~\eqref{eq:u_r1} and Eq.~\eqref{eq:u_r2} are identical to the standard QR-factoring process using two-level rotations~\cite{cybenko2001reducing}. 

QR decomposition is proposed in the context of linear optic system, where each computational basis $|j\rangle$ represents one spatial mode of the photon, and $T_{j,k}$ are realized by phase shifters and interferometers. 
It requires a total of $N(N-1)/2$ Givens rotations arranged as $2N-3$ layers. The optical circuit depth can be further improved~\cite{clements2016optimal}, but the standard approach is more friendly for further decomposition required by qubit systems.

For qubit systems, $T_{j,k}$ are equal to multi-qubit controlled rotations, and should be further decomposed into single-qubit and CNOT gates (see Sec.~\ref{sec:mqcg}). 
Along this line, circuit size of $O(n^3N^2)$~\cite{barenco1995elementary}, $O(n^2N^2)$~\cite{aho2003compiling}, and $O(nN^2)$~\cite{knill1995approximation} have been achieved. ~\cite{vartiainen2004efficient} further represented unitaries with Gray code basis instead of the conventional binary basis $|j\rangle$. Because only one bit changes between two adjacent basis of Gray code, this approach makes it possible to replace all C$^{n-1}$-$X$ gates involved in the decomposition schemes by two-qubit CNOT gates, and thus achieving the asymptotically optimal circuit size $\Theta(N^2)$. 

The CNOT count of~\cite{vartiainen2004efficient} is approximately $8.7\times4^n$. Below, we introduce a series of works, based on different recursive decompositions, which can provide further improvement.

\subsubsection{CS, Shannon and block-ZXZ decompositions}\label{sec:u_rd}
\citet{mottonen2004quantum} proposed a recursive unitary synthesis method based on the cosine-sine (CS) decomposition. Specifically, arbitrary unitaries can be decomposed as~\cite{paige1994history}

\begin{align}\label{eq:u_cs}
U = \begin{pmatrix}
U_{11} & \\
 & U_{12}
\end{pmatrix}
\begin{pmatrix}
C & S \\
-S & C
\end{pmatrix}
\begin{pmatrix}
U_{21} &  \\
 & U_{22}
\end{pmatrix},
\end{align}

\noindent where $U_{jk}$ are $N/2\times N/2$ unitaries, while $C$ and $S$
are real diagonal matrices satisfying $C^2+S^2=I$. Matrix $\begin{pmatrix}
C & S \\
-S & C
\end{pmatrix}$ represents some UCR along $y$-axes (see Eq.~\eqref{eq:Fn} in Sec.~\ref{sec:intro} for definition). Thus, $U$ can be decomposed into the following quantum circuit.

$\\$
\begin{centering}
\quad\Qcircuit @C=1em @R=1.2em {
&\qw &\ctrlo{1}  & \ctrl{1}&\gate{R_y} &\ctrlo{1}  & \ctrl{1}&\qw \\
&{/} \qw& \gate{U_{21}} & \gate{U_{22}}&\gate{}\qwx[-1]& \gate{U_{11}} & \gate{U_{12}} &\qw 
\inputgrouph{1}{2}{1.2em}{U=}{1.5em}}

\end{centering}
$\\$

\noindent Applying CS decomposition recursively on $U_{jk}$, one can decompose $U$ into a sequential implementation of UCRs. 
Using the optimized implementation of UCR based on Gray code~\cite{mottonen2004quantum} (see also Fig.~\ref{fig:sp_ucr}(d) and relevant text), the general unitary can be constructed with $4^n$ single qubit gates and $4^n-2^{n+1}$ CNOT gates. The single qubit gate count is optimal, while the CNOT gate count is better than the result obtained by QR decomposition~\cite{vartiainen2004efficient}.

In a subsequent work,~\cite{shende2005synthesis} shows that matrices $\begin{pmatrix}
U_{j1} &  \\
 & U_{j2}
\end{pmatrix}$, called \textit{multiplexor}, can be further decomposed into two $(n-1)$ qubit unitaries and UCR along $z$-axes. Combining with the CS decomposition,  $U$ can be constructed with the following circuit

$\\$
\begin{centering}
\quad\Qcircuit @C=.5em @R=1.2em {
&\qw &\qw&\gate{R_z}  & \qw&\gate{R_y} &\qw  & \gate{R_z}&\qw&\qw \\
&{/}\qw &\gate{U_{1}}& \gate{}\qwx[-1] & \gate{U_{2}}&\gate{}\qwx[-1]& \gate{U_3} & \gate{}\qwx[-1] &\gate{U_{4}} &\qw
\inputgrouph{1}{2}{1.2em}{U=}{1.5em}}

\end{centering}
$\\$

\noindent This method is considered as a quantum generalization of the Shannon decomposition of Boolean functions. 
This quantum Shannon decomposition can be applied recursively to $U_{j}$ by the above circuit, and finally achieve the CNOT gate count $(23/48)4^n-(3/2)2^n+4/3$ that is half that of the CS decomposition method.  

Another approach is the block-ZXZ decomposition \cite{de2016block,krol2024beyond}, which decomposes $U$ into the following circuit

$\\$
\begin{centering}
\quad\Qcircuit @C=.5em @R=1.2em {
&\qw &\ctrl{1}&\gate{H}   &\ctrl{1}&\gate{H}  & \gate{}\qwx[1]&\qw&\qw \\
&{/}\qw &\gate{C}& \qw&\gate{B} & \qw&\gate{A}& \qw&\qw
\inputgrouph{1}{2}{1.2em}{U=}{1.5em}}

\end{centering}
$\\$

\noindent for some unitary $A,B$ and $C$.  This method is similar to the Shannon decomposition, and naive implementation gives a similar circuit complexity. Recent work~\cite{krol2024beyond} showed that during the decomposition of $C$ and $A$, two of the CNOT gates can be merged into one. So it improves the leading term of CNOT count to 
$$\frac{22}{48}4^n.$$
Yet, there is still a gap between this best-known result and the best-known CNOT count lower bound $\lceil\frac{1}{4}(4^n-3n-1)\rceil$~\cite{shende2004smaller}.

\subsection{Ancilla-assisted multi-qubit unitary synthesis
\label{sec:unitary_st}}

Similar to quantum state preparation, the circuit depth of unitary synthesis can be reduced by adding ancillary qubits. Formally, the target $n$-qubit unitary transformation $U$ is embedded in a larger unitary $W_{U}$ applied at $n+m$ qubits, such that 
\begin{align}\label{eq:u_wu}
W_U(|\psi\rangle\otimes|0\rangle_{\text{anc}})=U|\psi\rangle\otimes|\text{garb}\rangle_{\text{anc}}
\end{align}
for all quantum states.
The ancillary system is initialized as  all-zero state, and can end up as any garbage state $|\text{garb}\rangle$ that is separable to $U|\psi\rangle$. 

In practice, we may allow the unitary synthesis up to an error $\varepsilon$. The target preparation process $\tilde{\mathcal{W}}_U$ then satisfies
\begin{align}\label{eq:u_wu_err}
\tilde{\mathcal{W}}_U(|\psi\rangle\otimes|0\rangle_{\text{anc}})=|\tilde{\phi}\rangle\otimes|\text{garb}\rangle_{\text{anc}}
\end{align}
where the distance between $|\tilde{\phi}\rangle$ and the ideal state $U|\psi\rangle$ is at most $\varepsilon$.

\subsubsection{Depth optimization}
One possible approach is to reduce unitary synthesis to quantum state preparation using Householder reflections \cite{kliuchnikov2013synthesis,low2024trading}. Let $|u_k\rangle=U|k\rangle$ be the columns of the target unitary \(U\), and define $|w_k\rangle=\frac{1}{\sqrt{2}}\left(|1\rangle|k\rangle-|0\rangle|u_k\rangle\right).$  The states $\{|w_k\rangle\}$ are orthonormal, and it can be verified that
\begin{align}
|0\rangle\langle1|\otimes U+|1\rangle\langle0|\otimes U^\dag=\prod_{k=0}^{N-1}R_{|w_k\rangle},
\end{align}
where $R_{|\psi\rangle}=I-2|\psi\rangle\langle\psi|$ is the Householder reflection associated with $|\psi\rangle$. Let \(P_{|\psi\rangle}\) be a state-preparation unitary satisfying $P_{|\psi\rangle}|0\rangle=|\psi\rangle$, the corresponding reflection can be implemented as
\begin{align}
    R_{|\psi\rangle}
    =
    P_{|\psi\rangle}
    \left(
        I-2|0\rangle\langle0|
    \right)
    P_{|\psi\rangle}^{\dagger}.
    \label{eq:rpsip}
\end{align}
The operator \(I-2|0\rangle\langle0|\) is a multi-qubit controlled-phase gate, which will be discussed in Sec.~\ref{sec:mqcg}. $P_{|w_k\rangle}$ can be constructed from a controlled-\(P_{|u_k\rangle}\), together with $O(n)$ additional Clifford gates. The target unitary  can therefore be implemented on the system register using one additional ancillary qubit initialized in \(|1\rangle\):
\begin{align}
    \left(
        \prod_{k=0}^{N-1}R_{|w_k\rangle}
    \right)
    |1\rangle|\phi\rangle
    =
    |0\rangle U|\phi\rangle.
\end{align}
The circuit complexity is mainly determined by  $P_{|\psi\rangle}$ and their inverses. For example, with two-qubit Clifford and $T$ gates,~\cite{kliuchnikov2013synthesis} shows that using two ancillary qubits, arbitrary quantum state can be prepared with $O(Nn(n+\log(1/\varepsilon)))$ Clifford$+T$ gate count, where $\varepsilon$ is the error measured in Frobenius distance. So the unitary synthesis can be realized with $O(N^2n(n+\log(1/\varepsilon)))$ Clifford$+T$ count. 
Alternatively, if one uses the depth-optimal approach with $O(n)$ circuit depth of single- and two-qubit gates and $O(N)$ ancillary qubits (see Sec.~\ref{sec:qsp_st}), $U$ can be synthesized with $O(nN)$ circuit depth and $O(N)$ ancillary qubits. This result is quadratically better than the circuit depth lower bound $\Omega(N^2/n)$ for ancillary-free unitary synthesis.

\citet{rosenthal2023quantum} further showed that the circuit depth can be improved via Grover search~\cite{rosenthal2023quantum}. Their method is based on unitary-column-constructor ($U$-CC). Specifically, $U$-CC is defined as the following unitary transformation 
\begin{align}\label{eq:u_uqram}
|k\rangle\otimes|0\rangle\xrightarrow[]{U-\text{CC}}|k\rangle\otimes U|k\rangle.
\end{align}
Because $U|k\rangle=|\psi_k\rangle$, $U$-CC can be considered as a uniformly-controlled state preparation, and can be constructed with $O(n)$ circuit depth and $O(N^2)$ ancillary qubits. The remaining task is to uncompute $|k\rangle$  while keeping $U|k\rangle$ unchanged. To do so, it suffices to construct the Grover iteration 
$G=(R_{|+\rangle^{\otimes n}}\otimes I_n)O$ for some unitary oracle $O$ satisfying
\begin{align}\label{eq:u_kou}
O|j\rangle\otimes U|k\rangle=
(-1)^{\delta_{jk}}|j\rangle\otimes U|k\rangle.
\end{align}
By a similar argument to the Grover search, one can  realize $|k\rangle\otimes U|k\rangle\longrightarrow|0\rangle\otimes U|k\rangle$ with $O(\sqrt{N})$ queries to $G$.  
To construct $O$, a flag qubit $|b\rangle$ is introduced, and we define $A$ as a $2n+1$ qubit unitary satisfying 
\begin{eqnarray}\label{eq:u_axyb}
A|x,y,b\rangle=
 \left\{
\begin{array}{lcl}
|x\rangle\otimes U|x\oplus y\rangle\otimes|0\rangle  &    &  b=0\\
U|x\rangle\otimes|y\rangle\otimes|1\rangle  &  &  b=1
\end{array} \right.
\end{eqnarray}
It can be verified that $A$ is a $U$-CC up to a layer of controlled swap gates. Then, oracle $O$ can be constructed as 
\begin{align}\label{eq:u_oaa}
O=A(I_n\otimes R_{|0^{n},0\rangle}))A^\dag\,
\end{align}
which is equivalent to  
$O=\prod_{x\in\{0,1\}^n}R_{|x\rangle\otimes U|x\rangle\otimes|0\rangle}$. 
It can then be verified that 
\begin{align}\label{eq:unitary_ojou}
O|j\rangle\otimes U|k\rangle\otimes|0\rangle  =(-1)^{\delta_{jk}}|j\rangle\otimes U|k\rangle\otimes|0\rangle.
\end{align}
So Eq.~\eqref{eq:unitary_ojou} satisfies the criteria in Eq.~\eqref{eq:u_kou} after tracing out the flag qubit. Because $O(\sqrt{N})$ queries to $G$ is sufficient, unitary synthesis can be realized by $O(n2^{n/2})$ circuit depth and $O(n4^{n})$ ancillary qubits~\cite{rosenthal2023quantum}. Alternatively, one can apply intermediate space-time trade-off implementation of $A$~\cite{Yuan.22}. With ancillary qubit number $\Omega(2^n/n)\leq m\leq O(4^n/n)$, one can achieve circuit depth $O\left(\frac{n^{1/2}2^{3n/2}}{m^{1/2}}\right)$.

The best-known circuit depth lower bound for unitary synthesis is $\Omega(n)$ according to the light-cone and counting argument~\cite{Yuan.22}, which is the same as the one for state preparation.  Thus, there is an exponential gap between the lower and upper bound, which can be formalized as the following open problem. 

\begin{challenge}\label{prob:para}
For every $n$-qubit unitary $U$, does there exist a quantum circuit with depth $O({\rm poly}(n))$ to implement $W_U$?
\end{challenge}
\noindent Challenge~\ref{prob:para} shows the complication of parallelizing unitaries. As a comparison, we have known result for state preparation problem, whose optimal circuit depth is $\Theta(n)$ (Sec.~\ref{sec:qsp_st}).

\subsubsection{T count optimization}
Optimizations have also been considered for reducing $T$ count of the Clifford$+T$ circuits. A naive Clifford$+T$  construction is to decompose all single-qubit unitaries with methods in Sec.~\ref{sec:u_ds}. Because $O(4^n)$ single-qubit rotations are required for ideal construction, this results in $O(4^nn)$ $T$ count. Besides, recall that unitary synthesis can be reduced to quantum state preparation, so low $T$ count state preparation methods can be naturally adapted to unitary synthesis. Based on the select-swap approach (see also Sec.~\ref{sec:sp_ss}), $T$ count can be reduced to $O(2^{3n/2}n)$ with $O(2^{n/2})$ ancillary qubits~\cite{low2024trading}. 

\citet{tan2025unitary} further improved the $T$ count to $$O(2^{4n/3}n^{2/3}).$$ Based on a careful analysis on the structure of CS decomposition (Sec.~\ref{sec:u_rd}), it is shown that the unitary can be decomposed as 
$U=W_0\prod_{j=1}^{2^{n-k}-1}F_jW_j$.
Here, $F_j$ is some UCR. Each $W_j$ is the product of $2^k-1$ UCRs in the form of (neglecting index $j$)
\begin{align}
W_j=\prod_{i=1}^{2^k-1}\sum_{x=0}^{2^{n-k}-1}|x\rangle\langle x|\otimes  D_{i,x}R_{i,x}.
\end{align}
Here, $R_{i,x}$ and $D_{i,x}$ are, respectively, a single-qubit $\text{SU}(2)$ gate and a phase gate, each applied to one of the last $k$ qubits. 
Each $R_{i,x}$ and $D_{i,x}$ can achieve a certain precision with no more than $L$ Hadamard gates and $L$ $T$ gates. 
With $x$ as input and gate sequences for all $R_{i,x}, D_{i,x}$ as output, one can then introduce a Boolean function  
$g: \{0,1\}^{n-k}\rightarrow\{0,1\}^{2(2^k-1)\cdot 2^k\cdot(2L)}$
corresponding to $W_j$. Then,
$W_j$ can be constructed by quantum oracle for $g$ and a few extra elementary gates. The advantage is that any Boolean function $\{0,1\}^{s}\rightarrow\{0,1\}^{r}$ can be realized by $O(\sqrt{r\cdot 2^s})$ $T$ count~\cite{low2024trading}. So $T$ count of the total circuit can be optimized by choosing appropriate $k$. After optimizing $k$ and sequence length $L$, the claimed  $T$ counts of $O(2^{4n/3}n^{2/3})$ can be achieved. Furthermore, tradeoff between $T$ count and space complexity can be realized by choosing different $k$. 

A more recent preprint also reports a further improved $T$ count of $\widetilde{O}(2^{5n/4})$~\cite{yuan2026quantumcircuitgeneralunitary}.

\subsection{Query complexity\label{sec:unitary_query}}
Similar to the parallelism problem raised in Challenge~\ref{prob:para}, the query complexity for unitary synthesis is significantly  different from the one for quantum state preparation. As discussed in Sec.~\ref{sec:sp_mq}, single query to Boolean function is already sufficient to efficiently prepare arbitrary quantum state. In contrast, the number of queries for unitary synthesis remains unknown, and could range from $2$ to $O(2^{n/2})$. There is a famous open problem raised by~\citet{aaronson2007quantum}.

\begin{challenge}[Unitary synthesis problem]\label{prob:usp}
Is it true that all $n$-qubit unitaries can be implemented in $\mathsf{BQP}^{A}$ for some Boolean oracle $A:\{0,1\}^*\rightarrow\{0,1\}$?
\end{challenge}

 Note that if one relaxes the restriction of polynomial, there exists unitary synthesis circuits with single query to oracle $A$, due to the Bernstein-Vazirani algorithm~\cite{bernstein1993quantum}\footnote{For detailed construction, see Lecture 6 of COMS 6998: Frontiers of Quantum Complexity and Cryptography, Henry Yuen.}.  However, this does not give an answer to Challenge~\ref{prob:usp} because their circuit size is exponential and thus not contained in $\mathsf{BQP}^A$.

It is conjectured that the answer to Challenge~\ref{prob:usp} is negative~\cite{aaronson2007quantum,aaronson2016complexityquantumstatestransformations}, i.e.~even if solving arbitrary classical function is considered easy, there are still quantum problems that are not efficiently solvable. Thus, a negative answer to Challenge~\ref{prob:usp} can serve as strong evidence of the existence of quantum advantage. Below, we introduce some progress related to Challenge~\ref{prob:usp}. 

Challenge~\ref{prob:para} and~\ref{prob:usp} are related. Indeed, since any Boolean oracle can be implemented in linear depth using exponentially many ancillas~\cite{nie2026nearlyoptimalquantumcircuits}, a positive answer to Challenge~\ref{prob:usp} would imply a positive answer to Challenge~\ref{prob:para}. Contrapositively, a negative answer to the latter implies one to the former. In this cense, Challenge~\ref{prob:para} is stronger than Challenge~\ref{prob:usp}, if one believe that the answer to both are negative. 

\subsubsection{Aaronson-Kuperberg bound}
\citet{aaronson2007quantum} ruled out the possibility of constructing universal unitary using single query to oracle $A$ applied  nontrivially only at the first $n$-qubit, i.e.~those satisfying $A|\psi\rangle\otimes|0\rangle=(U|\psi\rangle)\otimes|\text{garb}\rangle$. This result is achieved by counting argument: there exist only $O(4^{N})$ distinct oracles in this architecture, while $\Omega(c^{N^{2}})$ unique unitaries are required to cover the set of $\text{SU}(N)$ for some constant $c$.

\citet{lombardi2024one} 
generalized this counting argument to allow $t>1$ queries, and $m>0$ ancillary qubits, and limited accuracy. Yet, this counting argument becomes not useful when $t$ and $m$ are slightly larger. 

\subsubsection{Rosenthal bound}
With a similar idea to the oracle-free algorithm in Sec.~\ref{sec:unitary_st}, \citet{rosenthal2023quantum} showed that $$O(2^{n/2})$$ queries to oracles in the form of unitary QRAM in Eq.~\eqref{eq:u_uqram} is sufficient to synthesize arbitrary unitary. This is the first nontrivial upper bound for unitary synthesis, but still exponentially far away from polynomial as raised in Challenge~\ref{prob:usp}. \citet{rosenthal2023quantum} also ruled out the possibility of giving positive answer to Challenge~\ref{prob:usp} using a special type of unitary QRAM. Specifically, the author considered $2n$-qubit  oracles $A$ satisfying $A|x,y\rangle=|x,y\oplus\sigma(x)\rangle$, where $\sigma$ is some permutation operator. It is shown that when restricted to this specific type of oracles, $\Omega(2^{n/2})$ queries are required for almost all unitaries.

\subsubsection{Lombardi-Ma-Wright bound}

Lower bounds above contain extra structural assumptions of the quantum oracles.~\citet{lombardi2024one} derived a one-query lower bound for \textit{any} quantum oracles. Specifically, they showed that there is no efficient quantum circuit for arbitrary unitary, with single query to quantum oracles of Boolean function $f:\{0,1\}^{l}\rightarrow\{0,1\}$ with $l=o(2^n)$.  Note that with $l=O(2^{2n})$, one query is sufficient~\footnotemark[4]. This result can be generalized to allowing unbounded number of non-oracle gates and ancillary qubits, and $f$ with $O(\text{poly}(n))$ output bits. 

The Lombardi-Ma-Wright bound is based on an \textit{oracle state distinguishing game}, aiming to distinguish pseudorandom state from the \textit{uniformly random binary
phase state} $|\psi\rangle=1/\sqrt{N}\sum_{x\in\{0,1\}^{n}}h(x)|x\rangle$, where $h:\{0,1\}^n\rightarrow\{\pm1\}$ is a uniformly random Boolean function. \citet{lombardi2024one} derived a one-query circuit lower bound for oracle state distinguishing game, which implies the lower bound for unitary synthesis.

\subsection{Resource minimisation in fault-tolerant architectures}\label{sec:rm}
In fault-tolerant quantum computing architectures, Clifford gates are typically less expensive 
than non-Clifford gates, e.g., in the surface code~\cite{Fowler2012Surface}, Pauli $X$ and $Z$ 
can be realised virtually while CNOT gates can be implemented via
lattice surgery~\cite{horsman2012surface} or topological braiding~\cite{raussendorf2007fault,Fowler2012Surface}.
In contrast, non-Clifford gates introduce "magic" into quantum circuits by rendering them classically hard to simulate, while purely Clifford circuits are classically efficiently simulable~\cite{gottesman1998heisenberg}.
Unfortunately, typical quantum circuits use a large number of continuous rotations, which
are considered expensive resources~\cite{zimboras2025myths} given 
they need to be first decomposed into Clifford and $T$ gates approximately as detailed in the previous subsection

The cost model assumption above has been conventional in the literature. But we also note that it has been recently challenged 
by the introduction of cheap magic resources, such as $T$ states or $CCZ$ states, via magic state cultivation \cite{gidney2024magic}.

As detailed in Sec.~\ref{sec:u_ds}, approximating $R_{Z} (\theta )$
to a precision $\epsilon$ using an optimal ancilla-free direct synthesis approach requires
$\approx 3.02\log_{2} (1/\epsilon )+1.77$ $T$ gates~\cite{ross2016optimal}. For practically relevant parameter settings, e.g., $\epsilon < 10^{-6}$, a single rotation
may require more than $60$ $T$ gates. Below, we introduce progress made in further minimising these costs by exploiting higher-level structure in quantum circuits, which is particularly important for resource frugal, early-fault-tolerant implementations~\cite{katabarwa2024early, zimboras2025myths}.

\subsubsection{Repeat-Until-Success}

Repeat-Until-Success (RUS) is a generalisation of $T$ state teleportation to applying an arbitrary rotation angle. One proceeds by preparing a family of resource states fault tolerantly $\ket{\theta } := R_{Z} (\theta )|+\rangle$ for various values of $\theta$ and then iteratively teleporting them. First, applying gate teleportation 
using the resource state $\ket{\theta }$ yields with probability $1/2$ a measurement outcome $+1$ which indicates that the qubit is correctly rotated
as $R_{Z} (\theta )|\psi \rangle $. However, with equal chance, it yields a measurement outcome  $-1$ which indicates an inverse rotation $R_{Z} (-\theta )|\psi \rangle $. In the latter case, one needs to apply a rotation gate with twice the angle $R_{Z} (2\theta )$ in order to obtain the desired $R_{Z} (\theta )|\psi \rangle $ net effect. The approach is therefore repeated iteratively using resource states with angle settings $2^k \theta $ until a $+1$ outcome is achieved, which in general requires on average $\sum _{i=1}^{\infty }\frac{i}{2^{i}} =2$ number of trials.

The idea of RUS is first introduced in~\cite{paetznick2014repeat}, but requires classical exhaustive search.~\cite{bocharov2015efficient,bocharov2015fallback} further improved the synthesis algorithms with efficient classical runtime, and achieved expected $T$ count $\approx1.03\log_2(1/\varepsilon)+5.75$ 
for $R_z(\theta)$ with random angles, estimated through linear fit~\cite{kliuchnikov2023shorter}.

\subsubsection{Catalyst circuits}
Based on the efficient adder circuits in~\cite{hamming-Gidney}, catalyst towers were developed
in~\cite{sun2025low,kiumi2025te} building on~\cite{hamming, gidney2019efficient}.
The central object is a so-called catalyst circuit which consumes two $\ket{+}$ states, a resource state $\ket{\Delta}$ and a rotation gate $R_z(2\Delta)$, and outputs three resource states $\ket{\Delta}$, thus, in effect applies two $R_z(\Delta)$ rotations at the cost of consuming one $R_z(2\Delta)$ rotation and 4 $T$-states.~\cite{sun2025low} then stacked these catalyst circuits so that the overall circuit prepares
a family of resource states $\ket{2^k \Delta}$ by catalysis consuming only
a single rotation $R_z(2^{h} \Delta)$, where $h$ is the height of the tower. This allows preparing a large number of
resource states at the cost of approximately 4 $T$ states per resource state.

\subsubsection{Phase gradient states and operations}

The phase gradient state of $b$ qubits  
is a superposition of all basis states with a linearly increasing phase as
\begin{equation}
	| \phi \rangle = \sqrt{2^{-b}} \sum_{k=0}^{2^b-1} \omega_b(k) |k \rangle, \quad
	\omega_b(k):= \exp{(- i \frac{ 2 \pi  k}{ 2^b} )}.
\end{equation}
This state can be prepared efficiently using a series of single-qubit phase rotations 
as $| \phi \rangle = \prod_{l=0}^{b-1} P_l(\theta_l) |+^{\otimes b} \rangle$
with exponentially decreasing rotation angles $\theta_l = -\pi 2^{-l}$,
where $P_l(\theta) = \mathrm{diag}(1,e^{i \theta})$ is the
phase shift gate acting on the $l^{th}$ qubit~\cite{sanders2020compilation}.

This state has the special property that arithmetic addition, defined via the oracle
$U_{\text{add}} |k \rangle |q \rangle {\rightarrow }| k \rangle  | k {+} q \rangle$,
applies a phase gate rotation to the joint state of the system~\cite{hamming-Gidney,kitaev2002classical}
whereby the angle of rotation is determined
by the computational state $|k \rangle$ in the ancilla register as
\begin{equation}\label{addition}
	U_{\rm add} |k \rangle |\phi \rangle = \omega_b(-k) |k \rangle |\phi \rangle.
\end{equation}
As a consequence, the phase gradient state can be ``copied" efficiently by loading the Hadamard state $|+^{\otimes b} \rangle |\phi \rangle$
into the ancilla register and applying subtraction rather than addition yielding $|\phi \rangle |\phi \rangle$.

Phase-gradient states are very useful in practice to catalyse phase gate rotations such that the state $|\phi\rangle$ is recovered. For example,
suppose that possible rotation angles $\theta_k$ are discretized into $2^b$  ``notch settings'', as is natural for many hardware-level instruction sets~\cite{pai2024}, such that
\[
\theta_k = \frac{2\pi k}{2^b}, \qquad k = 0,1,\dots,2^b-1.
\]
Let $U_k$ denote a circuit of Pauli-$X$ operators that prepares the computational basis state encoding the desired rotation angle, such that
$U_k |0 \rangle^{\otimes b} = |k\rangle.$
Applying $U_k$ controlled on an arbitrary input qubit state $\alpha |0\rangle + \beta |1\rangle$, and then performing the addition operation from Eq.~\eqref{addition}, gives
\begin{align*}
	(\alpha |0\rangle + \beta |1\rangle) |0^b\rangle |\phi\rangle
	&\rightarrow
	\alpha |0\rangle |0^b\rangle |\phi\rangle
	+ \beta |1\rangle |k\rangle |\phi\rangle \\
	&\rightarrow
	\alpha |0\rangle |0^b\rangle |\phi\rangle
	+ \omega_b(-k)\,\beta |1\rangle |k\rangle |\phi\rangle.
\end{align*}
Uncomputing the controlled $U_k$ then yields the separable, rotated state
$\alpha |0\rangle + \omega_b(-k)\beta |1\rangle$,
which is equivalent to an $R_z(\theta_k)$ rotation up to a global phase. Using the efficient adder circuits of \cite{hamming-Gidney}, this approach requires only $(b-2)$ Toffoli gates per continuous rotation~\cite{sanders2020compilation}. It is therefore substantially more efficient than Clifford+$T$ synthesis when multiple rotation angles are provided in superposition as quantum data (for example, via QROM)~\cite{sanders2020compilation}, and this advantage has been exploited in resource-estimation studies~\cite{low2025fast}.

\subsubsection{Hamming weight phasing}
A broad range of practical applications in, e.g., quantum simulation require 
applying a number $n$ of continuous rotations $(R_z(\theta))^{\otimes n}$
with equal angles simultaneously to a set of $n$ qubits.
The approach detailed in~\cite{hamming-Gidney,hamming} relies on the main observation that rotating an arbitrary state
with $n$ identical-angle rotations applies a phase
\begin{equation}
	(R_z(\theta))^{\otimes n} \sum_{k=0}^{2^n-1} \alpha_k |k\rangle
	=
	\sum_{k=0}^{2^n-1} \alpha_k  e^{i\theta (k)} |k\rangle,
\end{equation}
where the phase angle $\theta(k) =  \theta[H(k) - n/2]$ depends only on the Hamming weight $H(k)$, defined as the number of $1s$ in the bitstring $k$.

Assuming for ease of notation that $n$ is a power of $2$,
the approach uses a small ancillary register of size $(1{+}\log_{2} n)$ into which 
the Hamming weight is calculated to yield the state $\sum_{k=0}^{2^n-1} \alpha_k |k\rangle  | H(k) \rangle$.
Then, a small number $(1{+}\log_{2} n)$ of continuous rotations with exponentially increasing rotation angles
as $R_z(\theta), R_z(2\theta), R_z(4\theta) \dots$
are applied to the individual qubits in the Hamming-weight register. Alternatively a phase-gradient addition can be used to apply the desired phase, as in Eq.~\eqref{addition}.
Uncomputing the Hamming weight register yields the desired rotated state.

The approach requires a total of  $4(n{-}1)$ $T$ gates~\cite{hamming-Gidney} 
and an additional $\log_{2} n$ continuous rotations, or addition of a phase gradient of size $b = 1+\log_{2} n$.
The approach led to substantially reducing resource requirements for quantum simulation tasks~\cite{gunther2025phase,hamming,kiumi2025te}.

\subsubsection{Reducing total surface code footprint}
Rather than minimising $T$ gates only, recent works also shifted focus to directly reducing the surface code space-time footprint of continuous rotations  \cite{sun2025space,huggins2025fluid}, which is particularly relevant for early fault-tolerant settings of low to medium code-distances. These works showed that the optimal implementation requires  co-design of gate synthesis and architecture-aware scheduling, and trading qubit overhead for reduced circuit depth can lower overall spacetime cost.

Finally, we also note that the above techniques often approximately synthesise unitary gates, up to a controllable error $\varepsilon$. Still, exact synthesis,
in the sense that target observables can be estimated unbiasedly, can be achieved by quantum circuit randomization as we detail further in Sec.~\ref{sec:cr}.

%% file: secs/QRAM_0809.tex
\section{Memory in quantum computing}\label{sec:QRAM}

Since the early days of quantum algorithms, coherent access to classical data has played an important conceptual role. For instance, Grover search~\cite{grover1996fast} assumes coherent access to an oracle encoding the search space (e.g., a database). Much of the early quantum-algorithms literature similarly use abstract oracle models where the implementation of the oracle is not considered. In 2008, \citet{Giovannetti.08, Giovannetti.08_2} proposed concrete QRAM architectures for accessing general data, showing how such oracles could potentially be implemented. Notably, given a single input address, their bucket-brigade architecture only activates a polylogarithmic number of routing elements, contributing to favorable error-scaling properties which have motivated substantial subsequent study (e.g.,~\cite{arunachalam2015robustness, hann2021practicality}).

Moreover, the importance of quantum memory has grown with the development of more quantum algorithms for processing classical data.
Building on the quantum linear systems algorithm of Harrow, Hassidim, and Lloyd (HHL)~\cite{Harrow.09}, a number of quantum linear algebra based algorithms have been developed which, when given access to an efficient QRAM, appeared to potentially offer exponential speedups in classical data-processing~\cite{kerenidis2016quantum, lloyd2014quantum, rebentrost2018quantum, van2017quantum}. 
This motivated closer examination of the input-access assumptions underlying such speedups.

Indeed, related literature developed dequantization techniques that, when classical algorithms are granted comparable input assumptions to their quantum counterparts, many of the claimed exponential speedups disappear~\cite{tang2019quantum,tang2021quantum,tang2022dequantizing,chia2022sampling}. Nevertheless, super-quadratic quantum speedups often remain, so efficient data access could still be important for practical quantum advantage in such tasks.
See Table 1 of~\cite{chia2022sampling} for a comparison of quantum and dequantized asymptotic complexity for a number of algorithms, see~\cite{babbush2021focus} for a discussion on the importance of super-quadratic speedup for practical quantum-advantage, and see~\cite{aaronson2015read} for an overview of some end-to-end considerations necessary to achieve speedups for quantum linear based algorithms.

A distinct limitation arises from the opportunity cost of the physical resources required to implement QRAM. The relevant quantity is not only the cost to initialize the QRAM and its circuit depth, but also how much energy input, quantum control, and classical processing it requires to enact a query. To analyze these constraints, we adopt and extend the taxonomy of \textit{active} and \textit{passive} QRAM as systematized in~\cite{jaques2025qram}, which builds on earlier work including~\cite{aaronson2015read,steiger2016racing, ciliberto2018quantum}.

A QRAM whose total per-query opportunity cost is $\Omega(N)$ (i.e., grows at least linearly with $N=2^n$) for a memory storing $N$-bits is called \textit{active}. 
The opportunity cost associated with active QRAM can eliminate quantum speedups even in settings where dequantization leaves a polynomial advantage, because the resources consumed by a QRAM query could instead be used to execute a parallel classical algorithm. 
This holds for a broad class of QSVT-based algorithms that use QRAM to store large unstructured matrices~\cite{jaques2025qram}, and also holds for some algorithms related to quantum cryptanalysis. 

This motivates the need for a passive QRAM whose total opportunity cost is sub-linear in the size of the memory, or even a strongly-passive QRAM, one whose total opportunity cost is polylogarithmic. However, if a passive QRAM is fully-error corrected with standard methods, $O(1)$ classical co-processors per qubit are required, so the architecture becomes active. Consequently, passive hardware QRAM proposals should have favorable error-resistance properties to avoid error-correction. The error-scaling of bucket-brigade provides one possible route, but interfacing a noisy QRAM with a fault-tolerant quantum computer remains an important problem. However, passive QRAM is not required for every application. Active circuit implementations, including QROM, are already useful in tasks including quantum chemistry, cryptanalysis and state preparation.

The remainder of this section is organized as follows. 
In Sec.~\ref{sec:qram:subsec:qram_taxonomy}, we situate QRAM within the broader memory landscape, depending on whether the quantum device and stored data are quantum or classical. 
In Sec.~\ref{sec:qram:subsec:resource_taxonomy} we discuss the overall resource accounting for implementing QRAM. 
In Sec.~\ref{sec:qram:subsec:active_qram} we provide an overview of the research into active QRAM systems, including circuit QRAM implementations such as quantum read-only memory (QROM).
Next, we present the literature investigating passive QRAM in Sec.~\ref{sec:qram:subsec:passive_and_strongly_passive_qram} and their variances, including those between fully active and fully passive in Sec.~\ref{sec:qram:subsec:other}.

\subsection{Memory primitives and terminology}\label{sec:qram:subsec:qram_taxonomy}

This section summarizes the common types of memory found in the literature, and we adapt the memory categorization presented in~\cite{kuperberg2013another} which classifies memory according to whether the accessing device and the stored data are quantum or classical.
This includes four categories: CRACM (classical random access classical memory), QRACM (quantum random access classical memory), CRAQM (classical random access quantum memory) and QRAQM (quantum random access quantum memory).

It is also useful to decompose a memory access into two conceptual tasks: routing and read-out. During routing, the address determines which memory cell is coupled to the output register; after read-out, the routing operation is undone. During read-out, the selected memory value is written into the output register. If each memory cell contains one bit, for classical memory this can be viewed as copying the selected bit. For quantum memory it can be implemented by a Toffoli gate for classical data, or controlled SWAP gate for quantum data. In~\Cref{fig:sec:qram:example_circuit_qram}, boxes (a) and (c) correspond to routing, while box (b) is the read-out.

\subsubsection{Classical Random Access Quantum Memory}\label{sec:qram:subsec:quantum_memory}

Diverging from~\cite{kuperberg2013another}, a functional memory-focused definition of Classical Random Access Quantum Memory (CRAQM) is any quantum computer with a distinction between computational qubits and memory qubits, where a classical computer controls the routing between memory and computation. It is becoming an increasingly important part of fault-tolerant quantum algorithms, and it potentially has utility in implementing certain error-corrected active QRAM in Sec.~\ref{sec:qram:subsec:active_qram:active_qram_outside_circuit_model}.

Early research focused on preserving quantum states encoded in physical qubits, for example~\cite{kielpinski2001decoherence}. Other early research proposed splitting quantum computers into computational regions, and storage regions~\cite{kielpinski2002architecture}.  
This same memory-qubit and computational-qubit split can also be seen in more recent work on error-correcting codes. For example, when a qubit's logical state needs to be preserved,  but operations do not need to be (easily) applied to it, the ratio of logical qubits to physical qubits (the rate) of the error-correcting code can often be much higher, see for example~\cite{bravyi2024high, pattison2025hierarchical, low2026denser}. 
Error-correcting codes designed to preserve logical qubit states at high rate, but which do not necessarily allow for easy logical operations, are often called \textit{quantum memories}.  In a yoked surface code architecture~\cite{gidney2025yoked}, idle logical qubits can be stored with fewer physical qubits per logical qubit. When a gate is to be applied to an idle qubit, it must be made accessible to the active surface code/lattice-surgery procedure, effectively increasing the error-correction overhead and enabling logical operations. The distinction between storage-optimized and compute-optimized logical qubits is already useful in fault-tolerant resource estimates. For example,~\cite{gidney2025factor} applies the yoked surface code to store idle qubits more efficiently as one of several optimizations resulting in a more efficient integer factorization algorithm. 

Other recent work also make distinction between computational regions and memory region~\cite{cain2026shor}. In~\cite{cain2026shor}, when operations must be applied to a logical qubit stored in memory, it is teleported from the high rate memory to a lower rate computational region. In~\cite{ransford2025helios}, qubits encoded in trapped ion are routed from storage region to the logic regions through a ``cache'' region. These examples make the classification as a classical-access quantum memory clear: the classical control hardware essentially enacts a ``memory request'', as dictated by the algorithm, and transfers the necessary logical qubit from the memory region to the computational region (which can be viewed as routing).

\subsubsection{Quantum Random Access Classical Memory}\label{sec:qram:subsec:qracm}

Quantum Random Access Classical Memory (QRACM) stores classical data while allowing coherent access to that data. 
Throughout this review, unless stated otherwise, we refer to QRACM as QRAM. This convention agrees with much (but not all) of the quantum algorithms literature~\cite{lloyd2013quantum,biamonte2017quantum,ciliberto2018quantum, martyn2021grand, abbas2024challenges}. 
A QRAM with $n$ address bits, storing binary bitstrings of length $d$, is any process implementing the following mapping,
\begin{align}\label{sec:qram:definition:general_qram}
    U_{\text{QRAM}}\ket{i}_n\ket{0}_d = \ket{i}_n\ket{x_i}_d,
\end{align}
where $i \in \{0,1\}^n$, and $x_i \in \{0,1\}^d$.
The first $n$-qubit register is the address register, and the second is the $d$-qubit output register.
The defining characteristic of QRAM is that it stores its data in classical memory, and the memory contents can only be updated classically, but it can be accessed by a superposition of addresses.

QRAM can also be used to efficiently implement the mapping $\ket{i}_n\ket{0}_d \mapsto \ket{i}_n\ket{\psi_i}_d$, where $\{\ket{\psi_i}_d\}_i$ are a set of arbitrary quantum states whose amplitudes are stored as classical discretized values. Importantly, this is different from QRAQM in Sec.~\ref{sec:qram:subsec:qraqm} where the stored data are coherent unknown states. 
In the block-encoding literature, QRAM-backed data structures for such access are often called ``quantum-accessible data structures''~\cite{chakraborty2018power}. 
There is also some overlap with quantum state-preparation~\Cref{sec:gra}. Outside of the context of QRAM, the map is sometimes also known as controlled quantum state preparation, e.g.,~\cite{Yuan.22}. 

Finally, we note that the definitions are not identical in the literature. For example, QRACM has also been called QCRAM~\cite{van2020quantum}.~\citet{gilyen2019quantum, apers2025randomized} refer to general QRACM as QROM. Importantly, this should not be confused with how  QROM is defined later, with our definition being consistent with the recent literature using QROM as a resource optimization tool for fault-tolerant algorithm design (and which is a special subcategory of QRACM), see~\Cref{sec:qram:subsec:active_qram:qrom}. QROM has also been called quantum lookup table~\cite{mukhopadhyay2025quantum} (qLUT), and has a subcategory called QROAM.

\subsubsection{Quantum Random Access Quantum Memory}\label{sec:qram:subsec:qraqm} 

A Quantum Random Access Quantum Memory (QRAQM) primitive gives coherent quantum-access to memory registers which may themselves contain quantum states.  There are multiple definitions of QRAQM found in the literature, but we adopt the one from~\cite{belovs2014applications}, called SWAP-QRAQM in~\cite{jaques2025qram}. Consider a memory storing $N$ arbitrary quantum states $\{\ket{\psi_i}_d\}_{i\in [N]}$, with $\ket{\psi_i} \in \mathbb C^{D}$ ($D = 2^d$). We put the address register first, the output register second, and the $N$ data registers last. 
SWAP-QRAQM implements the following mapping:
\begin{align}\label{sec:qram:def:swap_qraqm}
 &\ket{i}_n\ket{\phi}_d\bigotimes_{j=0}^{N-1}\ket{\psi_{j}}_d
    \xrightarrow{\mathrm{SWAP-QRAQM}} \\
    &\ket{i}_n\ket{\psi_i}_d\left(\bigotimes_{j<i}\ket{\psi_j}_d\right)\ket{\phi}_d\left(\bigotimes_{j>i}\ket{\psi_j}_d\right).
\end{align}
For example, with three memory registers, $\ket{1}_n\ket{\phi}_d\ket{\psi_0}_d\ket{\psi_1}_d\ket{\psi_2}_d\mapsto \ket{1}_n\ket{\psi_1}_d\ket{\psi_0}_d\ket{\phi}_d\ket{\psi_2}_d$. This definition applies to superpositions of address registers by linearity. 

The key distinction from QRAM is that the stored data are unknown quantum states rather than classical binary values. As a consequence, a query to a QRAQM may leave the memory registers entangled with the address and output registers, whereas a QRAM query will not. 

A circuit implementation of QRAQM can often be obtained from a circuit QRAM by swapping non-routing CX gates with SWAP gates (as noted in a related context in~\cite{Giovannetti.08_2}). For instance, the fan-out circuit QRAM depicted in~\Cref{fig:sec:qram:example_circuit_qram} would also implement SWAP-QRAQM by replacing non-address-routing CX gates with SWAP gates, and by storing quantum states in the memory registers. ``Quantum random access gate'' (QRAG) is also a popular name for SWAP-QRAQM~\cite{aaronson2019quantum, belovs2024taming, allcock2024constant}. QRAQM can instead be specified by two operations, QRAM-R (read) and QRAM-W (write)~\cite{bonnetain2023finding}. Sandwiching QRAM-W between QRAM-R yields the same transformation as SWAP-QRAQM~\cite{Jeffery_2025}. Other names for QRAQM also exist, see \cite{naya2020optimal,apers2025randomized} for example.

For an example set of quantum algorithms utilizing QRAQM, or related input assumptions, see~\cite{kachigar2017quantum, chailloux2021lattice, naya2020optimal, akmal2022near}. Notably, QRAQM-like primitives are particularly prevalent in algorithms which dynamically update data structures during quantum walks, see e.g.,~\cite{aaronson2019quantum}.

We will focus on QRAM (i.e., QRACM) in the remainder of this section. Since QRAQM can implement QRAM by storing individual computational basis states in memory, the limitations of QRAM also apply to QRAQM.

\subsection{Per-query resource accounting for QRAM}\label{sec:qram:subsec:resource_taxonomy}

The utility of a QRAM architecture does not only depend on its time cost (e.g., in the circuit model, circuit depth), but also on all the resources consumed to perform a query. These resources have an opportunity cost which could have alternatively been used to run a parallel classical algorithm. Example resources include energy input, classical control, magic state distillation, decoding, etc.
This opportunity cost perspective underlies the \textit{passive} vs \textit{active} classification for QRAM. 
The need for passive QRAM was already noted in the context of quantum linear-system solvers~\cite{aaronson2015read}, and related opportunity cost or classical parallelization arguments also appear in later work~\cite{steiger2016racing, ciliberto2018quantum}.  
We build on the formulation of active and passive QRAM in~\cite{jaques2025qram}, extending it to allow for a more fine-grained analysis. 

\subsubsection{Why QRAM is harder than classical memory}\label{sec:qram:subsubsec:qram_harder_than_classical_memory}
One may naturally wonder why these opportunity cost arguments don't also apply to classical memory. We now summarize arguments made in~\cite{jaques2025qram} motivating why QRAM is harder to efficiently realize than classical memory.

Consider the following  simplified model of classical memory. Given an $n$-bit address $x = x_0x_1\hdots x_{n-1}$ and a memory with $N=2^n$ cells, one can imagine a binary tree with $n$ levels routing the address one bit at a time down the tree. The number of components which need to be activated to perform this routing then scale with the height of the tree, i.e., logarithmic with the size of the memory. The read-out operation can be idealized similarly: only the selected memory cell needs to be copied to the output. 

On the other-hand, QRAM can be queried with a superposition of address registers which can have support over a generally unknown number of memory elements. For a uniform superposition, all $\Omega(N)$ routing elements will be used (whether they can route without external energy input or control is a separate question). Even more generally, as noted in~\cite{jaques2025qram, arunachalam2015robustness}, the QRAM controller usually cannot apply any operations or control conditioned on the support of the address superposition. Learning that support would require extracting information from the address register (or any systems entangled to it), thereby decohering the quantum state making the query. Consequently, except for in limited cases where information about the querying state is known a priori, operations must always be applied as though all addresses have support. If enacting each routing or read-operation requires constant externally supplied energy, control, error correction, or classical processing (as non-exhaustive examples), then the per-query opportunity cost scales linearly with the size of the memory.

\subsubsection{QRAM categorization based on query opportunity cost }

Rather than costing a QRAM query by the circuit depth (or time cost), the total resource overhead of a QRAM query is better measured by the opportunity cost: the classical compute that could otherwise have been performed with the resources allocated to the QRAM. For instance, the energy required to power $O(1)$ lasers could instead be used to power $O(1)$ CPUs. 
As a more concrete example, a QRAM circuit with $\tilde{\Theta}(N)$ constant-energy gates incurs a total opportunity cost of $\tilde{\Theta}(N)$, regardless of whether the gates are executed sequentially or in parallel.  We further define the \textit{total query cost} by excluding any one-time cost to initialize the QRAM (e.g., loading data into memory). Such one-time costs are often a property of the algorithm invoking the QRAM rather than of the QRAM itself. 
One can also make a further distinction between the quantum query cost (any costs to run the quantum aspects of the QRAM), and the classical query cost (any separate classical co-processing), depending on whether the resources are spent in classical co-processing or in enacting quantum operations. 

We can then define the different resource-based categories for a QRAM with $n = \log_2 N$ address bits. An \textbf{active QRAM} (Sec.~\ref{sec:qram:subsec:active_qram}) has $\Omega(N)$ total query cost. It is useful in specific applications, but will often lose the speedup of the quantum algorithm using it.  A \textbf{passive QRAM} (Sec.~\ref{sec:qram:subsec:passive_and_strongly_passive_qram}) has a total cost per query of $o(N)$. Many quantum algorithms instead require a strongly passive QRAM, i.e., one with total query cost $O(\mathrm{polylog}(N))$. A \textbf{quantum-passive classical-active QRAM} (Sec.~\ref{sec:qram:subsec:quantum_passive_classical_active_qram}) is one whose total quantum query cost is passive, but whose total classical query cost is active.\footnote{This could be generalized to a ``hybrid-cost'' QRAM, to allow different asymptotic pairings of quantum and classical resources, but we are not aware of any relevant results outside of the quantum-passive classical-active hybrid-cost regime.} This regime potentially has important implications, merits additional research. A \textbf{practically passive QRAM} (Sec.~\ref{sec:qram:subsec:practically_passive_qram}) refers to a QRAM that is asymptotically active, but with constant overheads so small that it functions as a passive QRAM in practice. This regime represents one of the most feasible avenues toward general QRAM-dependent practical quantum advantage.

\subsection{Active QRAM}\label{sec:qram:subsec:active_qram}

\begin{figure*}[t]
    \centering
    \includegraphics[width=1\linewidth]{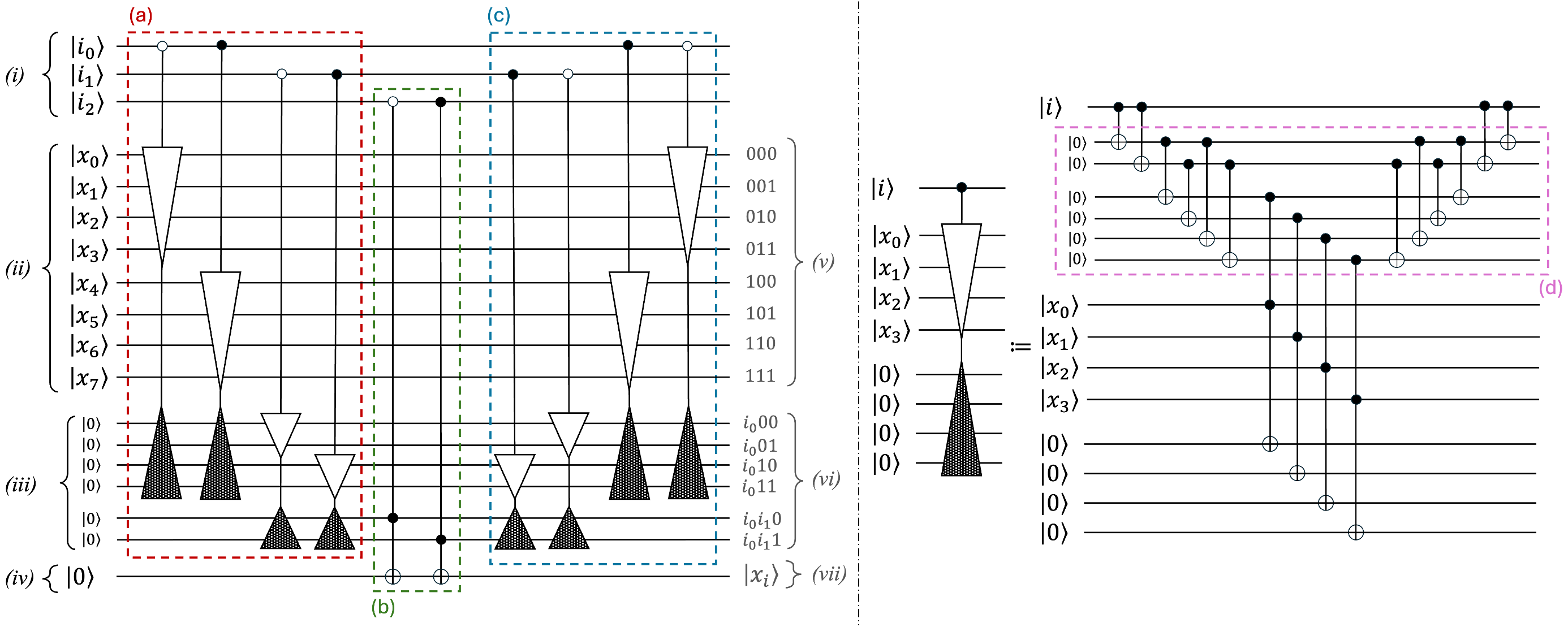}
    \caption{\textbf{Simple example circuit QRAM with 8 memory qubits.} The panel on the left shows a fan-out style circuit QRAM implementation with polylogarithmic depth (in the number of memory qubits). The panel on the right depicts a circuit which copies, in order, the bit values (in the standard basis) in a set of source qubits to a set of $\ket{0}$ initialized target qubits. 
    \textit{(i)} Address qubits, split into single qubits, the binary address $i = i_0i_1i_2$. \textit{(ii)} The memory qubits. \textit{(iii)} Temporary ancilla registers used for routing. \textit{(iv)} Output data register (i.e., this corresponds to the second register in the usual QRAM oracle definition $U_{\text{QRAM}}\ket{i}\ket{0} = \ket{i}\ket{x_i}$). \textit{(v)} The address of each memory qubit as a bitstring. \textit{(vi)} The address of the memory qubit whose data the labeled ancillary qubit contains after box (a) is executed. \textit{(vii)} After boxes (a), (b), and (c) are executed, the output register correctly contains the value $\ket{x_i}$, and the ancillary registers are uncomputed. In the panel on the right, box (d) contains temporary routing qubits used to split the control signal to allow for log depth. Since the ancillary qubits are uncomputed after routing, they may be reused (and the routing can instead be performed once at the start and end of the QRAM call, to asymptotically reduce the circuit depth). Note that control register in copy circuit is single qubit. Consistent with quantum circuit diagrams, white control circles indicate a gate application conditional on a $\ket{0}$ state, while black circles indicate conditional application based on a $\ket{1}$ state. 
    }
    \label{fig:sec:qram:example_circuit_qram}
\end{figure*}

\subsubsection{Active QRAM often loses quantum speedup}
\label{sec:qram:subsec:circuit_qram_is_active_and_active_loses_speedup}

Importantly, a QRAM implemented in the circuit model is active if its stored data does not have special structure, and either (1) the circuit is fully error-corrected, or (2) each gate execution requires a constant amount of energy (e.g., a laser pulse). This is a consequence of Theorem 5.2 of~\cite{jaques2025qram}, which shows that a QRAM capable of storing $N$ arbitrary bits necessitates $\tilde\Omega(N)$ gates. I.e., a generic QRAM cannot be represented, even approximately, by an asymptotically compact circuit.\footnote{This proof does not cover circuits with measurements and measurement-conditioned gate applications. Extending this lower-bound to account for such dynamic circuits is an open problem, however, we believe that the result will still hold.}
Consequently, if the gates require constant energy, the total energy input per query will be $\tilde\Omega(N)$. A similar asymptotic argument holds for a fully error-corrected QRAM (using standard error-correction techniques), considering the total cost to operate the classical control hardware.

We will now summarize a line of research which shows that the opportunity cost of an active QRAM often results in quantum speedup being lost.
First, translating the observation of~\cite{aaronson2015read} into the terminology we use here, linear system solving loses speedup when implemented with active QRAM.
Subsequently, \citet{steiger2016racing} observe that a number of such algorithms, using quantum PageRank as an example, lose speedup when using QRAM when compared to a parallel classical computer with a number of processors proportional to the size of the QRAM. 
\citet{ciliberto2018quantum} observe that these arguments can be made more general and apply to other algorithms as well. In~\cite{jaques2025qram}, they extend these arguments to general QSVT-based algorithms using active QRAM with dense or sparse data. Under realistic physical assumptions they show that comparable parallel classical algorithms usually eliminate quantum speedup.
As, QSVT is a powerful framework unifying many quantum algorithms~\cite{martyn2021grand}, these results highlight how opportunity cost arguments constrain the speedup of a broad class of algorithms. Informally, we expect that similar issues will arise for algorithms whose core subroutine consists of multiplying large unstructured matrices and vectors which are stored in an active memory.
Moreover,~\cite{jaques2025qram} also points to some examples relevant to quantum cryptanalysis where using an active QRAM loses speedup, including~\cite{bernstein2009cost, jaques2019quantum}. Additionally, algorithms like database search will also clearly lose their speedup when using an active QRAM.

\subsubsection{Example: constructing circuit QRAM}\label{sec:qram:subsec:active_qram:constructing_circuit_qram}

In~\Cref{fig:sec:qram:example_circuit_qram}, we present an example implementation of QRAM in the circuit model.
For simplicity, this diagram assumes each memory cell is a single qubit rather than a register containing multiple qubits, but the underlying idea is the same.\footnote{An expanded treatment of this pedagogical construction also appears in the PhD thesis of one of the authors~\cite{Rattew2025Thesis}}.
For a memory storing $N$ qubits, this circuit has $O(\log^2 N)$ total depth, $O(N)$ parallel operations, and $O(N\log N)$ total qubits. By performing the routing for each address qubit once at the start and end of the QRAM circuit, the circuit depth can be brought down to $O(\log N)$.  
The panel on the right of~\Cref{fig:sec:qram:example_circuit_qram} shows a ``controlled copy'' circuit (only basis vectors are copied via controlled CX gates, so this does not violate no-cloning). On the panel on the left, stage (a) shows the routing of the data towards the output register, conditioned on the address bit. In particular, a controlled copy gate is executed twice for each address bit, once with a control activated by $\ket{0}$ and once with a control activated by $\ket{1}$. This has the effect of moving the data with the satisfying address at that bit to the next stage of ancillary registers (of which there are half as many after each address bit is processed). Panel (b) shows the read-out, where the data with the associated input address is routed to so that it can be deterministically XOR'd onto the output register. Panel (c) shows the routing being uncomputed.

Importantly, this figure helps to understand the difference between a bucket-brigade style QRAM and a fan-out style QRAM as discussed in the original proposals~\cite{Giovannetti.08,Giovannetti.08_2}. For a \textit{single}, but unknown, input address basis vector (i.e., not in superposition), the circuit shown in the figure needs to perform $\Omega(N)$ (parallelizable) gates. In contrast, the original bucket-brigade proposals count only $O(\text{polylog(N))}$ routing elements as active for a single basis vector address query, although whether this notion of ``activation'' corresponds to an equivalent reduction in physical resource consumption is implementation-dependent. 

Finally, it is interesting to observe the similarity between a circuit QRAM implementation, and a general state-preparation circuit (see e.g.,~\Cref{sec:sp_bta}).

\subsubsection{Active circuit QRAM}\label{sec:qram:subsec:active_qram:qrom}

In this section we outline the history, directions of research, and open problems in creating circuit QRAM implementations. 
As a note, the first explicit circuit proposal of QRAM that we are aware of was given in~\cite{arunachalam2015robustness}, where they analyze bucket-brigade in the circuit model.

\paragraph{Unary Iteration}
The SELECT operation (see e.g.,~\cite{Childs.18}) covered in~\Cref{sec:sp_ss} is related to circuit-QRAM.
Given a set of unitary matrices $V_j$, and an $n$-qubit control register (with $N=2^n)$, SELECT applies each $V_j$ to an output register, conditioned on the control register being in the $\ket{j}$ basis state, i.e., it is $\sum_{j} \op{j}{j} \otimes V_j$. \citet{babbush2018encoding} independently develop ``unary iteration'' which performs the same mapping but optimizes T-gate count by reusing redundant control patterns (and $\log N$ temporary ancillary qubits) across different controlled gates. As such, they reduce the control-circuitry T-gate count from the $6N-4$ of~\cite{Childs.18}, to $4N-4$ and also consider cases where the number of controlled-operators are not an exact power-of-2. Additionally, they note that their T-gate cost for electronic spectra encoding is dominated by unary iteration.

\paragraph{QROM}
~\citet{babbush2018encoding} also introduce quantum read-only memory (QROM) as a special case of unary iteration, where the matrices $V_j$ are replaced with single qubit $X$ gates. The memory contents are encoded in the presence or absence of an $X$ controlled on the corresponding address basis state. 
For a QROM with $N$ bits the procedure of~\cite{babbush2018encoding} consumes $\log N$ ancilla qubits and has $O(N)$ circuit depth. The leading term of T-count is $4N$, which is better than bucket-brigade approach. 

\paragraph{QROM with lower $T$-count}
Originally posted as a preprint in 2018,~\cite{low2024trading} was the first proposal to achieve a QROM T-gate cost sublinear in $N$, specifically, $O(\sqrt{N})$.\footnote{The total gate count is still  $\Omega(N)$, so this is consistent with the gate lower-bound.} See Sec.~\ref{sec:sp_ss} for more discussions.
In~\cite{berry2019qubitization}, this was referred to as advanced QROM (QROAM). 

A general analysis of various circuit QRAM implementations under the surface code is provided in~\cite{di2020fault}.
In~\cite{haner2022space} they also consider QROM and unitary-iteration like constructions under the constraints of a 2D surface code, with related work in~\cite{xu2023systems}.
\citet{zhu2025unified} provides an overview of the circuit complexities of different circuit QRAM implementations, and recovers existing asymptotics (up-to polylogarithmic factors) while assuming 2D local connectivity. They also investigate error resilience. In,~\citet{mukhopadhyay2025quantum} a circuit QRAM implementation is given with $O(\log\log N)$ T-depth and $O(\sqrt{N})$ T-count. 
\citet{khattar2025rise} provide optimizations for address control routing improving constants for both QROM and unary iteration circuits.

\paragraph{Applications}
Circuit QRAM has been applied to range of applications to optimize resource costs. Examples include optimizing circuits for quantum chemistry~\cite{babbush2018encoding,berry2019qubitization,georges2025quantum,low2025fast}, optimizing arithmetic for elliptic curve discrete logarithm~\cite{haner2020improved}, optimizing general quantum arithmetic implementations~\cite{gidney2019windowed}, optimizing the Toffoli count in sparse state preparation~\cite{rupprecht2026sparse}, and resource analysis frameworks~\cite{harrigan2024expressing}. Additionally,~\citet{jaques2020low} provide an example where an active circuit QRAM still provides an asymptotic speedup over a QRAM-free variant.

\paragraph{Miscellaneous}
Finally, some work explores circuit QRAM implementations
and applications separately from QROM.
Some examples sit at the interface between QRAM and state-preparation, e.g.,~\cite{Yuan.22,Park.19,Veras.20}. Some consider circuit QRAM implementations with different physical gate sets~\cite{allcock2024constant}.

\subsubsection{Error correction for active QRAM}\label{sec:qram:subsec:active_qram:active_qram_outside_circuit_model}

QRAM is generally active when fully error-corrected (unless passive error-correction is feasible), but the constants and overheads may still vary substantially across architectures. Some work studies fault-tolerant circuit QRAM implementations~\cite{Hann.21}, and the error-resilience of bucket-brigade-like routing may allow QRAM-specific reductions in overhead compared with treating QRAM like an arbitrary circuit. Additional work has explored heterogeneous error-correction in the context of QRAM~\cite{singal2025heterogeneously}.

An interesting open direction is to combine active QRAM with the quantum-memory model discussed in~\Cref{sec:qram:subsec:quantum_memory}. This could allow different parts of the QRAM to receive different ``amounts'' of error-correction, accounting for the intrinsic error-resistance of certain QRAM proposals, thereby reducing overall error-correction overhead. For example, in a bucket-brigade architecture, nodes closer to the leaves might be able to use higher rate error-correcting codes than those closer to the root.

Another direction is to bypass error correction and realize active QRAM directly on physical hardware. The earliest proposal is based on atom-photon interaction assisted by cavities~\cite{Giovannetti.08_2}. There are other proposals based on neural atoms~\cite{hong2012robust}, superconducting circuits~\cite{Hann.19,wang2025quantum,shen2026bucket} and photonic integrated circuits~\cite{chen2021scalable}. Due to the intrinsic robustness of the bucket-brigade architecture, these approaches are expected to have reasonable accuracy, at least for a small scale case. Nevertheless, the challenge of interfacing such a noisy QRAM with an error-corrected quantum computer, a challenge which is covered in more detail for passive QRAM, would remain.

\subsection{Passive QRAM}\label{sec:qram:subsec:passive_and_strongly_passive_qram}

\subsubsection{Requirements for an end-to-end passive QRAM}
By definition, a passive QRAM with $N = 2^n$ bits in memory has a total query cost of $o(N)$.
I.e., the total access cost is sublinear in the size of the memory. 
For a QRAM to be fully passive, the entire stack must be passive. This includes at the physical device-level, at the architectural-level, and the system-level fault-tolerant integration. At the physical level, the device must be able to enact a noisy QRAM query with $o(N)$ total cost per query (e.g., total energy input, time, etc). Moreover, unless passive error-correction becomes possible, the QRAM cannot be fully error-corrected without becoming active (thereby losing utility for the broad set of algorithms outlined in~\Cref{sec:qram:subsec:circuit_qram_is_active_and_active_loses_speedup}). Consequently, a passive QRAM architecture likely needs to be intrinsically robust to errors (ideally, with per-component error rates scaling like $O(1/\mathrm{polylog}(N))$). Finally, this physical noisy QRAM device needs to somehow be queried by an error-corrected quantum computer.

\subsubsection{Challenges for constructing interaction-free passive QRAM}

If a QRAM is completely free from external intervention during a query, it is ``ballistic'', and its evolution must be described by a time-independent Hamiltonian $H$, i.e., $U_{\mathrm{QRAM}} = e^{i H t}$. Lemma 6.1 of~\cite{jaques2025qram} shows that for an $N$-bit QRAM implemented by a Hamiltonian summed from $n$, $W$-local $m$-qubit Pauli, with coefficient vector $\bm a = (a_0, ..., a_{n-1})$, then the maximum general table size which can be implemented is bounded by $N \in \tilde O(W n \lnorm{a}_1 t)$. However, this proof only applies to the initial construction cost of the QRAM, and so this restriction only makes the QRAM active if this cost must be paid upon every query. Moreover, this proof does not apply to Hamiltonians with external interventions.

\subsubsection{Architectural noise resilience}\label{sec:qram:subsec:architectural_noise_resilience}

Since a passive QRAM architecture cannot be fully actively error-corrected without becoming active, it likely needs to possess substantial intrinsic noise resilience. In particular, if a QRAM with $N = 2^n$ bits in memory requires physical error-rates on the order of $O(1/N)$, scaling such an architecture would be extremely challenging. 
However, the original bucket-brigade proposal~\cite{Giovannetti.08, Giovannetti.08_2} fortunately has very favorable error-scaling properties~\cite{arunachalam2015robustness,hann2021practicality,jaques2025qram}. A key difference between the bucket-brigade architecture and a fan-out architecture (e.g., the one shown in~\Cref{fig:sec:qram:example_circuit_qram}) is that given a \textit{single} input basis vector, a fan-out architecture activates $O(N)$ nodes, while bucket-brigade can potentially only activate $O(\mathrm{polylog}(N))$ nodes, although from an end-to-end perspective this depends on the underlying physical implementation.

One of the first papers to examine this error-scaling in detail was~\cite{arunachalam2015robustness}. 
One of the important challenges they observe is that while bucket-brigade has polylogarithmic error-scaling in the memory size \textit{for a single query}, repeated queries can still cause error to accumulate. They show that for quantum search algorithms making $O(\sqrt{N})$ queries that the physical error-rate must scale as $o(1/\sqrt{N})$. More generally, with physical error rates inverse polylogarithmic in the memory size, only $O(\mathrm{poly}(n))$ queries can be made to such a bucket-brigade QRAM while preserving an overall constant error-rate. 
The purification process of~\cite{dalzell2025distillation} provides a way around this (discussed further in Sec~\ref{sec:qram:subsec:quantum_passive_classical_active_qram}), allowing for multiple QRAM resource states to be consumed to suppress the overall error-rate linearly in the number of resource states consumed (without requiring infeasible physical error-rates). However, this would likely still preclude utility for many quantum search style algorithms making $O(\sqrt{N})$ logical queries to a QRAM, as $\tilde O(N)$ queries to the physical QRAM device would still be made in total.

In~\cite{Hann.21} they substantially generalize the error model studied in~\cite{arunachalam2015robustness}, and find that the error-resistant properties of bucket-brigade hold for arbitrary error channels. A bucket-brigade architecture can be visualized as a binary tree routing from the address register (at the root) to the memory registers (at the leaves). Their analysis formalizes the following intuition. Essentially, the error-resistant properties follow from the fact that a router at the $i^{th}$ level of the tree (with $i=0$ the root) is only active in $N 2^{-i}$ of the $N$ branches of a superposition (consider e.g., a uniform superposition in the address register). Consequently, the entanglement entropy of an individual router in the tree decays exponentially with its depth in the tree, and so routers near the leaves are only weakly entangled, limiting the impact of local errors.

\subsubsection{Logical access }\label{sec:qram:subsec:passive_qram:subsubsec:logical_access_to_physical_qram}

As previously discussed, a passive QRAM cannot be protected by conventional fault-tolerant error-correction techniques across all of its components. Consequently, a passive QRAM will likely consist of at least some physical qubits, resulting in a challenge of how those physical qubits can be interfaced with a main error-corrected quantum processor. 
One possibility is to distill a QRAM resource state and then teleport it~\cite{jaques2025qram}. 
The first proposal suggesting QRAM teleportation was presented in~\cite{chen2021scalable}, although the architecture is active overall.

However, Theorem 7.1 of~\cite{jaques2025qram} establishes a no-go theorem for the teleportation-distillation paradigm, proving that to achieve a sufficiently small constant error, $d = \Omega(\sqrt{N})$ queries are required to the physical QRAM device. Fortunately, this no-go theorem assumes that the purification channels are independent of the stored data, and all queries target the same fixed QRAM. The protocol in~\cite{dalzell2025distillation} circumvents this by classically updating the QRAM memory contents after each teleportation attempt, conditioned on the measured output bitstring. While the total quantum cost of their approach is polylogarithmic in the size of the memory, the classical cost is linear (up-to polylogarithmic factors) in the memory size. We believe that this approach is currently the most promising candidate for the fault-tolerant access of a physical noisy QRAM system, and we discus it in greater depth in~\Cref{sec:qram:subsec:quantum_passive_classical_active_qram}.~\citet{cesa2025fast} instead precompute a QRAM resource state which is independent of the data, which is then consumed to enact a query. However, this approach requires enacting $O(N)$ Clifford gates per query, making it an active proposal. Moreover, the resource state of~\citet{cesa2025fast} is of size linear in $N$, whereas the resource state of~\cite{dalzell2025distillation} is of size $O(\mathrm{polylog}(N))$. 
Exploring additional ways to efficiently interface noisy QRAM devices with fault-tolerant quantum computers remains an interesting open problem.

\subsubsection{Weakly passive QRAM}\label{sec:qram:subsec:weakly_passive_qram}
The intermediate sublinear-cost regime, termed weakly passive, covers QRAM whose query cost is between $o(N)$ and $\omega(\mathrm{polylog}(N)$. Note that the distillation-teleportation no-go theorem covered in Sec.~\ref{sec:qram:subsec:passive_qram:subsubsec:logical_access_to_physical_qram}) does not rule out the possibility of $\Theta(\sqrt{N})$ query cost. If achieved, there may exist applications evading the loss of speedup outlined in~\Cref{sec:qram:subsec:circuit_qram_is_active_and_active_loses_speedup}.

We note that some work explores parallel access to QRAM, e.g.,~\cite{beals2013efficient}. Recent work also revisits parallel QRAM access as a route to amortizing query costs, for example in~\cite{xu2025fat}.
Interestingly, if an active QRAM can serve a query to $P$ quantum computers in parallel with minimal overhead,  the amortized cost per query is $\tilde O(N/P)$. For sufficiently growing $P$, and ignoring speed of light constraints, this could be considered a passive QRAM on an amortized basis. Moreover, if it is possible to efficiently serve $P \in \tilde\Theta(N)$ processors in parallel, then it could even be a strongly passive QRAM. The feasibility of achieving passive QRAM through such amortized sharing becomes an efficient routing problem, see Section 5.5 of~\cite{jaques2025qram} for more information.

\subsubsection{Strongly passive QRAM}
Most quantum algorithms that assume cheap QRAM implicitly rely on the strict condition of a strongly passive QRAM, one with a total query cost of $O(\mathrm{polylog}(N))$. However, simultaneously addressing the physical, architectural, and fault-tolerant issues in this strongly passive regime presents substantial challenges. For example, Theorem 7.1 of~\cite{jaques2025qram} rules out strongly passive QRAM for fixed, dataset-independent protocols. The adaptive protocol by~\cite{dalzell2025distillation} also incurs classical co-processing cost of $\tilde{O}(N)$.
A  theoretical question then follows.
\begin{challenge}\label{prob:qram}
Is a fully strongly passive QRAM inherently impossible?
\end{challenge}
Proving such a universal impossibility result, however, seems non-trivial. Consider a thought experiment involving arbitrary precomputation and ancillary space. By unrolling and pre-storing every possible adaptive correction table from Dalzell et al.'s protocol, one incurs an immense one-time precomputation cost in space and time: roughly $\tilde{O}(2^{n(n-1)/2})$. However, because this one-time cost is excluded from the query accounting, the marginal query cost, if speed of light delays and physical routing limits are neglected, becomes $O(\mathrm{polylog}(N))$ for both the quantum and classical processors. While practically absurd, demonstrates that any rigorous impossibility proof for strongly passive QRAM should formally account for  ``practicality''.

\subsection{Beyond passive and active QRAM}\label{sec:qram:subsec:other}
\subsubsection{Quantum-passive classical-active QRAM}\label{sec:qram:subsec:quantum_passive_classical_active_qram}

Not all QRAM proposals have the same asymptotic quantum and classical cost. Importantly, a quantum operation is expected to be orders of magnitude more expensive than a single classical operation~\cite{babbush2021focus}, and so transferring compute from an error-corrected quantum computer to a classical coprocessor is desirable when possible.

An important recent advancement was presented in~\cite{dalzell2025distillation}. Given a desired query error (in diamond norm) of $O(1/\mathrm{poly}(n))$, access to a physical QRAM device with $O(\mathrm{poly}(n))$ query cost and $\Omega(1/\mathrm{poly}(n))$ resource state preparation fidelity, and assuming a dataset independent noise-model, they show how a QRAM storing $N$ bits of data can be queried with $O(\mathrm{polylog}(N))$ total quantum cost, and $\tilde O(N)$ classical cost (i.e., strongly-passive quantum cost and active classical cost). 
Importantly, they sidestep the distillation-teleportation no-go theorem of~\cite{jaques2025qram} by adaptively updating the memory contents of the QRAM during a query. They use a slightly different QRAM query definition, which can be used to obtain a regular QRAM query. In particular,  given a state $\sum_x \alpha_x\ket{x}$ their goal is to obtain $\sum_x (-1)^{f(x)}\alpha_x\ket{x}$, where the QRAM table stores $f(x)\in\{0,1\}$ at address $x$. 
In summary, their procedure has two parts: distillation and teleportation. 
First, they assume access to a physical QRAM device capable of preparing a noisy resource state with $\Omega(1/\mathrm{poly}(n))$ fidelity with $\frac{1}{\sqrt{N}}\sum_x(-1)^{f(x)}\ket{x}$. In terms of the target error-parameter, for fixed input fidelity, their state-agnostic distillation requires $\tilde O(1/\epsilon)$ samples.
Akin to T-state teleportation, they then teleport this state, and readout an $n$-bit string $m$, which informs that the resulting state is $\sum_{x}(-1)^{f(x\oplus m)}\alpha_x\ket{x}$. They then observe that a QRAM table storing $f(x\oplus m)\oplus f(x)$ is the correction operation, so they update the classical table (at $O(N)$ classical cost), distill the new resource state, and teleport that. This process then repeats $n= \log_2(N)$ times, with each teleportation usually dropping the level of the Clifford hierarchy the correction operation (specified by the updated table) occupies by 1, proving that after $n$ rounds the quantum state will be $\sum_x(-1)^{f(x)}\alpha_x\ket{x}$, and thus that the QRAM query was successfully enacted. 

There are several remaining open problems. 
First, can this procedure be modified such that the total classical cost can be made $o(N)$ (while preserving the quantum efficiency), or can it be proven that this is impossible? Second, they prove that in a parallel processing model with $O(N)$ processors, sparse matrix-vector multiplication can be reduced to local arithmetic and $O(\mathrm{poly}(n))$ calls to their classical update rule, again introducing opportunity cost arguments for applications which may wish to use such a QRAM implementation. Consequently, can the classical update rule be made to have cost comparable to the active cost of classical DRAM (see~\cref{sec:qram:subsec:practically_passive_qram} for more information)? Third, can the distillation procedure be made to have $o(1/\epsilon)$ complexity scaling? Their distillation procedure is broadly optimal for state-agnostic distillation, but can the structure of the QRAM table, or structure similar to their dynamic table update for teleportation be used to improve the distillation complexity. 
Such an improvement could potentially allow for QRAM usage beyond the $\mathrm{polylog}(N)$ query regime (discussed in Sec.~\ref{sec:qram:subsec:architectural_noise_resilience}), and speculatively could be combined with table permutations intended to average over spatially localized or persistent component failures.
Finally, it would be worth exploring optimizing this procedure specifically for the purpose of being used as QROM-like lookup table for resource optimization. The authors investigate applications of their procedure in QROM-like settings for cryptanalysis and quantum chemistry, and find a mixed picture. Improvements are likely be necessary for broad applicability as a QROM replacement. 

A more speculative route towards a quantum-passive classical-active architecture may arise from QROAM's $O(\sqrt{N})$ T-gate cost. In architectures where logical Clifford operations can be absorbed into classical frame tracking, the incremental logical quantum cost could be sublinear. Nevertheless, the classical tracking and the physical quantum cost of error-correction would likely remain $\tilde {O}(N)$. 

\subsubsection{Practically passive QRAM}\label{sec:qram:subsec:practically_passive_qram}

The concept of practically passive QRAM was first introduced in~\cite{rattew2025accelerating}.
The motivation lies in the limitations of asymptotic definitions. For instance, a constant-time constant-energy laser pulse is asymptotically equal to fixed sized CPU performing a constant time calculation. But their difference in constants are of many orders of magnitude.

The goal of a practically passive QRAM is to reduce the total cost per query (e.g., energy cost) to a level comparable to that of classical memory (e.g., DRAM). Thus, the first step would be to characterize the total energy costs of DRAM. The memory cells in DRAM require periodic refreshes (typically every 32–64 ms) to maintain data, incurring an $\Omega(N)$ energy overhead per cycle~\cite{bhati2015dram}. Thus, if $k$ queries are issued within one refresh window, the average energy per query can be expressed as:  $\frac{E_{\mathrm{refresh}}}{k} + E_{\mathrm{access}}$, where $E_{\mathrm{refresh}}$ is the total refresh energy and $E_{\mathrm{access}}$ is the single-access energy. Here, $k$ depends on specific applications, and $E_{\mathrm{refresh}}, E_{\mathrm{access}}$  can be obtained by referencing DRAM manufacturer specification sheets. See, e.g.,~\cite{chandrasekar2011improved} for a more detailed model.

A promising candidate for a large,  error-corrected, practically passive QRAM is, again, the adaptive distillation-teleportation protocol of~\cite{dalzell2025distillation} discussed in~Sec.~\ref{sec:qram:subsec:quantum_passive_classical_active_qram}. From a practical standpoint, it would be important to reduce the overhead of their classical update rule to make it comparable to the cost of DRAM. The following illustrates the difference in cost.
In a simplified classical RAM model, RAM can have depth $O(n)$, width $O(2^n)$ and total wire length $O(n2^n)$, resulting in constant (w.r.t. $n$) wire density. However, in~\cite{dalzell2025distillation}, they show that their classical update rule has wire length $\Omega(2^{3n/2})$ and wire density $\Omega(2^{n/2}/\mathrm{poly}(n))$ -- exponentially worse than the density of RAM.\footnote{We present their result setting the number of spatial dimensions $d$ to $d=2$.} Moreover, in a parallel computation model, where they have $O(N)$ parallel processors with local memory and which can perform local compute, they show sparse matrix-vector multiplication can be reduced to their classical update rule. Ideally, such a classical update rule would have costs closer to a DRAM access than a sparse matrix-vector multiplication. 
Moreover, we believe that an interesting question would also be to reconcile the exact cost of their classical update (under certain implementations) in relation to the exact costs of a DRAM with similar size.

%% file: secs/sparse.tex
\section{Encoding structured data\label{sec:sparse}}
We have discussed state preparation, unitary synthesis, and QRAM in previous sections. Our previous sections focused on the most general settings, so the circuit size grows linearly with data dimension. This linear dependency may limit quantum speedup to at most polynomial. Fortunately, real-world data often exhibit structure, so the actual circuit complexity for data loading in practical applications can be significantly lower than the worst-case bounds. 

As a typical example, one of the common and useful structures is the sparsity. Many vectors/matrices of practical interest contain only a small number of nonzero entries relative to their dimension. This property can be exploited to design more efficient data‑loading circuits and to improve the overall efficiency of quantum algorithms. As a typical example, estimating the ground energy of local Hamiltonian is generally a $\mathsf{QMA}$-complete problem (i.e. quantum analogue of $\mathsf{NP}$-complete problem). However, if we are further provided a sparse guiding quantum state which has nontrivial overlap with the ground state, the ground energy estimation problem can be $\mathsf{BQP}$-complete (i.e. the class of hardest problems that are efficiently solvable by quantum computers) for inverse-polynomial-precision~\cite{gharibian2022dequantizing}.

This section is organized as follows. In Sec.~\ref{sec:sparse_sp}, we introduce sparse quantum state preparation. In Sec.~\ref{sec:sparse_saim}, we consider the sparse access input model, which is a commonly used query model for quantum simulation and linear algebra. In Sec.~\ref{sec:bool}, we consider Boolean unitaries. In Sec.~\ref{sec:sparse_iso}, we introduce the synthesis of isometry, a norm-preserving map from an $M$-dimensional input space to an $N$-dimensional output space.  In Sec.~\ref{sec:sparse_unitary}, we introduce sparse unitary synthesis. In Sec.~\ref{subsec:cnot_circuit_clifford}, we discuss Clifford circuit. In Sec.~\ref{sec:ecf}, we introduce the encoding of efficiently computable functions. In Sec.~\ref{sec:sqram}, we discuss QRAM for structured data. In Sec.~\ref{sec:initial_state}, we introduce initial states for ground state preparation applications. In Sec.~\ref{sec:str_other}, we introduce some other structured data.

\subsection{Sparse quantum states\label{sec:sparse_sp}}

We now consider the preparation of $S$-sparse quantum states for some integer $S\ll N$. Such sparse states can be expressed as  
\begin{align}\label{eq:sparse_s}
|\psi\rangle=\sum_{s=0}^{S-1}\alpha_s|q_s\rangle.
\end{align}
Here, $q_s$ represents the $s$-th computational basis with non-zero amplitude. 

\subsubsection{no- or few-ancilla case}
We first introduce some approaches for preparing sparse states with no or constant number of ancillary qubits~\cite{Malvetti.21,Gleinig.21,Veras.21,mao2024toward,li2025nearly}. 

The approach of~\cite{Malvetti.21,li2025nearly} is based on basis permutations. The protocols begin by preparing the first $\lceil\log_2S\rceil$ qubits to a quantum state $$\sum_{s=0}^{S-1}\alpha_s|s\rangle\otimes|0\cdots0\rangle.$$ Then, one performs a permutation gate that reorders the 0s and 1s in the computational basis 
\begin{align}\label{eq:sparse_switch}
|s\rangle\otimes|0\cdots0\rangle\Longrightarrow|q_s\rangle
\end{align} for all $s$.~\cite{Malvetti.21} performed this permutation for each $s$ sequentially, and achieves $O(S n+\log S)$ CNOT count.~\cite{li2025nearly} developed an improved quantum circuit for sparse permutation operators (Sec.~\ref{sec:sbu}). Combining with the optimal gate count for general state preparation used in the first step,~\cite{sun2023asymptotically} achieved a circuit size of
\begin{align}\label{eq:sparse_ocs}
O\left(\frac{nS}{\log(n)}+n\right).
\end{align}  

An alternative approach in~\cite{Gleinig.21} is based on a subroutine that reduces the nonzero amplitudes, i.e. transforms an $S$-sparse state to a  $(S-1)$-sparse state. Applying this subroutine iteratively and inverting the quantum circuit, their method requires $O(Sn)$ CNOT gates and $O(S\log S+n)$ single-qubit gates.~\citet{Veras.21} further considered quantum states that are both sparse in the number of non-zero amplitudes and have bounded Hamming weight, called double sparse states.
Based upon Eq.~\eqref{eq:sparse_s}, they further assume that the Hamming weights (number of bits with value $1$) of $|q_s\rangle$ are at most $k$. The algorithm iteratively prepares the state $\sum_{s=0}^t\alpha_s|q_s\rangle$, from $t=0$ to $t=S-1$. 
To reduce the total cost, the basis $q_s$ here is sorted according to the Hamming weight. This method, called CVO-QRAM, achieves circuit size $O(Sk)$ using one ancillary qubit. 
Based on CVO-QRAM,~\citet{mao2024toward}  observed that, for a  batch of bitstrings, many bit positions are identical across the batch. Namely, a batch of $\log(n)-\omega(1)$ strings has at most $o(n)$  distinct patterns among positions. This property can help reduce the cost of multi-qubit controlled rotations, a key step in CVO-QRAM. Their improved Batch Elimination (BE)-QRAM achieves circuit size in Eq.~\eqref{eq:sparse_ocs}, using two ancillary qubits.

When ancillary qubit number is a $O(1)$, the best known circuit size in Eq.~\eqref{eq:sparse_ocs} is optimal for large classes of circuits models~\cite{mao2024toward,li2025nearly}. These classes include circuits that are not amplitude-aware (i.e. the structure depends solely on the basis $|q_s\rangle$ rather than the amplitudes $\alpha_s$)~\cite{mao2024toward}, as well as circuits constrained in the number or type of single-qubit gates~\cite{li2025nearly}. However, the best-known unconditional lower bound, established via counting arguments, is only $\Omega(n+S)$, indicating that the optimal circuit size is still an open question.

\subsubsection{Space-time tradeoffs}
Similar to general state preparation and unitary synthesis, sparse state preparation also admits a space-time tradeoff~\cite{zhang2022quantum,zhang2024circuit,ramacciotti2024simple,li2025nearly,luo2025space}. 

~\cite{zhang2022quantum} realized Eq.~\eqref{eq:sparse_switch} for all $s$ in parallel, using a binary tree architecture. The result showed that sparse state can be prepared by optimal circuit depth 
\begin{align}\label{eq:sparse_ocd}
\Theta(\log(nS))
\end{align}  using $O(nS\log S)$ ancillary qubits, and total circuit size $O(nS)$.~\cite{zhang2024circuit} further considered the intermediate ancillary qubit number, showing that with $m$ ancillary qubits with $\Omega(n)<m<O(nS\log S)$, one can achieve $O(nS\log S\frac{\log m}{m})$ circuit depth. 

Ref~\cite{li2025nearly} improved the circuit size using unary encoding. With $m$ ancillary qubits, the circuit size can be reduced to 
\begin{align}
O\left(\frac{nS}{\log(n+m)}+n\right)
\end{align}
for $m\in O\left(\frac{nS}{\log(nS)}+n\right)$. This is close to the lower bound $\Omega\left(\frac{nS}{\log(n+m)+\log S}+n\right)$ they derived. In a subsequent work~\cite{luo2025space}, the space-time tradeoff is also considered using similar techniques, achieving  
\begin{align}
O\left(\frac{nS\log m}{m\log(m/n)}+\log(nS)\right)
\end{align}
 circuit depth for ancillary qubit number $m\geqslant6n$. In particular, the optimal circuit depth Eq.~\eqref{eq:sparse_ocd} is achieved with asymptotically minimum ancillary qubit number $\Theta(nS/\log S)$ . 
 
The approaches in~\cite{li2025nearly,luo2025space} are based on the $(n, r)$-unary encoding. Specifically, the $n$-qubit basis $q_s\equiv q_{s}^{(n)}q_{s}^{(n-1)}\cdots q_{s}^{(1)}$ is divided into $n/r$ parts, each of length $r$. By defining $q_s(a,b)\equiv q_s^{(b)}\cdots q_s^{(a+1)}$, the basis can be rewritten as $|q_s\rangle=|q_{s}(n-r,n)\rangle|q_{s}(n-2r,n-r)\rangle\cdots|q_{s}(0,r)\rangle$. The initial trivial all-zero state is first prepared to the following intermediate state
\begin{align}\label{eq:sparse_unary}
\sum_{s=0}^{S-1}\alpha_s|e_{q_s(n-r,n)}\rangle|e_{q_s(n-2r,n-r)}\rangle\cdots|e_{q_s(0,r)}\rangle,
\end{align}
where $e_{q_s(0,r)}$ represents the unary encoding of $q_s(0,r)$, and similar for other parts.~\footnote{For example, when $n=8$, $r=2$, the $(8,2)$-unary encoding for $q_s=11011000$ is defined as $e_{11}e_{01}e_{10}e_{00}=0001\,0100\,0010\,1000$.  } The target state is obtained by transforming each basis in Eq.~\eqref{eq:sparse_unary} to $|q_s\rangle|0\cdots0\rangle$. The unary encoding of each part requires $2^r$ qubits, so Eq.~\eqref{eq:sparse_unary} is an  $n2^r/r$-qubit quantum state. So one can tune the space/time complexity by choosing different $r$.

\subsection{Sparse access input model\label{sec:sparse_saim}}
Given a sparse matrix $H$ with entries represented by binary numbers, the sparse access input model for $H$ is a pair of unitaries satisfying 
\begin{subequations}\label{eq:saim}
\begin{align}
O_H|x,y\rangle|z\rangle&=|x,y\rangle|H_{x,y}\oplus z\rangle,\label{eq:saim_a}\\
O_F|x\rangle |k\rangle&=|x\rangle|F(x,k)\rangle,\label{eq:saim_b}
\end{align}
\end{subequations}
where $H_{x,y}$ is the binary entry at the $x_{\text{th}}$ row and $y_{\text{th}}$ column of $H$, $F(x,k)$ is the $k_{\text{th}}$ nonzero element at the $x_{\text{th}}$ row. We say that $H$ is S row-sparse/column-sparse if there are at most $S$ nonzero entries at each row/column of $H$, and we define $F(x,k)=0$ for all $k>S$.
SAIM 
is originally proposed for simulating sparse Hamiltonians~\cite{aharonov2003adiabatic}, then found applications in quantum walk, solving linear systems, etc~\cite{Harrow.09,childs2010on,childs2011simulating,childs2017quantum,gilyen2019quantum,chakraborty2018power,babbush2023exponential}. In all these tasks, quantum algorithms are promised to be efficient when Eq.~\eqref{eq:saim} can be realized by polynomial size quantum circuit, namely $H$ is efficiently computable.

Taking Hamiltonian simulation as an example, the sparse Hamiltonian lemma in~\cite{aharonov2003adiabatic} showed that when $H$ is S row-sparse and efficiently computable, it can be efficiently decomposed as 
\begin{align}\label{eq:sparse_hm}
H=\sum_{m=1}^{M}H_m,
\end{align} 
for some $M\leq(S+1)^2n^6$, and  $H_m$ is some sparse Hamiltonian whose evolution $e^{-iH_mt}$ can be well approximated by polynomial-size quantum circuit. Using Trotter decomposition~\cite{lloyd1996universal}, $e^{-iHt}$ can then be simulated efficiently with arbitrary accuracy. Comparing to the tensor product structure assumed in~\cite{lloyd1996universal}, SAIM generalizes the applicability of quantum simulation.

Mathematically, Eq.~\eqref{eq:saim} is structurally identical to QRAM (Sec.~\ref{sec:QRAM}), as the first register remains unchanged and the second register stores binary data. For sparse $H$ without extra assumptions, the construction methods in Sec.~\ref{sec:QRAM} are all applicable for SAIM. It is important to note, however, that the QRAM-based SAIM circuit size scales at least linearly with the dimension $N$. Furthermore, by counting argument, the worst-case circuit size lower bound for SAIM is $\Omega(SN)$~\cite{zhang2024circuit}. These results imply that sparsity alone is insufficient to guarantee an exponential quantum advantage. For this advantage to be feasible in practice, the matrix $H$ must possess additional structure that makes it efficiently computable.

\subsection{Boolean unitaries\label{sec:bool}}
The entries of Boolean unitaries are either $0$ or $1$. The corresponding circuits are also called \textit{reversible logic circuits}, due to their information-lossless nature. Moreover, 
 because of the unitarity, there is exactly one entry of $1$ in each row and column, while all other entries are $0$. For any input state, Boolean unitaries swaps the amplitudes of different bases, so they are also called \textit{permutation gates}. Some important subroutines fall under this framework, including CNOT circuits and $\text{C}^{n}$-$X$ gates, and SAT-oracle.

\subsubsection{CNOT circuits}\label{sec:b_CNOT}

Practical quantum circuit may contain blocks of CNOT gates, therefore optimizing these blocks can significantly reduce the overall circuit size. A typical example is \textit{Clifford circuits}, i.e.~those that contain CNOT, Hadamard and $S$ gates only. As will be discussed in Sec.~\ref{subsec:cnot_circuit_clifford}, any Clifford  circuit can be decomposed into an 11-stage canonical form 
$\mathrm{H}\text{--}\mathrm{C}\text{--}\mathrm{S}\text{--}\mathrm{C}\text{--}\mathrm{S}\text{--}\mathrm{C}\text{--}\mathrm{H}\text{--}\mathrm{S}\text{--}\mathrm{C}\text{--}\mathrm{S}\text{--}\mathrm{C}$. 
 This normal form makes it clear that the nontrivial entangling cost in Clifford synthesis is concentrated in the CNOT layers, and also indicates that the optimization of Clifford circuits largely reduces to optimizing CNOT circuits.

\paragraph{Circuit size} 
The  circuit size lower bound  for CNOT circuit can be established by counting argument~\cite{shende2003synthesis,patel2008optimal} as $s=\Omega\left(\frac{n^2}{\log n}\right)$. When qubit connectivity constraints are imposed, the circuit size lower bound becomes different. We can formalize the connectivity structure by a topological graph $G(V,E)$, where each vertex $v \in V$ corresponds to a qubit and each edge $e = (v_1,v_2) \in E$ specifies a qubit pair $(v_1,v_2)$ on which two-qubit gates may act on. The maximum degree of $G$ is denoted as $\Delta$. Then, the circuit size lower bound becomes $s= \Omega\!\left(\frac{n^2}{\log \Delta}\right)$, which holds for arbitrary two-qubit gates~\cite{wu2023optimization}.

\begin{figure}
    \centering
    \includegraphics[width=1.0\linewidth]{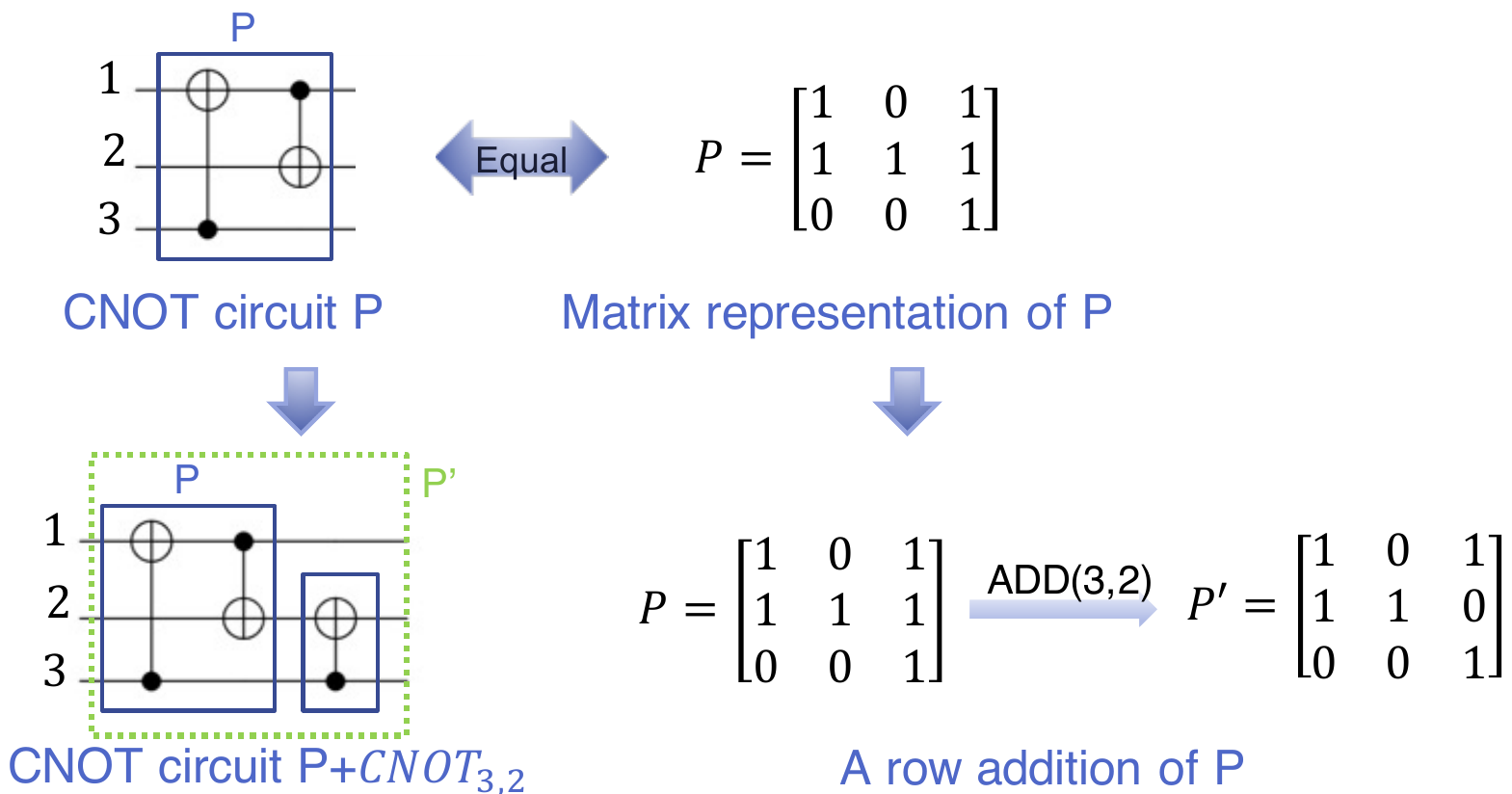}
    \caption{Ref.~\cite{wu2023optimization}. Example for CNOT circuit construction and row addition operation.}
    \label{fig:limited_structure_cnot_example}
\end{figure}

An important property of the CNOT gate is its linearity. For bit strings $a, b\in\{0,1\}^2$, we have $\text{CNOT}|a\oplus b\rangle=\text{CNOT}|a\rangle\oplus\text{CNOT}|b\rangle$, where $\oplus$ is modulo 2 addition. Since any CNOT circuit $\Ccal$ implements a linear invertible Boolean function $f_{\Ccal}:\cbra{0,1}^n \to \cbra{0,1}^n$, it can be associated with an invertible Boolean matrix $P \in \cbra{0,1}^{n \times n}$, as illustrated in Fig.~\ref{fig:limited_structure_cnot_example}.  
Specifically, the $j$-th column of $P$ is defined as the $n$-bit output string of $f_{\Ccal}(e_j)$, where $e_j$ denotes the $n$-bit basis vector with a single $1$ at the $j$-th position and zeros elsewhere. Although $P$ has a significantly smaller dimension than the corresponding unitary transformation matrix, it is a complete description due to the linearity of the CNOT circuit.

A key advantage of the above representation is the correspondence between the CNOT gate and row addition. As illustrated in Fig.~\ref{fig:limited_structure_cnot_example}, let $\mathrm{CNOT}_{i,j}$ be the CNOT gate with control qubit $i$ and target qubit $j$. If we append $\mathrm{CNOT}_{i,j}$ to a circuit represented by $P$, the resulting new matrix $P'$ is obtained by modulo 2 adding 
the $i$-th row of $P$ to the $j$-th row. 
 Thus,  constructing a CNOT circuit for $P$ is equivalent to reducing $P$ to identity via row additions and then performing inversion. The corresponding circuit size is equivalent to the number of row-addition operations. 
 
Mathematically, reducing $P$ to identity can be realized by Gaussian elimination or LU-decomposition using $O(n^2)$ row-additions, yielding $O(n^2)$ circuit size~\cite{beth2001quantum}.~\citet{patel2008optimal} further optimized the circuit size to 
$$\Theta\left(\frac{n^2}{\log n}\right)$$
 by appropriately grouping different rows together and eliminating them collectively. This idea has been further generalized to arbitrary connectivity~\cite{wu2023optimization}, obtaining the optimal circuit size of
 $\Theta\left(\frac{n^2}{\log\Delta}\right).$

\begin{figure}
    \centering
    \includegraphics[width=1.0\linewidth]{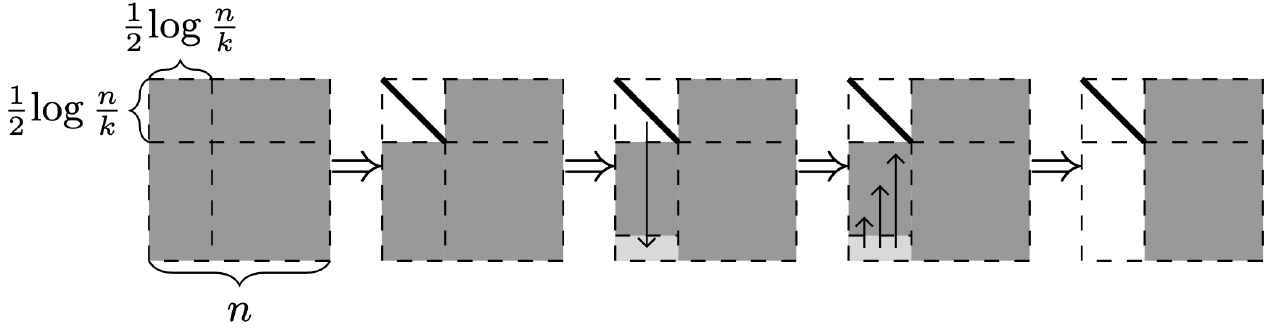}
    \caption{Ref.~\cite{wu2023optimization}. Illustration for the elimination of the first $s=\frac{1}{2}\log \pbra{\frac{n}{k}}$ columns of a CNOT circuit represented by the matrix $P$.}
    \label{fig:cnot_size_opt_alg}
\end{figure}

 We now briefly introduce the optimal construction for the fully connected case.  
Let $d_i$ be the degree of the $i$-th vertex, and define $k :=\min\{v:  \sum_{j=1}^v d_j \geq n\}$. The key idea is that for arbitrary $M$, one can eliminate the first $s = \tfrac{1}{2}\log(n/k)$ columns using $O(n)$ row additions. Here, elimination  means transforming the $j$-th row to $e_j$.
Because $\Delta \geq n/k$, applying this group elimination recursively results in $ O\!\left(n\times n/s\right) = O\!\left(n^2/\log \Delta\right)$ circuit size. See also Fig.~\ref{fig:cnot_size_opt_alg} for illustration.

\paragraph{Circuit depth.}
\citet{jiang2019optimal} showed that, in the all-to-all connectivity model, the depth of any CNOT circuit on $n$ data qubits with $m$ ancillas can be optimized to
\begin{equation}
\Theta\!\Big(\,\log n \;+\; \frac{n^2}{(m+n)\,\log(m+n)}\Big).\notag
\end{equation}
The lower bound follows from a counting argument combined with a truncation bound applied to general two-qubit gates. The idea of the optimal construction is to partition $P$ into multiple sub-blocks. These sub-blocks are implemented either sequentially or in parallel, depending on the number of ancillary qubit number $m$.

\subsubsection{$\text{C}^{n}$-$X$ gates and generalization}\label{sec:cnxg}

Another important reversible Boolean primitive is the multi-controlled NOT gate, denoted by $\mathrm{C}^{n}\text{-}X$. Acting on $n$ control qubits and one target qubit, it maps computational-basis states according to
\begin{align}\label{eq:bool_cnx}
|x_1\cdots x_n\rangle|t\rangle
\longmapsto
|x_1\cdots x_n\rangle\bigl|t\oplus (\wedge_{j=1}^{n}x_j)\bigr\rangle ,
\end{align}
where $\wedge$ denotes Boolean conjunction. 

The case $n=2$ is the Toffoli gate.
Even this smallest nontrivial instance is already expensive to realize exactly. Over the gate set consisting of arbitrary single-qubit gates and CNOT, the Toffoli gate requires six CNOT gates, and this bound is tight~\cite{shende2009cnot}. In the fault-tolerant Clifford+$T$ setting,  \citet{amy2013meet} gave a realization with $T$-depth $3$, while \citet{Gosset2014algorithm} later proved that no exact Clifford+$T$ implementation can use fewer than seven $T$ gates. For example, the following ancilla-free circuit has $T$-count $7$ and $T$-depth $3$:
\[
\Qcircuit @C=.5em @R=0.5em @!R {
&\ctrl{1} &\qw &&&\qw& \gate{T} & \targ & \qw & \ctrl{2} & \qw & \ctrl{1}&\gate{T^\dag}&\qw&\ctrl{2}&\targ&\qw\\
&\ctrl{1}  &\qw & \push{\rule{.3em}{0em}=\rule{.3em}{0em}}& &\qw& \gate{T} & \ctrl{-1} & \targ & \qw & \gate{T^\dag} & \targ &\gate{T^\dag} &\targ&\qw&\ctrl{-1}&\qw\\
&\targ &\qw &&&\gate{H}& \gate{T} & \qw & \ctrl{-1} & \targ & \qw & \qw &\gate{T} &\ctrl{-1}&\targ&\gate{H}&\qw
}
\]
\citet{Selinger2013quantumcircuit} further showed that the $T$-depth can be reduced to $1$ by using four clean ancillas.

A useful relaxation is to allow a relative-phase implementation. For $\mathrm{C}^{n}\text{-}X$, this means that the gate induces the same permutation of computational-basis states as in Eq.~\eqref{eq:bool_cnx}, but may attach basis-state-dependent phases. Equivalently, one allows
\begin{align}
|x_1\cdots x_n\rangle|t\rangle
\longmapsto
e^{i\phi(x_1,\ldots,x_n,t)}
|x_1\cdots x_n\rangle\bigl|t\oplus (\wedge_{j=1}^{n}x_j)\bigr\rangle ,
\end{align}
for some real-valued phase function $\phi$. Such variants can be implemented much more economically: relative-phase Toffoli gates with four $T$ gates and as few as three CNOT gates are known~\cite{Maslov2016advantages}. More importantly, these relative-phase gates remain highly useful in the synthesis of larger $\mathrm{C}^{n}\!\text{-}X$ circuits, since exact Toffoli gates that appear inside compute--uncompute patterns can often be replaced by relative-phase versions without changing the overall functionality~\cite{Maslov2016advantages}.

For general $\text{C}^{n}\text{-}X$,~\citet{barenco1995elementary} proposed an ancillary-free approach based on the following decomposition 
\[
\Qcircuit @C=.5em @R=0.5em @!R {
&{/}\qw& \ctrl{1} & \qw & & &&{/}\qw& \qw & \ctrl{1} & \qw & \ctrl{1} & \ctrl{2} & \qw\\
&\qw& \ctrl{1} & \qw & \push{\rule{.3em}{0em}=\rule{.3em}{0em}} & & &\qw& \ctrl{1} & \targ & \ctrl{1} & \targ & \qw & \qw\\
&\qw& \gate{X} & \qw & & &&\qw& \gate{V} & \qw & \gate{V^\dag} & \qw & \gate{V} & \qw
}
\]
\noindent with $V^2=X$. Applying this decomposition recursively gives a construction with $O(n^2)$ circuit size and depth. For ancillary-free preparation, one can improve the circuit size to $O(n)$~\cite{gidney_2015}, the circuit depth to $O(n)$~\cite{Saeedi2013linear} and subsequently $O(\log^2(n))$~\cite{nie2024quantum}. However, methods in~\cite{Saeedi2013linear,gidney_2015,nie2024quantum} require single-qubit rotations whose precision increases exponentially with $n$. This means that in the fault-tolerant setting, the $T$ count is suboptimal.

A more practical approach is to introduce a single ancillary qubit. For example,~\citet{barenco1995elementary} proposed the following recursive decomposition
\[
\Qcircuit @C=.5em @R=0.5em @!R {
&{/}\qw& \ctrl{1} & \qw & & &&{/}\qw& \ctrl{3} & \qw &\ctrl{3}& \qw& \qw \\
&\qw& \ctrl{1} & \qw & & &&    \qw& \qw      & \ctrl{1} & \qw& \ctrl{1}& \qw \\
&\qw& \ctrl{2} & \qw &\push{\rule{.3em}{0em}=\rule{.3em}{0em}} & &&   \qw& \qw      & \ctrl{1} & \qw& \ctrl{1}& \qw \\
&\qw& \qw & \qw &  & & &\qw& \targ & \ctrl{1}&\targ&\ctrl{1}& \qw  \\
&\qw& \targ & \qw & & &       &\qw&\qw    & \targ & \qw&\targ& \qw }
\]
resulting in both linear circuit size/depth and $T$ count/depth. There are some subsequent works improving circuit size by constant factors~\cite{iten2016quantum,Maslov2016advantages,zindorf2025efficient}. For example,~\citet{zindorf2025efficient} achieved CNOT count $12n-20$ and $T$ count $16n-32$ for $n\geq6$, with one dirty ancillary qubit. Then methods above saturate the asymptotic $T$ count lower bound $\Omega(n)$~\cite{Beverland2019LowerBounds,Gosset2025multi_qubit}.

~\citet{nie2024quantum} further showed that the multi-controlled $X$ gate can be implemented with asymptotically optimal depth $O(\log n)$ using only a single ancillary qubit. The central idea is to employ an intermediate Toffoli-like gadget in which the control register is allowed to change temporarily during the computation and is restored only at the end. Conceptually, for a control register $x\in\{0,1\}^m$, an ancilla initialized to $|0\rangle$, and a target qubit $|t\rangle$, this gadget may be written as
\begin{equation}\label{eq:toft}
\widetilde{\mathrm{TOF}}\ket{x}\ket{0}\ket{t}
=
\begin{cases}
\ket{\tilde{x}}\ket{0}\ket{t}, & x\neq 1^m,\\[2mm]
\ket{\tilde{x}}\ket{1}\ket{\bar t}, & x=1^m,
\end{cases}
\end{equation}
where bit string $\tilde{x}$ need not be equal to $x$. Thus, the gadget computes the conjunction of the control bits onto the ancilla and flips the target when all controls are $1$, while postponing the restoration of the control register to a later uncomputation step.

\begin{figure}[t]
    \centering
          \includegraphics[width=1\columnwidth]{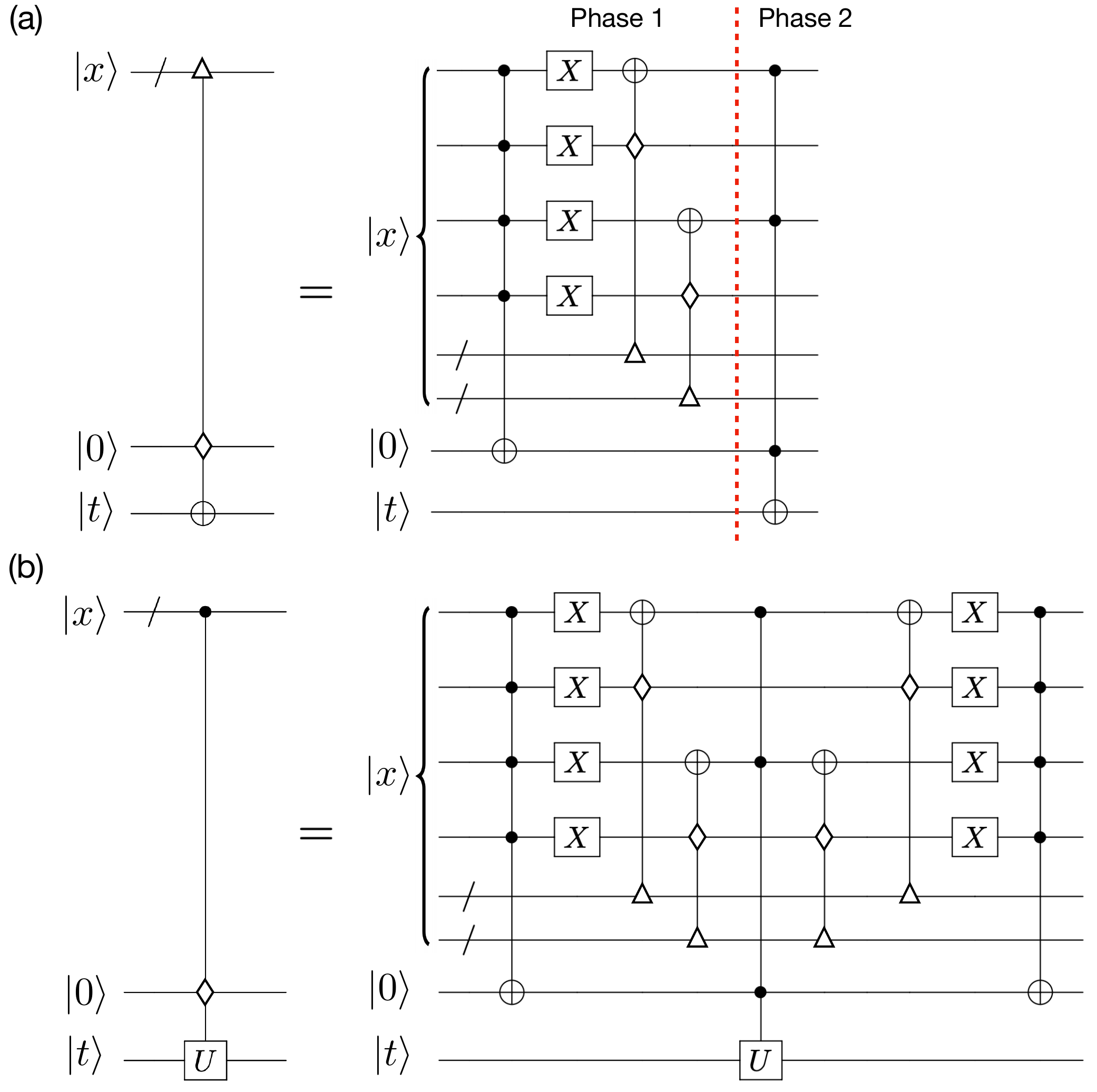}
       \caption{Quantum circuit for general multi-qubit controlled unitary with $O(\log n)$ depth and $O(n)$ circuit size in~\cite{nie2024quantum}. (a) Recursive quantum circuit for $\widetilde{\text{TOF}}$. (b) Quantum circuit for general $\text{C}^{n}\text{-}U$. Diamonds represent  ancillas within the corresponding recursive calls, while triangles indicate the target wires of those calls.
} \label{fig:mcu}
\end{figure}

As illustrated in Fig.~\ref{fig:mcu} (a), when the number of controls is even, this construction admits a recursive decomposition in which two subcircuits of half the size are executed in parallel. This gives the depth recurrence $D(n)=D(n/2)+O(1)$, and hence $D(n)=O(\log n)$, while the circuit size remains $O(n)$. Finally, by applying the inverse of the first stage, both the modified control register and the ancilla are uncomputed, yielding an exact implementation of $\mathrm{C}^{n}$-$X$. \citet{dutta2025spacedepth} subsequently refined this framework by analyzing the space--depth trade-off in greater detail, showing that additional clean ancillas can further reduce the constant factor in the Toffoli depth, although the asymptotically optimal $O(\log n)$ scaling remains unchanged.

\paragraph{Generalization to $\text{C}^{n}\text{-}U$:}\label{sec:mqcg}
We now introduce the generalization to multi-qubit controlled gate $\text{C}^{n}\text{-}U$ for arbitrary  $U\in\text{U}(2)$, as their techniques and circuit complexities are largely the same as $\text{C}^{n}\text{-}X$. 
When there are no ancillary qubits,~\citet{barenco1995elementary} showed that the circuit size lower bound for exact preparation is $\Omega(n^2)$. This lower bound can be achieved by a quantum circuit similar to the one for ancillary-free $\text{C}^{n}\text{-}X$ above, with $X$ gate replaced by $U$, and $V^2=U$. Moreover, if approximation error $\varepsilon$ is allowed, one can achieve circuit size $\Theta(n\log(1/\varepsilon))$~\cite{barenco1995elementary}.~\citet{da2022linear} then proposed an exact construction with circuit size $O(n^2)$ and depth $O(n)$. 

 With one ancillary qubit, exact multi-controlled gates admit linear-size decompositions.
In particular, \citet{barenco1995elementary} showed that a $\mathrm{C}^{n}\text{-}U$ gate can be
implemented with $\Theta(n)$ elementary gates given one clean ancilla.
For the special case $\mathrm{C}^{n}\text{-}X$, \citet{nie2024quantum} gave an exact construction
with size $O(n)$ and depth $O(\log n)$ using one ancilla; moreover, when $U^{2}=I$, the ancilla may be taken dirty. Hence, in the standard single- and two-qubit gate model, $\mathrm{C}^{n}\text{-}X$ with one ancilla has optimal asymptotic size $\Theta(n)$ and optimal asymptotic depth $\Theta(\log n)$. See also~\cite{zindorf2025efficient} for further one-ancilla constructions, and \citet{Maslov2016advantages} for constant-factor Clifford+T optimizations.

\subsubsection{SAT-oracle}\label{sec:ecnfg}
A SAT-oracle for quantum computation is a unitary satisfying  $U_{\text{SAT}}|x\rangle|c\rangle \;\mapsto\; |x\rangle|c \oplus f(x)\rangle$ for some function $f(x)$ defined as

\begin{align}\label{eq:bool_sat}
f(x)=\bigwedge_{i=1}^m\left(\bigvee_{j\in\mathcal{S}_i}l_j\right),
\end{align}
where each $l_j$ is a literal, i.e.~either a variable $x_p$ or its negation $\neg x_p$.  
If $|\mathcal{S}_i|\leqslant k$, we call $f(x)$ a $k-$CNF, denoted as $\text{CNF}_{n,m}^k$. SAT-oracle is important for practical applications related to Boolean satisfiability~\cite{mcmillan2003interpolation,kunz2013reasoning}. From the standpoint of expressive power, the SAT-oracle model is at least as general as QRAM: the Boolean relation implemented by a QRAM lookup can, in principle, be encoded as a CNF formula and hence realized in the form of Eq.~\eqref{eq:bool_sat}. This universality is only representational, however, since the CNF encoding may be highly inefficient and can incur exponential overhead in the worst case.

A straightforward way of constructing SAT-oracle is to introduce $m$ ancillas,  store the result of clause $\vee_{j\in\mathcal{S}_i}l_j$ to the $i$-th ancilla through a $k$-qubit controlled-not gate, perform transformation $|c\rangle \;\mapsto\; |c \oplus f(x)\rangle$ through a $m$-qubit controlled-not gate, and then uncompute the ancillary qubits~\cite{Kole2024qSAT}. But this approach has suboptimal circuit complexity and a fixed number of ancillary qubits. \citet{yang2024efficient} proposed algorithms for SAT-oracle with near-optimal circuit size/depth. Specifically, given $l$ ancillary qubits, their method achieves a circuit size
\[
   O\!\left(n \Bigl(\tfrac{km}{n}\Bigr)^{1 + \log_{l/2+1}4}\right)
\] 
for exact construction. This scaling can be compared to the lower bound $\Omega(km)$, which holds even when approximation error $\varepsilon\leqslant1/\sqrt{2}$ is permitted. An alternative depth-oriented algorithm achieves circuit depth
\[
O\!\left(k \Bigl(\tfrac{mS}{l}\Bigr)^{1+c} \log l\right)
\] 
{at the cost of a slightly increased circuit size.} Here, $S = \max(k/\log l,1)$ and $c = \log_{l/S} 4$. In particular, with $l=O(m)$ ancillary qubits, the depth scales logarithmically with $m$ as $\widetilde O(\log m)$.  

Beyond CNF-structured SAT oracles,~\cite{nie2026nearlyoptimalquantumcircuits} considered the synthesis of arbitrary Boolean oracles $f:\{0,1\}^n\rightarrow\{0,1\}^b$. They obtained circuit size $O\left(\frac{b2^n}{\log(n+a)}\right)$ and depth
$O\left(\frac{b2^n}{n+a}\right)$ using \(1\leq a\leq\Theta(2^n/n)\) ancillary qubits, which is closed to the lower bounds. They also discussed the quantum circuits for partial Boolean functions and sparse Boolean functions.

\subsubsection{General Boolean unitary}\label{sec:gbu}

For the synthesis of arbitrary reversible Boolean functions, equivalently, permutation unitaries on the computational basis, \citet{shende2003synthesis} gave a constructive decomposition with gate set $\{\text{NOT}, \text{CNOT}, \text{Toffoli}\}$. More precisely, for $n>3$, every even permutation can be realized without ancillary qubits using at most $n$ NOT gates, $n^2$ CNOT gates, and $3(2^n+n+1)(3n-7)=O(n2^n)$ Toffoli gates; arbitrary permutations can be handled with at most one additional ancilla. Subsequent work developed alternative synthesis paradigms, including library-based and cycle-based methods~\cite{saeedi2010library,saeedi2010reversible}. In a different direction, \citet{Malvetti.21} showed that a Householder-reflection-based synthesis can also be specialized to permutation gates, yielding an implementation with one dirty ancilla and at most $(18n-26)(2^n-1)$ CNOT gates. Thus, while later methods improve constants and often perform better on structured instances, the worst-case circuit size of general-purpose constructions for arbitrary reversible Boolean functions remains $O(n2^n)$.

\citet{wu2024Asymptotically} later showed that arbitrary $n$-bit reversible functions can be synthesized without ancillary bits using $O(2^{n}n/\log n)$ elementary gates, thereby matching the worst-case lower bound $\Omega(2^{n}n/\log n)$ established in~\citet{shende2003synthesis}. The main idea is not to synthesize the target permutation $\pi$ directly, but instead to construct an auxiliary permutation $\pi'$ that itself admits an $O(2^{n}n/\log n)$-gate implementation and is chosen so that the residual permutation $\pi' \circ \pi$ is sparse, namely, it differs from the identity on only a small subset of inputs. This sparse remainder can then be synthesized separately at lower additional cost, leading to an overall asymptotically optimal circuit size of $\Theta(2^{n}n/\log n)$.

\subsection{Isometry}\label{sec:sparse_iso}
An $N\times M$ dimensional isometry can be represented by a non-square matrix $V=[|\psi_0\rangle,|\psi_1\rangle,\cdots,|\psi_{M-1}\rangle]$,
where $|\psi_j\rangle$ are some $N$-dimensional vectors that are orthogonal to each other. It can be verified that $V^\dag V=I$. We say that a quantum circuit $U_V$ implements $V$, if
\begin{align}\label{eq:sparse_uvp0}
U_V=[|\psi_0\rangle,|\psi_1\rangle,\cdots,|\psi_{M-1}\rangle,|\psi_{M}\rangle\cdots,|\psi_{N-1}\rangle],
\end{align}
where $|\psi_{j\geqslant M}\rangle$ may be chosen as any orthonormal completion of the specified columns. Equivalently,  $U_V|j\rangle=|\psi_j\rangle$ for any $0\leqslant j\leqslant M-1$. Constructing $U_V$ is simpler than universal unitary synthesis, because for input basis $|j\rangle$ with $j\geqslant M$, there is no restriction on the circuit output.  For example,  when $M=1$, the isometry reduces to quantum state preparation. 

\citet{knill1995approximation} demonstrated that an isometry can be realized using $O(M)$ quantum state preparation operators together with some extra multi-qubit controlled phase gates. This approach has been subsequently improved by~\cite{iten2016quantum} and~\cite{Malvetti.21}. 
The idea is to find a quantum circuit diagonalizing the isometry. The inverse of this quantum circuit is the target $U_V$.  As an example, we define 
\begin{align}
|u_0\rangle=\frac{|\psi_0\rangle-e^{i\theta_0}|0\rangle}{\left\||\psi_0\rangle-e^{i\theta_0}|0\rangle\right\|}
\end{align}
 with $\theta_0=\text{arg}(\langle0|\psi_0\rangle)$, and introduce a multi-qubit phase gate $\text{Ph}_0=e^{-i\theta_0}|0\rangle\langle 0|+\sum_{j\neq 0}|j\rangle\langle j|$. It can be verified that 
\begin{align}\label{eq:sparse_ph0}
\text{Ph}_0R_{|u_0\rangle}V\equiv U^{(0)}=\left[|0\rangle,|\psi^{(0)}_1\rangle,\cdots,|\psi^{(0)}_{M-1}\rangle\right],
\end{align}
where $R_{|u_0\rangle}$ is reflection operator (see definition in Eq.~\eqref{eq:rpsip}) that can be realized by multi-qubit controlled-Z gate, state preparation of $|u_0\rangle$ and its inverse. 
 $|\psi^{(0)}_j\rangle$ is some new quantum state. It turns out that we can eliminate other columns in a similar way. Suppose at the $k$th step, we have the isometry 
\begin{align}\label{eq:sparse_uk}
U^{(k)}=\left[|0\rangle,|1\rangle,\cdots,|k\rangle,|\psi^{(k)}_{k+1}\rangle,\cdots,|\psi^{(k)}_{M-1}\rangle\right].
\end{align}
We define $\text{Ph}_{k+1}$ and $R_{|u_{k+1}\rangle}$ in similar ways to $\text{Ph}_0$ and $R_{|u_0\rangle}$, with $|\psi_0\rangle$ replaced by $|\psi_{k+1}^{(k)}\rangle$. Then, it can be verified that  
\begin{align}\label{eq:sparse_phk}
\text{Ph}_{k+1}R_{|u_{k+1}\rangle}U^{(k)}=U^{(k+1)}.
\end{align}
By reversing the circuit for diagonalization, the following 
\begin{align}\label{eq:ru0}
U_V=R_{|u_0\rangle}^\dag\text{Ph}_0^\dag\cdots R_{|u_{M-1}\rangle}^\dag\text{Ph}_{M-1}^\dag
\end{align}
 implements the target isometry $V$. For a general isometry, Eq.~\eqref{eq:ru0} has circuit size $O(NM)$. Moreover,~\cite{iten2016quantum} further proposed two alternative protocols, called column-by-column decomposition and cosine-sine decomposition, with circuit sizes $O(NM)$ and $O(N^2+M^2)$ respectively.
 
Ref~\cite{Malvetti.21} considered special isometries when $V$ is column-wise sparse, i.e. most of the entries of $V$ are zero. In this case, quantum circuit in Eq.~\eqref{eq:ru0} is not preferred, because the elimination of one column (i.e. Eq.~\eqref{eq:sparse_ph0},~\eqref{eq:sparse_phk}) will fill in some of the nonzero entries in other columns, thus destroys their sparsity.~\cite{Malvetti.21} resolved this limitation by introducing two ancillary qubits. They showed that by performing an appropriate row permutation, the fill-in can be avoided. So sparse state preparation methods (Sec.~\ref{sec:sparse_sp}) can be adapted to construct $R_{|u_j\rangle}$. For isometry with totally $S$ nonzero entries, the total circuit size of their no-fill-in approach scales as $O(nS+(n+\log M)M)$.

\subsection{Sparse unitary \label{sec:sparse_unitary}}
Sparse unitary can be considered as an $N\times N$ dimensional sparse isometry, so one can use the approach in~\cite{Malvetti.21} for construction. For a general sparse unitary with $S$ nonzero entries, the circuit size is $O(n(S+N))$. Below, we further introduce the synthesis of some sparse unitaries with extra structures.  

\subsubsection{Efficiently computable sparse unitary}
~\citet{jordan2009efficient} considered sparse unitary under assumption that $U$ is efficiently computable (see definition in Sec.~\ref{sec:sparse_saim}).   We first introduce a Hermitian matrix
\begin{align}\label{eq:sparse_HU}
H=\begin{pmatrix}0&U\\U^\dag&0\end{pmatrix}.
\end{align}
Because $H^2=\mathbb{I}$, it follows that $e^{i H\theta}=\cos(\theta) \mathbb{I}+i\sin(\theta) H$. Setting $\theta=\pi/2$ and applying this evolution to $|1\rangle|\psi\rangle$ yields 
\begin{align}
e^{iH\pi/2}|1\rangle|\psi\rangle=i|0\rangle U|\psi\rangle.
\end{align}
The desired unitary evolution $U|\psi\rangle$, up to a global phase, is obtained by tracing out the first qubit. Therefore, synthesizing a sparse unitary is reduced to simulating the dynamics of the sparse Hamiltonian $H$. As discussed in Sec.~\ref{sec:sparse_saim}, $e^{iH\pi/2}$ is efficiently simulable when $H$ is efficiently computable. According to Eq.~\eqref{eq:sparse_HU}, this requirement can be ensured when $U$ is efficiently computable.

\subsubsection{Boolean sparse unitary}\label{sec:sbu}
A special case of practical relevance is the sparse Boolean unitary, or permutation unitary~\cite{Malvetti.21,ramacciotti2024simple,li2025nearly}, which is useful for preparing sparse quantum states (Sec.~\ref{sec:sparse_sp}). Given a general permutation unitary $\pi$, we define $\text{size}(\pi)$ as the number of basis $j$, such that $\pi(j)\neq j$.~\cite{Malvetti.21} showed that a sparse permutation unitary requires $O(n\text{size}(\pi))$ circuit size. In~\cite{li2025nearly}, the given permutation $\pi$ is decomposed into multiple permutations $\pi_i$ with bounded number of pairwise disjoint transpositions, and each $\pi_i$ is executed successively.  Their method improves the circuit size to 
$$O\left(\frac{n\text{size}(\pi)}{\log n}+n\log\min\{\text{size}(\pi),\log n\}\right).$$

\subsection{Clifford circuit}\label{subsec:cnot_circuit_clifford}

The Clifford circuit plays an important role in quantum error correction and the study of entanglement. 
Any Clifford  circuit admits an 11-stage canonical form consisting of alternating single-qubit layers and CNOT blocks~\cite{aaronson2004improved}, e.g., the sequence
$$\mathrm{H}\text{--}\mathrm{C}\text{--}\mathrm{S}\text{--}\mathrm{C}\text{--}\mathrm{S}\text{--}\mathrm{C}\text{--}\mathrm{H}\text{--}\mathrm{S}\text{--}\mathrm{C}\text{--}\mathrm{S}\text{--}\mathrm{C}$$
where $\mathrm{H}$ and $\mathrm{S}$ denote Hadamard and $S=\begin{pmatrix}
1&0\\
0&i
\end{pmatrix}$ gate layers, $\mathrm{C}$ denotes  CNOT blocks which will be introduced in Sec.~\ref{sec:b_CNOT}.
Beyond this result,~\citet{maslov2018shorter} introduced a shorter layered normal form based on  Bruhat decomposition. With CZ gates, they obtained a $7$-stage decomposition in the form of 
$$\text{--}\mathrm{C}\text{--}\mathrm{CZ}\text{--}\mathrm{S}\text{--}\mathrm{H}\text{--}\mathrm{S}\text{--}\mathrm{CZ}\text{--}\mathrm{C}\text{--}.$$
 This improves the two-qubit gate depth from $25n$ to $14n-4$, and is executable in the linear nearest neighbor architecture. Using the gate library $\{H,S,\text{CNOT}\}$, their method gives a $9$-stage decomposition 
$$\text{--}\mathrm{C}\text{--}\mathrm{S}\text{--}\mathrm{C}\text{--}\mathrm{S}\text{--}\mathrm{H}\text{--}\mathrm{C}\text{--}\mathrm{S}\text{--}\mathrm{C}\text{--}\mathrm{S}\text{--}$$
which is also shorter than~\cite{aaronson2004improved}.

Except for the normal form decomposition, there are other optimization techniques for Clifford circuits. Specifically,~\cite{kliuchnikov2013optimization}  and~\cite{bravyi2022sixqubit} exhaustively obtained optimal implementation (i.e. with minimum CNOT gates) for arbitrary Clifford circuits with up to four and six qubits, respectively.~\cite{Schneider2023sat} translates the Clifford circuits optimization to a satisfiability problem, and their method can obtain optimal implementation for up to 26 qubits within hours. A similar SAT-based idea has also been proposed in~\cite{shaik2025cnotoptimalcliffordsynthesissat}. 

When the qubit number is large, there are other heuristic methods that are useful to find suboptimal implementations. For example,~\cite{bravyi2022sixqubit} proposed an optimization method based on template matching and peephole optimization.~\cite{duncan2020graphtheoretic} proposed a diagrammatic ZX-calculus method (e.g., PyZX) to shrink Clifford circuits, which can be combined with the reinforcement learning/GNN techniques.~\cite{de2025graph} proposed a graph-state-based framework, which can recover many normal forms and proves two-qubit depth $7n-2$ on linear-nearest-neighbour architectures.

\subsection{Quantum state for continuous functions}\label{sec:ecf}
When the amplitudes of target quantum states are some functions
\begin{align}
|\psi_f\rangle=\frac{1}{\|f\|_2}\sum_{x}f(x)|x\rangle, \label{eq:st_cont}
\end{align}
where $f(x)$ has some extra properties, the circuit complexity of state preparation can be reduced.

\cite{Rattew.22} considered $f(x)$ that are efficiently computable. Equivalently, the quantum query access $O_f|x\rangle|0\rangle=|x\rangle|f(x)\rangle$ is realizable by polynomial-size quantum circuit. The state preparation is realized by adiabatic evolution. Let  $H_f\propto|\psi_f\rangle\langle\psi_f|$ be the target Hamiltonian. We define intermediate function $f_s=(1-s)f_0+sf$ with $f_0$ a constant function, and time-dependent Hamiltonian $H(t)=H_{f_{t/T}}$. It can be verified that $H(t)$ always has large spectral gap for $t\in[0,T]$. Because $f_s$ is also efficiently computable, one can efficiently approximate the adiabatic evolution $U=\mathcal{T}\text{exp}\left(-i\int_0^TH(t)dt\right)$ by Trotter decomposition, and obtain the target state $|\psi_f\rangle$ with high accuracy. This method has query complexity $O(\mathcal{F}^{-4}/\varepsilon^2)$, where $\mathcal{F}$ is the filling ratio, i.e. the integral of $|f(x)|$ divided by the area of the bounding box of $|f(x)|$. 
Subsequent work~\cite{rattew2023non} has generalized state preparation to the nonlinear transformation of state amplitudes $\psi_j|j\rangle\rightarrow f(\psi_j)|j\rangle$. In this framework, preparing Eq.~\eqref{eq:st_cont} can be considered as a special case of the nonlinear transformation.

\cite{marin2023quantum} considered $f(x)$ with upper bounded second derivative $|\partial^2f(x)/\partial x^2|\leqslant\eta$. Their simplification is based on the uniformly controlled rotation methods. Due to the continuity of $f(x)$, consecutive rotation angles are close to each other. Thus, many controlled rotations can be clustered into one (e.g. $|0\rangle\langle0|\otimes R_0+|1\rangle\langle1|\otimes R_1\approx I\otimes R_0$ when $R_0$ and $R_{1}$ are close to each other). Given a tolerable infidelity $\epsilon$, they can approximate the ideal target state with $2^{k}$ circuit size, where $k=\max\{\lceil-\frac{1}{2}\log_2(4^{-n}-\frac{96}{\eta^2}\log(1-\epsilon))\rceil,2\}$.

There are other schemes for specific functions, such as Gaussian function~\cite{iaconis2024quantum,manabe2025state}, polynomial functions~\cite{gonzalez2024efficient}, and Fourier/Chebyshev series and their generalization to multivariable functions \cite{rosenkranz2025quantum}.

\subsection{Initial state for ground state preparation algorithms}\label{sec:initial_state}

In typical ground state preparation algorithms, the initial state determines the overall efficiency of the quantum algorithm. 
Different from other scenarios that have well-defined target states, the initial state should has a nontrivial overlap with the target state. Then, with projective algorithms like quantum phase estimation (QPE)~\cite{kitaev1995quantum} and quantum singular value transformation (QSVT)~\cite{lin2020near}, the target state component is amplified. These initial states are usually obtained by first performing classical optimization, and then prepared by efficient quantum circuits. Below, we introduce some typical schemes. 

For quantum chemistry applications,  a simple choice of the initial state, is the Hartree-Fock state, i.e. the lowest-energy single-Slater-determinant state within the Hartree–Fock ansatz~\cite{reiher2017elucidating,von2021quantum,o2016scalable,o2019quantum}. Classically obtaining Hartree-Fock states is simple, and they can be prepared by only one layer of $X$ gates.  
However, due to the orthogonality catastrophe~\cite{chan2012low,tubman2018postponing,lee2023evaluating}, this overlap becomes exponentially small for large systems. So nontrivial initial states with entanglement are required. To classically optimize the initial state, typical approaches include the sum of multiple Slater determinants~\cite{tubman2018postponing,wang2008quantum,veis2010quantum,babbush2015chemical,sugisaki2016quantum,fomichev2024initial}, and MPS states~\cite{fomichev2024initial,huggins2025efficient,berry2025rapid}. These states can provide  larger overlaps, and at the same time are still efficiently implementable. Alternatively, there are also heuristic quantum optimization techniques, such as adiabatic preparation~\cite{aspuru2005simulated,veis2014adiabatic,lee2023evaluating}, and variational quantum eigensolver~\cite{peruzzo2014variational,yung2014transistor,cerezo2021variational}.

\subsection{Structured QRAM}\label{sec:sqram}
We now summarize some QRAM circuits specialized for structured data. In~\cite{sanders2020compilation}, they optimize QROM in the case where multiple addresses are associated to the same output value (data collisions). Some work explores approximate QRAM implementations, e.g., as parameterized quantum circuits~\cite{niu2022entangling,phalak2023trainable}.
As the QRAM gate lower-bound of~\cite{jaques2025qram} still applies in such settings, such techniques likely require highly structured data to yield notable savings. See~\cite{jaques2025qram} for a critical discussion of variational QRAM approaches. Moreover, there is notable overlap between structured QRAM circuits and quantum arithmetic circuits. For a deterministic Clifford-only QROM (no T-gates), a general QRAM map of the form $\ket{x}_n\ket{0}_m \mapsto \ket{x}_n\ket{f(x)}_m$ (for $x \in \{0,1\}^n, f(x) \in \{0,1\}^m$), must be an affine transformation, i.e., $f(x) = Ax \oplus b$ for some binary matrix $A$ and binary vector $b$ (see e.g.,~\cite{dehaene2003clifford} for relevant binary Clifford background, and~\cite{li2025stab} for a concrete discussion in the context of QRAM).

\subsection{Other structured data}\label{sec:str_other}

Below, we discuss the encoding of some other structured data that are not covered by previous sections. 

\textbf{Dicke/Fixed-Hamming-weight states:} There are quantum states with fixed Hamming weight as follows
\begin{align}\label{eq:prac_hw}
|\psi\rangle=\sum_{\text{HW}(j)=k}\alpha_j|j\rangle,
\end{align}
where $\text{HW}(j)$ represents the number of $1$s in the bitstring $j$. 
In particular, the Dicke state is a special case of Eq.~\eqref{eq:prac_hw} with identical nonzero amplitudes  $\alpha_j=\binom{n}{k}^{-1/2}$, which has broad applications in both physics and data science~\cite{childs2000finding,stockton2004deterministic,hume2009preparation,bartschi2019deterministic,mukherjee2020preparing,mukherjee2020actual,wang2021preparing,bartschi2022short,aktar2022divide,mozafari2022efficient,stojanovic2023dicke,yu2024efficient,yuan2025depth}. For example, $W$-state is the Dicke state with Hamming weight $1$. 

The best known asymptotic circuit size for Dicke state is $O(kn)$~\citet{bartschi2019deterministic}. In terms of circuit depth,~\citet{bartschi2022short} proposed a method with depth $O(k\log (n/k))$ (or $O(k\sqrt{n/k})$) for all-to-all (or two-dimensional grid) connectivity, which does not require ancillary qubits. \citet{yuan2025depth} further improve the circuit depth for different connectivities. For example, the circuit depth for  all-to-all connectivity has been improved to $O(\log(k) \log(n/k) + k)$. 
Moreover, the circuit depth can be further reduced to constant if measurement and feedback are allowed.~\cite{buhrman2024state} achieved constant circuit depth for $k=O(\sqrt{n})$, using $O(n^2\log n)$ ancillary qubits. Note that the classical depth is still logarithmic. 
The circuit complexity can be further refined with different trade-offs in resources or infidelity~\cite{piroli2024approximating,yu2024efficient,vasconcelos2026constant}.

For general fixed-Hamming-weight state,~\cite{luo2025optimal} and~\cite{li2025preparation} independently showed that Eq~\eqref{eq:prac_hw} can be constructed with circuit size $O(\binom{n}{k})$. The method in~\cite{luo2025optimal} requires $\max\{0,n-3\}$ ancillary qubits. The method in~\cite{li2025preparation} requires $O(\binom{n}{k})$ ancillary qubits and achieves circuit depth $O(\log\binom{n}{k})$, which achieves the fundamental lower bound for both circuit size and depth.

\textbf{Low data density states:}
For a non-sparse quantum state $|\psi\rangle=\sum_{j=0}^{N-1}\alpha_j|j\rangle$, if the amplitudes are concentrated on a small number of basis, the cost of preparation can be largely reduced. Let $\rho=\frac{1}{N}\sum_{j=0}^{N-1}(|\alpha_j|/\max|\alpha_{j}|)^2$ be the data density,~\cite{pagni2025fast} showed that given $O(M\log N)$ ancillary qubits, $|\psi\rangle$ can be prepared with circuit depth  
$O\left(\frac{1}{\sqrt{\rho}}\frac{N}{M}\log(M+1)\right).$
Here, $M$ is a parallelization parameter that scales linearly with the number of ancillary qubits.

\textbf{Low rank state:}  As introduced in Sec.~\ref{sec:qsp_without}, the state preparation method in~\cite{Plesch.11} is based on the recursive Schmidt decomposition $|\psi\rangle=\sum_{j=0}^{2^{n/2}-1}\sigma_j|\tau\rangle_j|v\rangle_j$. Here, $\sigma_j$ is the Schmidt coefficient.~\cite{araujo2023low} showed that when the Schmidt rank of the state is low, i.e. there are only a few nonzero $\sigma_j$,  the unitary synthesis part can be drastically simplified, thus reducing the overall state preparation cost.

\textbf{States represented by decision diagram:}
A general quantum state can be represented by a decision tree with $2^n$ paths. For some structured states, the representation can be simplified into a decision diagram with much less paths.~\cite{mozafari2022efficient} showed that when the target $n$-qubit state can be represented by a decision diagram with $k$ paths, it can be prepared with CNOT count $O(kn)$. This decision diagram state serves as a generalization of the sparse state, because a state with $S$ nonzero entries can always be represented by a $S$-path decision diagram.

%% file: secs/tensor.tex
\section{Constructing tensor 
network states and operations\label{sec:tensor}}

\begin{figure}
    \centering
    \includegraphics[width=0.95\linewidth]{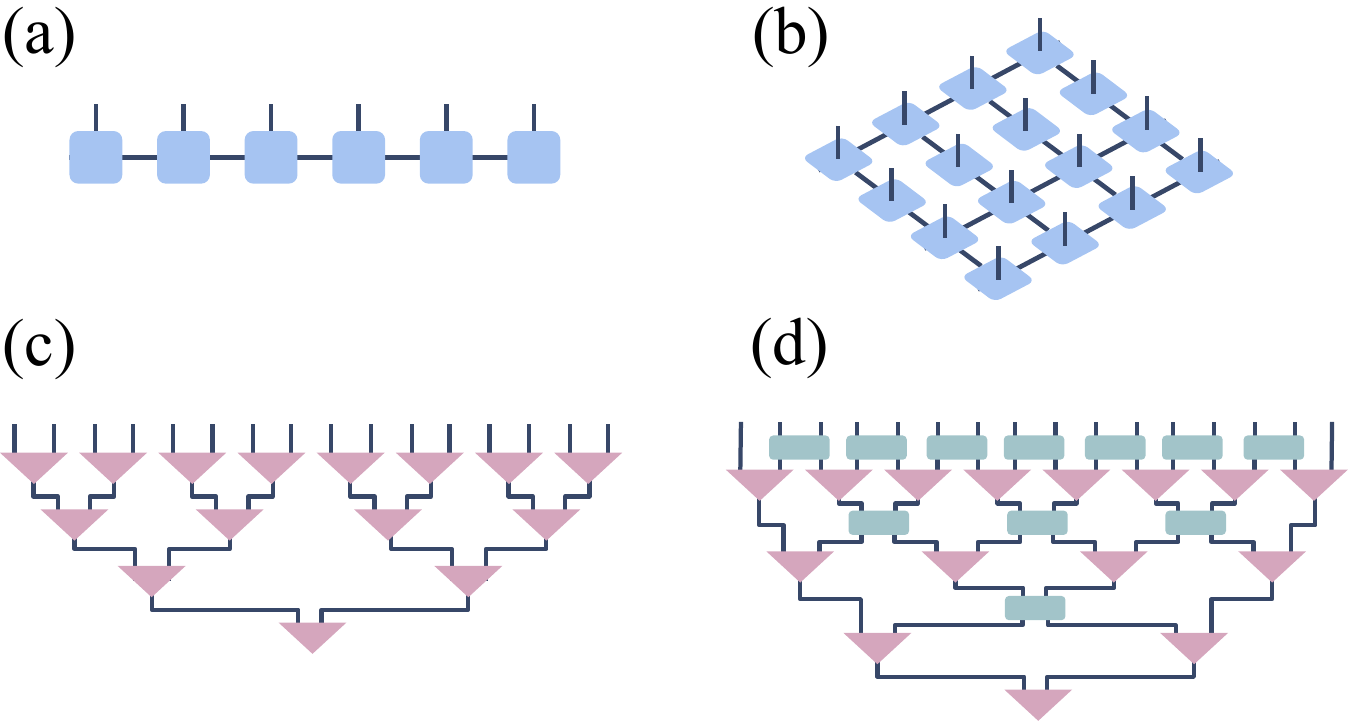}
    \caption{Tensor network geometries discussed in this section. Open legs denote physical (site) indices; contracted legs represent auxiliary (bond) degrees of freedom mediating correlations between sites. (a) Matrix-product state (MPS). (b) Projected entangled-pair state (PEPS) on the square lattice, generalizing MPS to two dimensions. (c) Tree tensor network (TTN); triangle tensors are isometries. (d) Multi-scale entanglement renormalization ansatz (MERA), which extends TTN by incorporating unitary gates (rectangles).}
    \label{fig_tensor_tns}
\end{figure}

Tensor networks provide an indirect, yet structured, representation of quantum states. Rather than explicitly enumerating all computational-basis amplitudes, the state is encoded in a set of tensors whose contraction, according to a prescribed network connectivity, yields those amplitudes (see Fig.~\ref{fig_tensor_tns}).

The advantage of the tensor-network description becomes apparent in systems composed of a large number of qubits. In this setting, a direct specification of the state via its amplitudes would typically require specifying an exponentially large number of coefficients. Tensor networks, by contrast, exploit the fact that these amplitudes are not necessarily independent and can often be compressed. This compression is beneficial whenever the entanglement in the state is low across the bipartitions defined by the network geometry, yielding a description whose cost scales at most polynomially in the number of tensors rather than exponentially in the system size. Fig.~\ref{fig_tensor_tns} illustrates the main tensor network types discussed in this section.

At the same time, tensors do not, in general, correspond to physically meaningful operations, such as unitary gates. Preparing a quantum state from its tensor-network description therefore requires an additional step: translating the Ansatz into a quantum circuit. This is the focus of the present section.

There are several recent review articles on tensor networks, covering both formal aspects~\cite{cirac2021matrix} and applications to quantum computing~\cite{berezutskii2025tensor}. Here, we focus specifically on formal results concerning the preparation of tensor-network states on quantum devices.

\subsection{Matrix-product states}\label{sec:tensor_MPS}

The paradigmatic tensor network in 1D is the matrix-product state (MPS) family. An $n$-qudit MPS is defined via a collection of rank-3 tensors $A_1,\dots,A_n$, each of them graphically represented as
\begin{align}
    \left(A_k^{i}\right)_{lm} =
        \begin{array}{c}
        \begin{tikzpicture}[scale=0.55,baseline={([yshift=-8ex] current bounding box.center)}]
            \ATensor{0,0}{\small $A_k$}{0}
		\draw (-1.4,0) node {$l$};
		\draw (1.4,0) node {$m$};
		\draw (0,1.4) node {$i$};
        \end{tikzpicture}
        \end{array}
        \in \mathbb C
        \;.
\end{align}
Here $k = 1,\dots, n$ labels the tensor, $i = 1,\dots,d$ is the \emph{physical} index, and 
$l = 1,\dots,D_{k-1}$, $m = 1,\dots,D_{k}$ are the left and right \emph{bond} indices. Thus each $A_k^{i}$ is a $D_{k-1} \times D_{k}$ matrix. By convention, $D_0 = D_n = 1$ and, therefore, $A_1$ and $A_n$ are row and column vectors, respectively.

Given a fixed computational basis on each site, the corresponding MPS is then defined and represented graphically as
\begin{align}
        \sum_{i_1,\dots,i_n} A_{1}^{i_1} \dots A_n ^{i_n} \ket{i_1\dots i_n}
        =
        \begin{array}{c}
        \begin{tikzpicture}[scale=0.55]
		      \foreach \x in {4,...,4}{
                \GTensor{(-\singledx*\x,0)}{1}{.5}{\small $A_2$}{0}
        }
		      \foreach \x in {3,...,3}{
                \SingleDots{-\singledx*\x,0}{\doubledx*.8}
        }
		      \foreach \x in {5,...,5}{
                \GTensor{(-\singledx*\x,0)}{1}{.5}{\small $A_1$}{-1}
        }
		      \foreach \x in {2,...,2}{
                \GTensor{(-\singledx*\x-.6,0)}{1}{.5}{\small $A_{n}$}{1}
        }
        \end{tikzpicture}
                \end{array}
        \;.
\end{align}
Evidently, the individual state amplitudes are obtained after matrix multiplication (a form of tensor contraction), justifying the name of the family.
The \emph{bond dimension} $D \coloneqq \max_k D_k$ will play an important role in the preparation complexity of MPSs.
We also assume MPSs to be normalized throughout -- computing the norm of an MPS is classically efficient, with cost $O(d n D^3)$, and any overall normalization factor can be absorbed into one of the tensors.

The expressivity of the MPS family is governed by the bond dimension $D$. In particular, each $D_k$ limits the entanglement of the underlying state across each bipartition ($1,\dots,k:k+1,\dots,n$) of the chain. More precisely, MPSs of \textit{uniform} bond dimension $D$ ($D_k = D$ for all $k=1,\dots, n-1$) can exactly express states with at most the same Schmidt rank across all bipartitions of the chain~\cite{vidal2003efficient}. Recall that the Schmidt rank is a pure-state entanglement measure, the logarithm of which coincides with the zeroth-order R\'enyi entanglement entropy $S_0$~\cite{horodecki2009quantum}. In general, MPSs can approximate well states with R\'enyi entanglement entropy $S_{\alpha} = O(\log n)$ (for some $\alpha < 1$) while maintaining $D = O(\mathrm{poly} (n))$, thus providing an efficient description of the quantum state~\cite{verstraete2006matrix,schuch2008entropy}. Note that representing such states by directly storing all their amplitudes, rather than using tensors, would result in a description that generally requires exponentially many parameters. This limited-entanglement setting is particularly relevant in the context of low-energy states of locally interacting 1D Hamiltonians with a constant spectral gap: such Hamiltonians admit a sublinear bond dimension MPS approximation for their ground state~\cite{hastings2007area,arad2013area}.

\begin{table}[t]
\centering
\begin{ruledtabular}
\begin{tabular}{p{0.22\columnwidth} p{0.7\columnwidth}}
\textbf{Concept} & \textbf{Definition and properties} \\ \hline
{\bf Blocking}
& Grouping neighboring tensors into a single effective tensor:
\begin{align*}
    \begin{array}{c}
        \begin{tikzpicture}[scale=0.4,baseline={([yshift=-3.1ex] current bounding box.center)}]
            \draw[thick,dotted] (0,0) to (2,0);
            \ATensor{(0,0)}{}{0};
            \ATensor{(3,0)}{}{0};
        \end{tikzpicture}
    \end{array}
    \mapsto
            \begin{array}{c}
            \begin{tikzpicture}[scale=0.4,baseline={([yshift=-2.9ex] current bounding box.center)}]
                \begin{scope}
		              \draw[thick] (-1.4,0) -- (1.4,0);
		              \draw[thick] (-0.6,0) -- (-0.6,1);
                    \draw[thick] (0.6,0) -- (0.6,1);
                    \draw[dotted, thick] (-0.3,.85) -- (0.3,.85);
                    \draw[thick, fill=tensorcolor, rounded corners=2pt] (-1,-0.6) rectangle (1,0.6);
	            \end{scope}
            \end{tikzpicture}
            \end{array}
\end{align*}
For MPSs, it preserves the bond dimension. \\

{\bf Normal \newline tensor} 
& Blocking a finite number of sites, the resulting blocked tensor becomes injective. Generic tensors are normal. \\

{\bf Injective \newline tensor}
& The map from virtual (bond) indices to physical indices is injective (one-to-one):
\begin{align*}
    \sum_{ilm}
        \begin{tikzpicture}[scale=0.3,baseline={([yshift=-1ex] current bounding box.center)}]
            \ATensor{0,0}{}{0}
		\draw (-1.6,0) node {$l$};
		\draw (1.6,0) node {$m$};
		\draw (0,1.6) node {$i$};
        \end{tikzpicture}
        \ket{i} \bra{l}\!\bra{m} \;.
\end{align*}
Correlations of an MPS built from a repeated injective tensor decay exponentially, e.g., $\langle A_i B_j \rangle = O[\exp(-|i - j|/\xi)]$ for all single-site observables $A_i,B_j$.
\\

{\bf Parent \newline{Hamiltonian}}
& Geometrically local and frustration-free Hamiltonian having the given MPS or PEPS as a ground state. For injective tensors, the parent Hamiltonian acts on nearest neighbor sites and has unique ground state. For injective MPS, it has a spectral gap.

\end{tabular}
\end{ruledtabular}
\caption{Summary of common tensor properties for MPS and PEPS composed of a single, repeated tensor with periodic boundary conditions. See ~\cite{cirac2021matrix} for more details.} \label{tab:tensor_defs}
\end{table}

\subsubsection{Sequential preparation of MPSs} \label{sec:tensor_seq_prep_mps}

To construct a quantum circuit that prepares a given MPS from a product state, one cannot directly interpret the MPS tensors as physical operations -- each individual tensor is not, in general, unitary. However, the representation of an MPS in terms of tensors is non-unique~\cite{fannes1992finitely,perez_garcia2007matrix}. This freedom can be exploited to turn any MPS into a sequential quantum circuit acting on a product state.

The first step is to rewrite the same physical state described by the MPS in terms of new tensors $B_k$ (of bond dimension $D_k$ which is equal or smaller than the original $A_k$ tensors) such that 
\begin{align} \label{eq:tensor_iso_mps}
    \sum_i (B_k^i)^\dagger B_k^i  = 
            \begin{array}{c}
            \begin{tikzpicture}[scale=0.5,xscale=-1,baseline={([yshift=-0.65ex] current bounding box.center)}]
                \draw[thick] (-1*\doubledx,-1) -- (-1*\doubledx,1);
                \draw[thick] (0.9-1*\doubledx,-.8) -- (-2*\doubledx+0.6,-.8);
                \draw[thick] (0.9-1*\doubledx,.8) -- (-2*\doubledx+0.6,.8);
    		      \foreach \x in {1,...,1}{
                  \ETensor{-1*\doubledx*\x,0}{0};
                 }
    		      \foreach \x in {2,...,2}{
                  \SideIdentityTensor{-\doubledx*\x,0}{}{}{-1};
                 }
            \end{tikzpicture}
            \end{array}
        =
            \begin{array}{c}
            \begin{tikzpicture}[scale=.5,xscale=-1,baseline={([yshift=-0.65ex] current bounding box.center)}]
                \SideIdentityTensor{0,0}{}{}{-2};
            \end{tikzpicture}
            \end{array}
    = I_{D_{k-1}} \;.
\end{align}
This representation can be obtained by successive singular-value decompositions~\cite{vidal2003efficient,perez_garcia2007matrix}. The idea is to now reinterpret Eq.~\eqref{eq:tensor_iso_mps} as $v_k^\dagger v^{}_k = I_{D_{k-1}}$, where $v_k \equiv \sum_i B^i_{k} \otimes \ket{i}$, that is, each $v_k$ is an \emph{isometry}. Therefore, it always can be completed to a unitary $u_k$, possibly with some fixed input partial state\footnote{Although the bond dimensions of the initial tensors $A_{k}$ can be arbitrary, from the Schmidt decomposition, the bond dimensions of neighboring sites of the $ B_k$ tensors satisfy $D_{k-1} \le d\, D_k \le d^2 D_{k-1}$. Thus the ancilla input required to complete $v_k$ to a unitary is at most $d$-dimensional.}, $v_k = u_k \ket{0}$.
Using this fact, we can directly reinterpret the MPS as a \emph{sequential circuit}~\cite{schon2005sequential}. Graphically, for $ d=D=2$,
        \begin{align*}
        \begin{array}{c}
        \begin{tikzpicture}[scale=0.55,baseline={([yshift=-4.5ex]current bounding box.center)}]
		      \foreach \x in {4,...,4}{
                \GTensor{(-\singledx*\x,0)}{1}{.5}{\small $B_2$}{0}
        }
		      \foreach \x in {3,...,3}{
                \SingleDots{-\singledx*\x,0}{\doubledx*.8}
        }
		      \foreach \x in {5,...,5}{
                \GTensor{(-\singledx*\x,0)}{1}{.5}{\small $B_1$}{-1}
        }
		      \foreach \x in {2,...,2}{
                \GTensor{(-\singledx*\x-.6,0)}{1}{.5}{\small $B_{n}$}{1}
        }
        \end{tikzpicture}
                \end{array}
\quad &\mathrel{\scalebox{2}{$=$}} \quad
		\begin{tikzpicture}[scale=.45,thick,baseline={([yshift=-0ex]current bounding box.center)}]
         \begin{scope}[xscale=-1]
		      \foreach \x in {2,...,5}{
      \draw [thick] (-\doubledx*\x,1) -- (-\doubledx*\x,5.6);
      \draw (-\doubledx*\x-.05,.5) node {\scriptsize $\ket{0}$};
        }
		\foreach \x in {2,...,2}{
        \gatevar{(-\doubledx*\x-\doubledx*0.5,\doubledx/2*\x)}{\scriptsize $u_1$};          
        }
		\fill[fill=white] (-3.1*\doubledx,2.8) rectangle (-2.9*\doubledx,2.2);
        \draw [thick, dotted] (-2.8*\doubledx,2.2) to (-3.2*\doubledx,2.8);
        \begin{scope}[shift={(0,1)}]
		\foreach \x in {3,...,3}{
        \gatevar{(-\doubledx*\x-\doubledx*0.5,\doubledx/2*\x)}{\scriptsize $u_{n-2}$};
        }
		\foreach \x in {4,...,4}{
        \gatevar{(-\doubledx*\x-\doubledx*0.5,\doubledx/2*\x)}{\scriptsize $u_{n-1}$};
        }
		\foreach \x in {4,...,4}{
        \gatevar{(-\doubledx*\x-\doubledx*0.5,\doubledx/2*\x)}{\scriptsize $u_{n-1}$};       
        }
        \end{scope}
        \end{scope}
        \begin{scope}[shift={(7.5,5)}]     
        \draw[ thick, fill=whitetensorcolor, rounded corners=1pt] (0,0.27) rectangle (\doubledx/2+0.2,-.27); 
	    \draw (\doubledx/2-0.25,0) node {\scriptsize $u_n$};
        \end{scope}
        \end{tikzpicture}
        \;.
    \end{align*}
Importantly, each $u_k$ satisfies $u_k^\dagger u_k = I$, ensuring that the preparation circuit is fully unitary and requires no post-selection.

For a general MPS over $n$ qubits with arbitrary bond dimension $D$, the resulting circuit retains its sequential structure, consisting of $O(n)$ unitary layers, with each layer acting simultaneously on $\lceil \log_2 D \rceil$ qubits~\cite{schon2005sequential}. These unitaries then need to be further decomposed into constant width gates (e.g., 2-qubit nearest-neighbor gates). The overall depth thus scales as $T = O(n D^2)$. The sequential preparation scheme thus provides a general recipe for constructing a quantum circuit for MPSs.

For the case of fixed local dimension $d$ and bond dimension $D$, the linear depth $T = O(n)$ achieved by the sequential scheme is, in fact, optimal. This is established by the following correlation argument~\cite{bravyi2006lieb}. On the one hand, the output of a depth-$T$ circuit composed of nearest-neighbor gates, acting over a product-state input, obeys a light-cone constraint, that is, distant marginals factorize. More precisely, $\rho_{AB} = \rho_A \otimes \rho_B$ for any consecutive subsystems $A,B$ with $\mathrm{dist}(A,B) > 2T$. On the other hand, the set of MPSs includes long-range correlated states, such as the GHZ state $\ket{{\mathrm {GHZ}}} = 2^{-1/2} (\ket{0}^{\otimes n} + \ket{1}^{\otimes n})$ (which has $D = 2$). This state violates the factorization property for all subsystems with $\mathrm{dist}(A,B) \ge \Omega(n)$, since it has correlations independent of $\mathrm{dist}(A,B)$. Combining these two facts, it follows that any local quantum circuit architecture capable of preparing all MPS with constant $D$, must have depth at least $T = \Omega (n)$. A similar argument for all-to-all connectivity gives a depth $T = \Omega (\log n)$, which is saturated for the tree-decomposition of MPS (see \cref{sec:tensor_tree}).

\subsubsection{MPSs with short-range correlations}

Many MPSs of interest exhibit short-range correlations, i.e., exponentially decaying with the distance of any two subsystems. It is thus natural to investigate whether more powerful preparation methods become possible for such short-range correlated MPS families which, for instance, exclude the GHZ state. Exponentially decaying correlations are generic, as they appear with overwhelming probability when the local tensors $A_k$ are chosen at random~\cite{lancien2022correlation}. This occurs, for instance, with unit probability for a tensor built from a random choice of matrices $A^i$, repeated at all sites with periodic boundary conditions\footnote{Note that, in general, the resulting state is not automatically normalized.}. In the translationally invariant setting, this behavior holds precisely for tensors that are \emph{normal} (see \Cref{tab:tensor_defs}).

Intuitively, the exponential decay of correlations restricts the range over which information has to propagate, which allows the parallelization of the sequential circuit, thus reducing its depth. This idea was first established in the context of classification of MPSs and phases of matter~\cite{chen2010local,schuch2011classifying}, which we discuss later. In the quantum circuit language, \citet{brandao2019finite} showed that an MPS generated by a repeated normal tensor (\Cref{tab:tensor_defs}) can be expressed, up to a small error in fidelity, as a constant depth circuit composed of $O(\log N)$-width gates. Using ideas from MPS renormalization~\cite{verstraete2005renormalization}, this result was subsequently strengthened~\cite{piroli2021quantum,malz2024preparation} to a preparation circuits with constant-width, nearest-neighbor gates and depth scaling as $O(\log (n) D^4 / \epsilon)$ ($\epsilon$ denotes the allowed error in global fidelity). This scaling of depth with system size is also optimal for this method, for normal, translation-invariant MPSs using circuits composed of nearest-neighbor gates. An analogous logarithmic-depth construction also holds for more general MPSs, consisting of site-dependent tensors, that satisfy an approximate factorization condition~\cite{malz2024preparation}. The bond-dimension dependence was recently improved to 
$\tilde{O}(D^2\log n + D^4)$ ($\tilde{O}$ suppressing polylogarithmic 
factors in $D$), now achieving exact rather than approximate 
preparation~\cite{murota2026exact}; the scheme is non-deterministic, 
implementing correction maps via post-selected block-encodings, with 
the constant success probability factored into the quoted expected depth.

The preparation cost can be reduced even further if one is only interested in reproducing local observables rather than the global wavefunction. In this case, for any 1D state, exponential decay of correlations alone implies the existence of an MPS that can be prepared by a circuit of constant depth, such that all local marginals on regions of fixed size are accurately reproduced~\cite{dalzell2019locally}.

The \emph{adiabatic theorem}~\cite{kato1950adiabatic} provides a complementary route to preparing MPSs, by realizing them as ground states of suitably chosen local Hamiltonians. Informally, the theorem states that a system initialized in the ground state of a Hamiltonian $H(0)$ will remain close to the instantaneous ground state of a slowly varying family $H(t)$, provided the variation is sufficiently slow compared to the inverse square of the spectral gap of that eigenstate. This yields a general state-preparation paradigm: one engineers an interpolation from a trivial Hamiltonian, whose ground state is a product state, to a target Hamiltonian for which the desired MPS is the unique gapped ground state, and then implements the corresponding adiabatic evolution. The resulting adiabatic algorithm can then be implemented on a digital quantum computer by realizing the unitary time evolution as a quantum circuit using quantum-simulation techniques.

In the context of MPSs, one can always explicitly construct a local \emph{parent Hamiltonian} whose ground space contains the given MPS~\cite{fannes1992finitely,nachtergaele1996spectral,perez_garcia2007matrix}. For MPSs composed of injective tensors, this parent Hamiltonian has a \emph{unique} ground state with a \emph{spectral gap} (\Cref{tab:tensor_defs}). Moreover, for injective MPSs one can explicitly construct a path of local Hamiltonians $H(s)$, $s \in [0,1]$, that linearly interpolates between a trivial parent Hamiltonian with a product-state ground state and the parent Hamiltonian of the target MPS, while the spectral gap remains uniformly bounded along the entire path~\cite{schuch2011classifying,chen2011complete}.
These properties make the adiabatic theorem particularly natural for this class of states: the desired MPS can be obtained as the endpoint ground state of a gapped path starting from an easy-to-prepare product state. However, a naive application of rigorous adiabatic-theorem bounds~\cite{jansen2007bounds}, combined with standard Hamiltonian-simulation techniques, leads to a polynomial in $n$ scaling, which does not improve over the sequential preparation scheme.

To overcome the unfavorable scaling, \cite{ge2016rapid} established a variant of the adiabatic theorem, utilizing the fact that the Hamiltonian path consists of local Hamiltonians, thus obeys the Lieb-Robinson bound~\cite{lieb1972finite,nachtergaele2010lieb}. The result implies the preparation of all translation-invariant injective MPS with depth $O(\log^2 n/\epsilon)$.
Although the method does not saturate the asymptotically optimal $\log N$ dependence, it is applicable to certain PEPS classes beyond 1D (see \cref{sec_tensor_inj_PEPS}).

The parent Hamiltonian, and hence the choice of the adiabatic path leading to a target MPS, is not unique. This freedom can be exploited to optimize the resulting state-preparation protocols, for example, by reducing the locality of the Hamiltonian~\cite{wei2023efficient}, or by numerically maximizing the spectral gap along the adiabatic path~\cite{rai2025spectral}.

Variational methods provide an alternative route to MPS preparation by approximating the target state using a parameterized quantum circuit. In this approach, the circuit parameters are trained to minimize a cost function that quantifies the distance to the desired MPS. Owing to the favorable structure of MPS, such methods often perform well in practice, e.g., \cite{ran2020encoding,rudolph2023decomposition,jaderberg2025variational}; however, they generally come without rigorous guarantees on convergence or circuit depth.

\subsubsection{Measurements and feedforward}

So far, we have restricted attention to unitary state-preparation schemes. Many quantum algorithms, however, employ a richer set of resources -- most notably in active quantum error correction, where the circuit is interspersed with mid-circuit measurements and subsequent operations are conditioned on their classical outcomes. Here we discuss how the same ingredients can reduce the circuit depth required to prepare certain tensor-network states. The idea of using measurements and feedforward to speed-up state preparation can be traced back to many different contexts~\cite{Gottesman.99,verstraete2004valence,raussendorf2002one,broadbent2009parallelizing}. As measurements outcomes are probabilistic, the central challenge in this setting is to avoid prohibitively small post-selection rates. This can be achieved by leveraging feedback, applying adaptive unitary corrections.

The intuition behind the advantage of such schemes is that feedforward effectively circumvents the light-cone constraints of purely unitary, local circuits: classical information about measurement outcomes can be propagated to any distance, allowing correlations to be established between distant regions even when the underlying quantum operations and measurements are local. To illustrate the key idea behind such schemes, we describe the constant-depth preparation of the GHZ state on $n$ qubits
\begin{align}
    \ket{\mathrm{GHZ}}_n = \frac{1}{\sqrt{2}}\left(\ket{0}^{\otimes n} + \ket{1}^{\otimes n}\right),
\end{align}
in constant depth using mid-circuit measurements and feedforward, following~\cite{watts2019exponential,piroli2021quantum,baumer2024efficient}. 

\begin{figure}[t]
    \centering
    \includegraphics[width=0.8\linewidth]{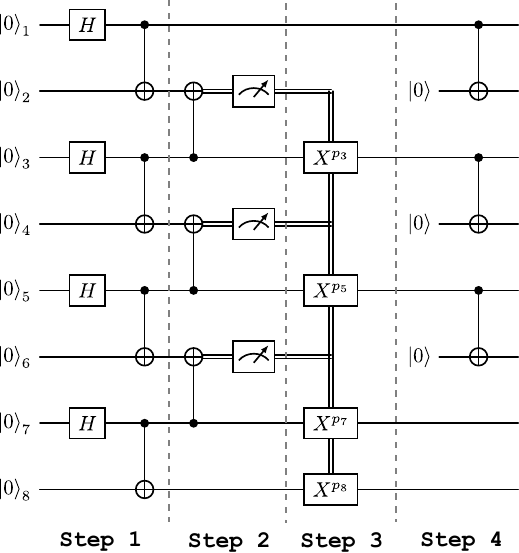}
\caption{Constant-depth GHZ preparation protocol on $n=8$ qubits.}
\label{fig:ghz_full}
\end{figure}

We consider a 1D chain of $n$ qubits (with $n$ even) initialized in $\ket{0}^{\otimes n}$. The protocol is depicted in Fig.~\ref{fig:ghz_full}.

\emph{Step 1: Create nearest-neighbor Bell pairs.}
Apply a Hadamard followed by a CNOT on each odd-even pair $(2k-1, 2k)$, creating $\bigotimes_{k=1}^{n/2} \ket{\Phi^+}_{2k-1,2k}$.

\emph{Step 2: Fusion with measurements.}
Apply a second CNOT layer with qubits $\{3,5,\dots,n-1\}$ as controls and the previous even qubit as target, $\bigotimes_{k=1}^{n/2 - 1} \mathrm{CNOT}_{2k, 2k+1}$,
and then measure all even qubits $\{2, 4, \ldots, n-2\}$ in the computational basis, recording outcomes $m_k \in \{0,1\}$. The combined action of the CNOT and the subsequent measurement can be expressed as a generalized measurement with Kraus operators
\begin{align}
    A_0 = \ket{0}\!\bra{00} + \ket{1}\!\bra{11}, \qquad
    A_1 = \ket{1}\!\bra{01} +\ket{0}\!\bra{10} \;.
\end{align}
Thus each measurement ``fuses'' adjacent Bell pairs: outcome $A_0$ everywhere yields a GHZ state on the unmeasured qubits, while outcome $A_1$ introduces a domain wall, i.e., a relative $X$ error on all subsequent unmeasured qubits.

\emph{Step 3: Correcting errors with feedforward.}
To account for the unwanted $A_1$ measurement outcomes, we may have to apply local flips to the unmeasured qubits. In particular, every time a $A_1$ outcome occurs all subsequent unmeasured qubits need to be flipped, i.e., we need to apply $X^{p}$ ($p \in \{0,1\}$) to qubits $\{3,5,\dots,n-1,n \}$ according to the parity rule $p_{2k+1} = \sum_{j=1}^{k} m_{2j} \mod 2$ (and $p_{n} = p_{n-1}$).

\emph{Step 4: Extend to all $n$ qubits.}
After correction, the GHZ state resides on the $n/2$ odd qubits, together with the $n^{\rm th}$ one. To extend it to all $n$ qubits, reset all measured qubits to zero and apply a final layer of CNOTs with each odd qubit $2k-1$ as control and its even neighbor $2k$ as target, $\bigotimes_{k=1}^{n/2} \mathrm{CNOT}_{2k-1, 2k}$. This produces $\ket{\mathrm{GHZ}}_n$ on all $n$ qubits. 

The total circuit consists of a constant ($n$-independent) number of nearest-neighbor gates, plus one layer of computational-basis measurements, whose outcomes are used to classically condition single-qubit corrections, achieving GHZ preparation in $O(1)$ depth. Note that no post-selection is needed and that no ancilla qubits are used, assuming that the measured qubits are reset to $\ket{0}$ after the measurement.

Beyond the GHZ state, speedups due to measurements and feedforward can be achieved to all MPSs consisting of a single tensor with periodic boundary conditions (regardless of injectivity), of which GHZ is a particular example. In~\cite{malz2024preparation}, it was shown that all such states can be prepared with depth $T = O(\log \log (n/ \epsilon))$, using the same number of rounds of measurements, where $\epsilon$ denotes the target infidelity. A variant of the scheme achieves preparation with depth $T = O(\log (n / \epsilon))$ and a single round of measurements.
Conceptually, the construction relies on decomposing the MPS into a fixed-point state, that encodes the long-range correlations~\cite{verstraete2005renormalization}, which has a GHZ-like structure and can be prepared efficiently using measurements and feedforward. Exact preparation ($\epsilon=0$) at depth $\tilde{O}(D^2\log\log n + D^4)$ 
was recently achieved by supplementing this construction with 
post-selected correction maps~\cite{murota2026exact}.

Even faster, constant-depth preparation can be achieved for certain restricted classes of MPS by exploiting underlying symmetries. This was first explicitly observed for the AKLT state~\cite{smith2023deterministic}, a particular example of a normal MPS (\Cref{tab:tensor_defs}). Constant-depth realization was later extended to broader classes of MPS satisfying suitable symmetry conditions~\cite{li2023symmetry,smith2024constant,stephen2024preparing,sahay2025classifying,zhang2024characterizing,gunn2025phases}. In these cases, the underlying MPS can be prepared exactly in constant depth using a single round of local measurements together with feedforward, without post-selection. The connection between symmetry and preparability is closely related to the measurement-based parallelization of Clifford circuits~\cite{raussendorf2002one}: every brickwork circuit in 1D, composed of Clifford gates, can be  turned into a constant-depth 2D circuit using measurements and feedforward. In tensor network diagrammatics, unwanted measurement outcomes can be interpreted as defects in the virtual degrees of freedom of the MPS. The imposed symmetry then guarantees that appropriate feedforward unitaries can deterministically remove these defects, yielding the desired state. It remains open whether all MPS can be prepared in constant time using measurements and feedforward.

\subsubsection{Dissipative} \label{sec:mps-dissipative}

The dissipative framework provides an alternative paradigm for state preparation, in which dissipation is treated as a resource rather than a drawback. In the digital setting, this framework can be viewed as replacing unitary gates with local quantum channels, thereby allowing continuous access to fresh auxiliary qubits and the ability to trace out subsystems, as compared to the standard unitary setting. By appropriately engineering the dissipation, the dynamics can be designed to drive the system toward a unique steady state, which in the present context is the desired MPS. The relevant notion of efficiency is then the \emph{mixing time}, i.e., the time required for the system to converge to this steady state up to a specified error. In continuous time, the corresponding evolution is described by a Lindblad master equation. A key conceptual difference from unitary schemes is that dissipative protocols can start from arbitrary initial states, which may endow them with intrinsic robustness against certain types of noise~\cite{cubitt2015stability,kashyap2025accuracy}. For instance, this follows for local observable expectation values in local, rapidly (logarithmic-time) mixing systems with unique steady state~\cite{cubitt2015stability}. 

The dissipative preparation of MPSs was first studied systematically in~\cite{diehl2008quantum,kraus2008preparation,verstraete2009quantum}. \citet{verstraete2009quantum} gave a construction of a local Lindblad generator whose unique steady state is a given injective MPS (\Cref{tab:tensor_defs}), with mixing time $T = O(n^{\log n})$. \citet{zhou2021symmetry} introduced a dissipative scheme that, combined with measurements and feedforward, prepares injective MPSs (\Cref{tab:tensor_defs}) with an on-site symmetry (e.g., the AKLT state) in mixing time $T = O(\log^2 n)$. More recently, \cite{baruah2026dissipative} constructed local dissipative schemes, in both continuous-time and discrete-time (quantum channel) formulations, that prepare any injective MPS in time $T = O(\log(n/\epsilon))$, which is asymptotically optimal.

\subsection{Projected entangled-pair states}

\begin{figure}[t]
    \centering
    \includegraphics[width=\linewidth]{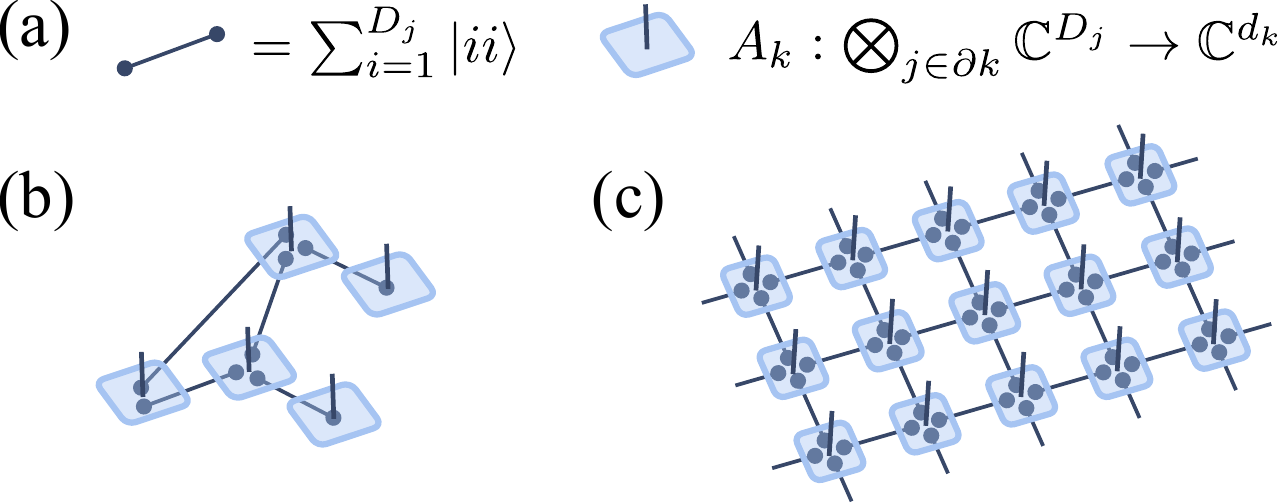}
    \caption{The PEPS construction. (a) Graph edges correspond to maximally entangled pairs (virtual degrees of freedom). At each graph vertex $k$, an operator $A_k$ is applied, mapping virtual to physical space. (b) A PEPS over a graph. (c) PEPS over the square lattice.}
    \label{fig_tensor_peps_constr}
\end{figure}

Projected entangled-pair states (PEPS) are the natural higher-dimensional generalization of MPSs~\cite{verstraete2005renormalization}. To define a PEPS on a given graph, each edge $j$ is associated with a maximally entangled pair of virtual bond dimension $D_j$. At each vertex $k$, one assigns a tensor $A_k : \bigotimes_{j \in \partial k} \mathbb C^{D_j} \to \mathbb C^{d_k}$, which maps the virtual degrees of freedom on the incident edges $\partial k$ to a physical degree of freedom of dimension $d_k$ (Fig.~\ref{fig_tensor_peps_constr}). Contracting all virtual indices according to the graph connectivity yields the many-body quantum state. In the following, we will mainly focus on PEPS defined on the two-dimensional square lattice with uniform physical dimension $d$ and bond dimension $D$. MPSs are recovered as the special case of PEPS on a one-dimensional lattice. By construction, PEPS with a uniform bond dimension $D$ satisfy an \emph{entanglement area law}: the Schmidt rank of the reduced state over any region $R$ is at most $D^{|\partial R|}$, where $|\partial R|$ is the size of the boundary of $R$. Thus the entanglement entropy scales as $S = O(|\partial R|)$ for all regions $R$. 

There are several fundamental differences in complexity between 2D PEPS and 1D MPSs. Most notably, access to an oracle that prepares a 2D PEPS with $D=d=2$ from its classical description, followed by a single-qubit measurement, would allow one to efficiently solve all problems in the complexity class $\mathsf{PP}$~\cite{schuch2007computational}. This class contains $\mathsf{QMA}$ (and hence also $\mathsf{NP}$) and is widely believed to be intractable even for quantum computers. Unless this belief is incorrect, this result rules out the existence of efficient, general-purpose preparation algorithms for 2D PEPS, even at the minimal bond dimension $D=2$. This stands in sharp contrast to the one-dimensional case, where explicit and efficient preparation schemes for MPSs are available.

Second, even classically estimating the normalization of a two-dimensional PEPS is $\#\mathsf{P}$-complete~\cite{schuch2007computational}. Since $\#\mathsf{P}$ is at least as hard as $\mathsf{PP}$, an efficient algorithm for computing the norm of PEPS would enable the solution of extremely hard combinatorial counting problems, which is considered highly unlikely. This hardness persists even in the average case and for multiplicative precision~\cite{haferkamp2020contracting}. In the translation-invariant setting, even the problem of deciding whether a given PEPS tensor defines the zero vector (i.e., has vanishing norm) for all $n\times n'$ periodic lattices is undecidable~\cite{scarpa2020projected,acuaviva2023minimal}.

These difficulties highlight that, in order to make progress on the problem of PEPS state preparation, it is necessary to restrict attention to subclasses of PEPS endowed with additional structure.

\subsubsection{Isometric Tensor-Network States}

For MPSs, we saw that any state admits a representation as a sequential circuit. Isometric tensor-network states (isoTNS)~\cite{haghshenas2019conversion,zaletel2020isometric,soejima2020isometric} extend this idea to higher spatial dimensions by defining a subclass of PEPS that, by construction, retain an analogous circuit interpretation. As a result, isoTNS admit efficient preparation schemes and avoid some of the fundamental difficulties associated with generic PEPS, such as the hardness of norm estimation. Nevertheless, computing local expectation values of a 2D isoTNS remains classically intractable in general --  evaluating local observables is $\mathsf{BQP}$-complete
\cite{malz2025computational}.

An isoTNS is defined by imposing an \emph{isometric} condition on its local tensors, i.e., that the operator
\begin{equation}
    V =
    \begin{gathered}
    \includegraphics[height=.8cm]{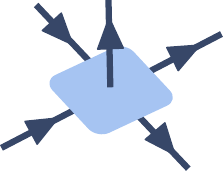}
\end{gathered}
= \sum_{ijrst} 
    \begin{gathered}
    \includegraphics[height=0.85cm]{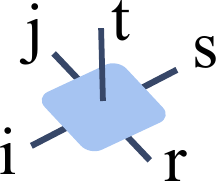}
\end{gathered}
\ket{rst} \bra{ij}
\end{equation}
is an isometry,
\begin{equation}
    V^\dagger V =
    \begin{gathered}
    \includegraphics[height=1cm]{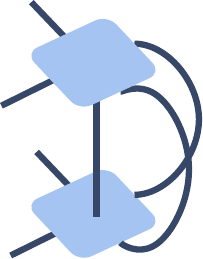}
\end{gathered}
=
    \begin{gathered}
    \includegraphics[height=1cm]{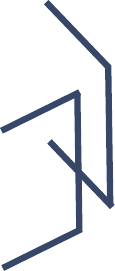}
\end{gathered}
= I \otimes I \;,
\end{equation}
directly generalizing the one-dimensional condition in Eq.~\eqref{eq:tensor_iso_mps}. In one dimension, such a condition can always be imposed without loss of generality and leads naturally to the sequential preparation of MPSs. In contrast, in two or higher dimensions the existence of an isometric structure is no longer generic and instead constitutes the defining assumption of the isoTNS class itself. While the resulting isoTNS preparation circuit is slightly more involved than in one dimension, it can still be implemented efficiently: for an $n\times n'$ square lattice, the preparation depth scales as $T = O(\max\{n,n'\})$~\cite{zaletel2020isometric,wei2022sequential}, with analogous constructions extending to higher spatial dimensions.

The key question is thus which classes of PEPS fall within the isoTNS framework. \citet{zaletel2020isometric} introduced a general procedure -- albeit relying on uncontrolled approximations -- to transform a PEPS into isoTNS form, potentially at the cost of an increase in bond dimension. While this construction is not guaranteed to be efficient or exact in general, it provides a systematic way of bringing PEPS into the isoTNS class and thereby directly yields a corresponding state-preparation scheme.

Beyond heuristic constructions, isoTNS are known to exactly represent nontrivial topological phases of matter. In particular, isoTNS exactly encompass all \emph{string-net models}~\cite{levin2005string} with constant bond dimension~\cite{soejima2020isometric}, a result that is obtained building on their PEPS representations~\cite{gu2009tensor,buerschaper2009explicit}. These translation-invariant states can realize distinct patterns of long-range entanglement. More precisely, string-net states belonging to different phases cannot be connected to one another by local unitary circuits of sub-linear depth. The fact that all such phases admit an isoTNS representation therefore directly implies that linear circuit depth (with respect to the linear size of the hexagonal lattice) suffices to prepare all string-net states starting from a product state. This scaling is also a lower bound in the local-gate model, due to the presence of long-range correlations, in close analogy with the GHZ state preparation discussed in \cref{sec:tensor_seq_prep_mps}.

One reason string-net states play a distinguished role is that a PEPS satisfying a \emph{strict area law} can be connected to a string-net state via a constant-depth local quantum circuit~\cite{kim2024strict,kim2024classifying}. Strict area law means that entanglement entropy satisfies $S(R) = \alpha |\partial R| - \gamma$ for $\alpha, \gamma$ constants independent of the region $R$.

\subsubsection{Injective PEPS} \label{sec_tensor_inj_PEPS}

In contrast to 1D MPS, PEPS in higher dimensions do not automatically lie in a trivial phase even when they satisfy an injectivity condition (\Cref{tab:tensor_defs}). Injective PEPS are defined by local tensors that are injective maps from virtual to physical indices, ensuring they are unique ground states of local, frustration-free parent Hamiltonians~\cite{perez_garcia2008PEPS}. However, injectivity in PEPS allows for critical systems, i.e., those with power-law-decaying correlations~\cite{verstraete2006criticality}. Because of this, preparing arbitrary injective PEPS may encounter the full complexity of many-body state preparation, and extra structural assumptions are needed to render the preparation protocols efficient.

A natural assumption to avoid these difficulties is to assume the existence of a gapped adiabatic path. Specifically, if there exists a smoothly varying family of local Hamiltonians with ground states interpolating between the trivial product state and the target injective PEPS, while remaining uniformly gapped along the entire path, then the injective PEPS preparation becomes efficient. Under this assumption, the target PEPS can be prepared by adiabatic evolution with the corresponding circuit depth scaling as $T = O [\polylog (n / \epsilon)]$ \cite{ge2016rapid}. This constitutes a near-exponential improvement over naive adiabatic bounds.

\subsubsection{Measurements and feedforward} \label{sec:peps-measurements}

As with MPS, the use of measurements and feedforward can speed up the preparation of certain classes of PEPS, going beyond the injective case. This was observed early on for the toric code~\cite{raussendorf2005long,aguado2008creation}, which can be prepared deterministically using a single round of local measurements followed by feedforward corrections, i.e., local unitary operations conditioned on the measurement outcomes. The underlying physical picture is that undesired measurement outcomes correspond to creating pairs of excitations of the toric code, which, due to their Abelian nature, can always be locally paired and annihilated (feedforward), leaving the system in the desired toric code ground state without the need for post-selection.

These ideas have more recently been extended to certain more general classes of topologically ordered states, as described by the string-net formalism. In such schemes, measurements are used to prepare a particular ground-state representative within a given topological phase, which can then be converted into other states in the same phase using shallow unitary circuits or adiabatic evolution, in case a gapped path can be constructed. In particular,~\cite{bravyi2022adaptive,tantivasadakarn2023hierarchy,ren2025efficient} showed that a single round of measurements suffices to deterministically prepare certain states with non-Abelian anyonic excitations. A key insight underlying these constructions is that measurements can be interpreted as implementing a gauging procedure~\cite{tantivasadakarn2024long}. Based on this perspective, the authors conjectured a classification of topological phases according to the minimum number of measurement rounds required for their preparation. \citet{lu2022measurement} realized that a logarithmic number of measurement rounds and layers of local gates are enough to prepare any string-net state -- the construction leveraged the fact that all such states admit a representation in terms of MERA.

Beyond the string-net formalism, certain symmetry properties of PEPS can also guarantee efficient preparation using a single round of measurements and feedforward. As in one dimension~\cite{smith2023deterministic}, the key connection between symmetry and feedforward is the correctability of unwanted measurement outcomes. These outcomes can be interpreted as defects living in the virtual (bond) degrees of freedom of the target PEPS. The imposed symmetry ensures that such defects can be systematically moved to the open boundary by appropriate feedforward unitaries, where they can be annihilated~\cite{zhang2024characterizing,sahay2024finite}. In this way, symmetry singles out a subclass of PEPS that admit efficient, measurement-assisted preparation schemes.

\subsubsection{Dissipative}

The dissipative framework introduced in \cref{sec:mps-dissipative} extends naturally to higher dimensions. For injective PEPS whose parent Hamiltonian is commuting, and under an additional technical condition, \citet{verstraete2009quantum} constructed a local Lindbladian having the target PEPS as its unique steady state; no bound on the mixing time is, however, available in general. Recently, \citet{baruah2026dissipative} removed these restrictions, constructing for any injective PEPS on a bounded-degree graph geometrically local dissipative processes, in both continuous-time (Lindbladian) and discrete-time (quantum channel) formulations, whose unique fixed point is the target state. When the tensors are, in addition, highly injective, the dynamics is rapidly mixing and prepares the state to error $\epsilon$ in time $T = O[\log(n/\epsilon)]$.

\subsection{Tree Tensor Networks and Multi-scale Entanglement Renormalization Ansatz} \label{sec:tensor_tree}

Tree tensor networks (TTNs) constitute a simple class of hierarchical tensor-network states, in which the tensors are connected according to a tree graph spanning multiple length scales~\cite{fannes1992ground,shi2006classical,murg2010simulating,silvi2010homogeneous} (see Fig.~\ref{fig_tensor_tns}). In their most general form, TTNs consist of arbitrary, not necessarily isometric, tensors. Nevertheless, any TTN can be brought into a canonical form in which every tensor is an isometry, up to an overall factor corresponding to the norm of the state, by successive singular-value decompositions proceeding from the leaves toward the root; this procedure can only compress, never increase, the bond dimensions. In the isometric form, the TTN directly specifies a quantum circuit that prepares the state starting from the root and acting successively toward the leaves, with each isometry completed to a unitary using ancillas.

As with sequential preparation (\cref{sec:tensor_seq_prep_mps}), any MPS can be recast as a TTN by successive singular-value decompositions, now arranged in a tree pattern, at the cost of increasing the bond dimension from $D$ to at most $D^2$. For an $n$-qubit MPS, the resulting tree form yields a circuit of two-qubit gates of depth $T = O(D^6 \log n)$, saturating for constant $D$ the $\Omega(\log n)$ lower bound for circuits with unrestricted connectivity (\cref{sec:tensor_seq_prep_mps}). The gates are not nearest-neighbor on a line but follow the tree connectivity. Even under locality constraints, however, logarithmic-depth preparation remains possible if measurements and feedforward are permitted, since long-range gates can then be implemented via standard quantum teleportation~\cite{lu2022measurement}.

The Multi-scale Entanglement Renormalization Ansatz (MERA)~\cite{vidal2008class} can be viewed as a refinement of TTNs tailored to efficiently capture quantum states with scale-invariant, long-range correlations, as typically encountered at criticality. Compared to a TTN, MERA enriches the tree-like structure of isometries by introducing additional local unitary gates, known as disentanglers, which act prior to each coarse-graining step and are designed to remove short-range entanglement (Fig.~\ref{fig_tensor_tns}). This additional structure yields a logarithmic number of layers and allows MERA, similarly to TTNs but with greater expressive power, to reproduce states whose entanglement entropy exhibits logarithmic corrections, a feature not accessible to, for example, constant-$D$ MPSs. In higher spatial dimensions, MERA is expressive enough to exactly capture all string-net models~\cite{konig2009exact,aguado2008entanglement,gu2009tensor}. Crucially for state preparation, a MERA also defines a sequence of isometric and unitary operations and therefore admits a direct realization as a quantum circuit. As a result, the same circuit-based preparation techniques developed for TTNs, including those relying on measurements and feedforward to effectively implement non-local gates, can be directly applied to MERA states.

\subsection{Mixed Tensor-Network States and Unitaries}

Tensor networks can also be used to represent mixed quantum states. In particular, MPS and PEPS admit a natural generalization to the operator setting by equipping each tensor with a bipartite physical index corresponding to the local input and output space. For instance, matrix product operators (MPOs) are operators of the form
\begin{align}
        \sum_{\substack{i_1,\dots,i_n \\ j_1,\dots,j_n}} A_{1}^{i_1 j_1} \dots A_n ^{i_n j_n} \ket{i_1\dots i_n} \bra{j_1\dots j_n}
\end{align}
with a graphical representation
\begin{align}
        \begin{array}{c}
        \begin{tikzpicture}[scale=0.55]
		      \foreach \x in {4,...,4}{
                \GDTensor{(-\singledx*\x,0)}{1}{.5}{\small $A_2$}{0}
        }
		      \foreach \x in {3,...,3}{
                \SingleDots{-\singledx*\x,0}{\doubledx*.8}
        }
		      \foreach \x in {5,...,5}{
                \GDTensor{(-\singledx*\x,0)}{1}{.5}{\small $A_1$}{-1}
        }
		      \foreach \x in {2,...,2}{
                \GDTensor{(-\singledx*\x-.6,0)}{1}{.5}{\small $A_{n}$}{1}
        }
        \end{tikzpicture}
                \end{array}
        \;.
\end{align}
When an MPO is positive semidefinite and has unit trace, it represents a valid mixed quantum state; such operators are referred to as \emph{matrix product density operators} (MPDOs), and constitute the mixed-state analogue of MPS \cite{verstraete2004matrix,zwolak2004mixed}. This construction extends naturally to higher dimensions via PEPS.

Perhaps the most natural strategy for preparing mixed tensor-network states is to purify them and apply a pure-state preparation strategy, subsequently discarding the environment. The central question is thus whether a purification exists, whether the joint pure state of system and environment admits a tensor-network description with favorable bond dimension scaling. This is not guaranteed in general: it has been shown that certain families of MPDOs over $N$ sites with $D_{\rm MPDO} = O(1)$ require a purifying MPS with bond dimension $D_{\rm MPS} = \Omega(\log n)$ \cite{cuevas2013purifications}. This demonstrates that, even in one dimension, there can be a system-size dependent overhead when passing to a purification. A convenient purification can also be sought numerically~\cite{hauschild2018finding}.

When a local purification is not available, it is therefore natural to seek methods that prepare the mixed state directly, either via a quantum circuit of channels or via a Lindbladian evolution. Such methods are currently known only in special cases, even in 1D. For instance, when the MPDO corresponds to a thermal state, for which several distinct preparation schemes are applicable,  or when it takes the form of a so-called fixed-point MPDO~\cite{cirac2017matrix2,liu2026parent}.

A closely related question is how, starting from the tensor network representation of a unitary, to turn it into a quantum circuit implementing its action over an arbitrary input. In 1D, the corresponding class is \emph{matrix product unitaries} (MPUs), i.e., MPOs that are unitary. Note that unitarity of the physical operator does not necessarily imply that each individual tensor can itself be brought into a unitary form. For MPUs composed of a single repeated tensor with periodic boundary conditions over the auxiliary space, which are unitary for all system sizes, it is known that they coincide with \emph{quantum cellular automata} (QCAs)~\cite{cirac2017matrix1,sahinoglu2018matrix}. These are, by definition, unitaries with a strict light cone, and can be implemented by finite-depth quantum circuits with ancillas~\cite{arrighi2011unitarity}. In higher spatial dimensions, it remains open if a similar correspondence between unitary tensor networks and QCA holds.

Beyond this restricted translation-invariant setting, MPUs can generate long-range correlations and, under a suitable technical condition, correspond to quantum circuits of at most polynomial depth~\cite{styliaris2025quantum}.
A special case of particular interest is that of \emph{dualities}, i.e., MPOs mapping between different local Hamiltonians that typically send local operators to nonlocal ones~\cite{lootens2024dualities}. A canonical example is the Kramers--Wannier transformation, which maps product states to GHZ-type states and can be implemented by constant-depth circuits with measurements and feedforward~\cite{tantivasadakarn2024long}. More generally, MPO dualities admit linear-depth circuit realizations, which under certain technical conditions can be parallelized to constant depth~\cite{lootens2025low}.

For a general MPO that is not unitary, as introduced in Sec.~\ref{sec:block_encoding}, one can embed it into a unitary with larger dimension through block encoding. For an MPO  with bond dimension $D$, the block encoding can be constructed with $O(nD^2)$ circuit size using $n+\lceil\log D\rceil$ ancillary qubits~\cite{nibbi2024block}. However, the worst-case normalization factor for method in~\cite{nibbi2024block} increases exponentially with $n$.

%% file: secs/Block_encoding.tex
\section{Quantum block-encodings\label{sec:block_encoding}}

The block-encoding framework is a powerful, flexible, and methodical approach to encoding and processing data in quantum algorithms~\cite{gilyen2019quantum}. Quantum block-encodings are deeply related, both historically and practically, to Quantum Singular Value Transformation (QSVT)~\cite{gilyen2019quantum}, Quantum Signal Processing (QSP)~\cite{low2018optimal}, and Qubitization~\cite{low2019hamiltonian}. 
Rather than directly encode data in a quantum state, the goal of a block-encoding is to construct a quantum circuit whose corresponding unitary matrix representation embeds the desired data matrix or vector in the top-left block. Once a suitable quantum circuit is available, a wide range of downstream primitives (e.g., polynomial transformations on matrices) can be executed to obtain new block-encodings, and thus new data encodings. In this way, block-encodings serve as a unifying data-encoding interface that cleanly connects data-input and quantum algorithms.

In this section we comprehensively review existing techniques for obtaining and manipulating block-encodings,
focusing on the utility of block-encodings as a framework for quantum data encoding.
In contrast, a number of complementary resources are available in the literature, e.g.,
Ref.~\cite{dalzell2023quantum}[Section 10] contains a brief summary of basic block-encoding and QSVT/QSP techniques,
while~\cite{martyn2021grand} provides a broad overview of QSVT and QSP,
demonstrating how the framework unifies all quantum algorithms
with a brief discussion on block-encodings. Moreover, there are a number
of pedagogical resources available for learning the block-encoding, QSVT and QSP framework, such
as~\cite{lin2022lecture}[Chapter 6-8] and~\cite{childs2017lecture}[Chapter 29].

This section is structured as follows. In~\Cref{section:block_encodings:subsection:history}, we provide a brief history both motivating and detailing the development of the quantum block-encoding framework. In~\Cref{section:block_encodings:subsection:definition_and_intuition}, we formally define block-encodings, and provide some intuition to help readers unfamiliar with the technique. In~\Cref{section:block_encodings:subsection:be_techniques}, we summarize existing techniques in the literature for obtaining block-encodings in a range of contexts. In~\Cref{section:block_encodings:subsection:operations_on_block_encodings} we outline operations which map block-encodings to block-encodings, and thus offer additional tools for quantum data encoding.

\subsection{Historical developments}\label{section:block_encodings:subsection:history}

Since quantum states are coherently evolved by unitary transformations, quantum algorithms must be enacted by unitary operations. However, many algorithms are more naturally solved by applying non-unitary operators to a given state vector. Consequently, quantum algorithms have a long history of utilizing ancillary qubits to create subspaces where the desired non-unitary operators are implicitly enacted, as we summarise in the following.

First, as an archetypal example, the Linear Combination of Unitaries (LCU) technique~\cite{childs2012hamiltonian}
allows non-unitary operators (constructed from a weighted sum of unitary operators) to be applied to a given quantum state.
By using a number of ancilla qubits logarithmic in the number of unitary operators being summed,
the non-unitary operator is implicitly constructed in a subspace of the larger unitary. 
Second, given a Hamiltonian $H$ and an inverse temperature $\beta$, one can prepare a Gibbs state
as $e^{-\beta H}$ via a non-unitary operator~\cite{poulin2009sampling,chowdhury2016quantum}.
In~\citet{poulin2009sampling}, a technique based on QPE~\cite{luis1996optimum, cleve1998quantum} is used,
while~\cite{chowdhury2016quantum} builds on LCU. In both cases, the desired non-unitary is encoded in a
larger unitary augmented with ancillary qubits.
Third, in quantum linear algebra, one can solve a linear system of equations by applying the inverse of a given matrix to the quantum state.
In the original proposal~\cite{Harrow.09}, as a simplification, the non-unitary matrix inverse
is applied to an input quantum state by augmenting the whole procedure with additional ancillary qubits,
and following a procedure akin to quantum phase estimation~\cite{luis1996optimum, cleve1998quantum}.
All these examples have common underlying structure which can be made explicit through the use of block-encodings.

An important advancement in quantum algorithm design came with Quantum Signal Processing (QSP)~\cite{low2018optimal}, where an optimal Hamiltonian simulation algorithm was presented. In summary, they show that a non-unitary operator can be embedded in a larger unitary operator, and then by adding an additional ancilla qubit, a sequence of phase operations (interleaved with the controlled unitary embedding) can yield a polynomial transformation of the eigenvalues of the encoded operator. They note that their sequence of phase operations has substantial similarity with techniques from quantum optimal control, see e.g., ~\cite{khodjasteh2010arbitrarily, caneva2009optimal}. 

Subsequently, the framework of Quantum Singular Value Transformation (QSVT) was presented~\cite{gilyen2019quantum}, generalizing the ideas presented in QSP allowing polynomial transformations to the singular values of encoded operators. Importantly,~\cite{gilyen2019quantum} introduced the formal framework of quantum block-encodings, making the ideas in prior work explicit and standardized. They then observe that a number of common algorithms can be easily (and often optimally) formalized through the lens of applying a polynomial function to the eigenvalues (or singular values) of a block-encoded operator (including, e.g., Hamiltonian simulation, matrix inversion, amplitude amplification, and Gibbs sampling).

\subsection{Definition and intuition}\label{section:block_encodings:subsection:definition_and_intuition}

\begin{figure*}[t]
    \centering
    \includegraphics[width=1.5\columnwidth]{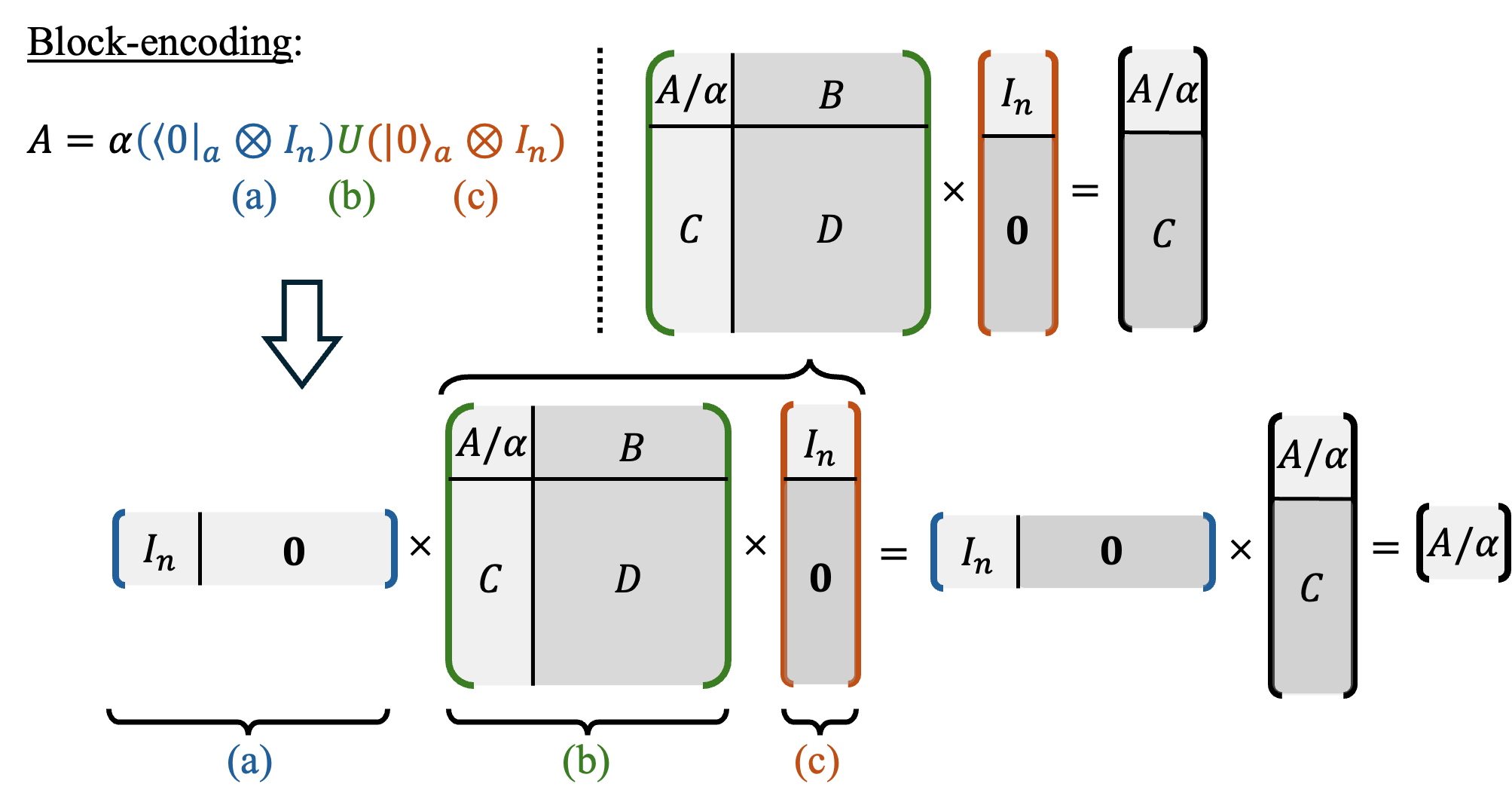}
    \caption{\textbf{Visualization of quantum block-encodings.} 
Here,  $U$ is an $(\alpha, a,0)$-block-encoding of the $n$-qubit matrix $A$. Blocks labeled $B, C$ and $D$ represent blocks whose entries are irrelevant. $\bm{0}$ blocks represent blocks with all-zero entries. View $\bra{0}_a$ as a $1\times 2^{a}$ matrix, and $\ket{0}_a$ as a $2^{a}\times 1$ matrix. Then, $\bra{0}_a\otimes I_n$ and $\ket{0}_a\otimes I_n$ just follow naturally by the Kronecker product definition for matrices (and are of dimensions $2^n\times 2^{a + n}$ and $2^{a + n}\times 2^n$, respectively). Blocks with matching shades of gray indicate blocks that connect in block-matrix multiplication.}
    \label{fig:visualization_of_quantum_block-encoding}
\end{figure*}

We begin by providing some intuition to help understand the subsequent formal definition of a quantum block-encoding.
We note that a definition similar to a block-encoding was also given in~\cite{low2019hamiltonian}, where they call it ``standard form''.

Regarding notation, in this section, we use subscripts on bras and kets to denote the number of qubits they act on, e.g., $\ket{\psi}_n$ is an $n$-qubit quantum state. 

Let $A$ be an arbitrary $n$-qubit matrix (not necessarily unitary), and let $U_A$ be an $n + a$ qubit unitary, with $a \ge 0$. At a high level, the goal of a block-encoding is to obtain a quantum circuit implementing $U_A$ whose top-left $2^n \times 2^n$ block contains $A$. To motivate the formal block-encoding definition, we need the mathematical language that allows us to pick blocks from a given matrix. Adopting a variant of the notation used in the literature, we can define the partial isometries $\Pi_R$ and $\Pi_C$ such that when applied to $U_A$ from the left, $\Pi_R$ yields the first $2^{n}\times 2^{n + a}$ block-matrix row of $U_A$, and when applied from the right, $\Pi_C$ yields the first $2^{n + a}\times 2^{n}$ block-matrix column of $U_A$. The subscript indicates whether it selects a column or row.
Using the usual Kronecker product definition for matrices, $\Pi_R$ and $\Pi_C$ can be formally defined as follows,
\begin{align}
    &\Pi_R := \bra{0}^{\otimes a}\otimes I_n,\\
    &\Pi_C := \ket{0}^{\otimes a}\otimes I_n.
\end{align}
For example,
\begin{align}
    \Pi_R &= \begin{pmatrix} 1 & 0 & \hdots & 0\end{pmatrix} \otimes I_n
    = \begin{pmatrix} I_n & \bm 0 & \hdots & \bm 0 \end{pmatrix}
\end{align}
Then, if the top-left block of $U_A$ is exactly $A$, it is clear that $\Pi_R U_A \Pi_C = A$. To view this mathematically, note that we can always write an arbitrary $n+a$ qubit matrix as a Kronecker product between the $a$-qubit ancilla register and the $n$-qubit main register,
\begin{align}
    U = \sum_{i = 0}^{2^a - 1}\sum_{j = 0}^{2^a-1} \op{i}{j}\otimes \mathbf{M}_{ij} = \begin{pmatrix}\mathbf{M}_{00} & \mathbf{M}_{01} & \hdots \\\mathbf{M}_{10} & \mathbf{M}_{11} & \hdots \\ \vdots & \vdots &  \ddots \end{pmatrix},
\end{align}
where each $\mathbf{M}_{ij}$ is a $2^n\times 2^n$ block of $U$, with $\mathbf{M}_{00}=A$.
Then, it is clear that $\Pi_R$ selects all the blocks across columns with row index $0$, $\Pi_R U =\sum_{j = 0}^{2^a-1} \bra{j}\otimes \mathbf{M}_{0j}$, while $\Pi_C$ selects all rows with column index $0$, $U\Pi_C = \sum_{i = 0}^{2^a-1} \ket{i}\otimes \mathbf{M}_{i0}$. Thus, $\Pi_R U\Pi_C = \mathbf{M}_{00}$, yielding the top left-block.
This is visualized in Fig.~\ref{fig:visualization_of_quantum_block-encoding}. 

However, if we demand that the top-left block-exactly encode the target operator, some generality is lost. Therefore, we generalize by allowing cases when $\lnorm{A}_2 > 1$ (and thus $A$ must be scaled down to admit a unitary embedding), and when we only have some approximation to $A$, $\Pi_R U_A \Pi_C \approx A$. The formal definition of a block-encoding handles both of these cases, introducing a tuple notation to precisely characterize a given block-encoding circuit. 
\begin{definition}[Block-Encoding~\cite{gilyen2019quantum}]\label{sec:block_encoding:def:block_encoding}
A unitary matrix $U_A$ is called an $(\alpha, a, \epsilon)$-block-encoding for the $n$ qubit operator $A$ if
\begin{align}
    \lnorm{A - \alpha (\bra{0}^{\otimes a}\otimes I_n)U_A(\ket{0}^{\otimes a}\otimes I_n)}_2 \le \epsilon.
\end{align}
Using the partial isometry definitions, this can be equivalently written as $\lnorm{A - \alpha \Pi_R U_A\Pi_C}_2 \le \epsilon$.
\end{definition}
A block-encoding is implemented by a quantum circuit, and is described by tuple notation. Given $U_A$ as defined in~\cref{sec:block_encoding:def:block_encoding}, it is an $(\alpha, a, \epsilon)$-block-encoding for $A$. Here, $\alpha$ is a renormalization factor (sometimes called a normalization or subnormalization factor) ensuring that $\lnorm{\Pi_R U_A \Pi_C}_2\le 1$. 
The second parameter in the tuple, $a$, specifies the number of ancilla qubits that the block-encoding uses. The final parameter, $\epsilon$, specifies the error obtained in the block-encoding from $A$. 

We note that $\alpha$ plays an important role in algorithmic complexity. After applying  a block-encoding to an initial state $|0\rangle|\psi\rangle_n$, we obtain $U_A|0\rangle|\psi\rangle_n=\alpha^{-1}|0\rangle A|\psi\rangle+|\text{garb}\rangle$ for some garbage state $|\text{garb}\rangle$. The useful information is encoded in the first term, so the overhead of extracting the useful information is proportional to $O(\alpha/\|A|\psi\rangle\|_2)$. For example, it takes $O(\alpha/\|A|\psi\rangle\|_2)$ rounds of amplitude amplification to boost the amplitude of the first term to a constant level. This motivates the desire for so-called \textit{good block-encodings} where $\alpha \approx \lnorm{A}_2$~\cite{nguyen2022block}, and also a more holistic cost benchmark for block-encodings~\cite{sunderhauf2024block,li2025binary,yang2025dictionary}: $\alpha \;\times$  circuit size of $U_A$.

\subsection{Block-encoding techniques}\label{section:block_encodings:subsection:be_techniques}

Below, we introduce techniques for constructing block-encodings of different target matrices.

\subsubsection{General matrices}\label{sec:bl_general_matrix}

The block-encoding of general matrix can be realized based on the query access to matrix $A$.
When $|A_{jk}|\leqslant1$, we can define query operation $O_A$~\cite{camps2024explicit,camps2022fable}, such that 
\begin{align}\label{eq:bloa}
O_A|0\rangle|j\rangle|k\rangle=(A_{jk}|0\rangle+\sqrt{1-|A_{jk}|^2}|1\rangle)|j\rangle|k\rangle.
\end{align}
Here, $O_A$ can be realized by QRAM. Alternatively, it is a uniformly controlled rotation, and thus can be realized by techniques in Sec.~\ref{sec:qsp_ucr} with $O(N^2)$ circuit size. It can be verified that
\begin{align}
U_A=(I_1\otimes H^{\otimes n}\otimes I_n)(I_1\otimes {\rm SWAP})O_A(I_1\otimes H^{\otimes n}\otimes I_n)
\end{align}
is a $(2^{n},n+1,0)$-block-encoding of $A$. The quantum circuit can be further compressed if one can tolerate a certain error $\delta_c$. Specifically, during the synthesis of $O_A$, one can neglect all rotation gates with angle smaller than $\delta_c$, resulting in a $(2^{n},n+1,N^3\delta_c)$-block-encoding~\cite{camps2022fable}.

The renormalization factor $2^n$ for above methods is large, and deviate from the desired $\alpha\approx\|A\|_2$ as explained in Sec.~\ref{section:block_encodings:subsection:definition_and_intuition}. An alternative method based on quantum state preparation can reduce the renormalization factor to the so-called Frobenius norm (square root of sum of squares of matrix entries).  We define $|\psi_j\rangle=\sum_{k=0}^{N-1}\frac{A_{jk}^*}{\|A_{j,\cdot}\|}|k\rangle$ and $|\phi_k\rangle=\sum_{j=0}^{N-1}\frac{\|A_{j,\cdot}\|}{\|A\|_F}|j\rangle$, where $\|A_{j,\cdot}\|=\sqrt{\sum_{k}|A_{jk}|^2}$ and $\|\cdot\|_F$ is the Frobenius norm of the matrix.  Let $U_R$ and $U_L$ be some (controlled) quantum state preparation unitary satisfying 
\begin{align}\label{eq:be_urul}
U_R|0\rangle|j\rangle&=|\psi_j\rangle|j\rangle,\quad U_L|0\rangle|k\rangle=|k\rangle|\phi_k\rangle.
\end{align}
It can then be verified that $U_R^\dag U_L$ is a $(\|A\|_F,n,0)$-block encoding of $A$~\cite{gilyen2019quantum,kerenidis2020quantum,chakraborty2018power}.

Based on this construction,~\citet{clader2023quantum} uses a  $T$-depth-optimal quantum state preparation method to achieve $O(\log(N/\varepsilon))$ $T$-depth. ~\citet{li2025binary} uses a multiplexor operations for state preparation, achieving nearly optimal circuit size and classical computation time. 
Further reduction of the renormalization factor can be made if one can prepare states proportional to $A_{jk}^{p}$ and $A_{jk}^{1-p}$ for some $p\in(0,1)$, instead of $A_{jk}$. Then, the renormalization factor can be reduced to $\mu_p(A)=\sqrt{S_{2p}(A)S_{2(1-p)}(A^T)}$, with the definition $S(A)=\max_{j}\|A_{j,\cdot}\|_p^p$~\cite{gilyen2019quantum,kerenidis2020quantum,chakraborty2018power,clader2023quantum,li2025binary}. However, the worst-case value of this renormalization factor remains exponential in the number of qubits (i.e. $\sqrt{N}$ for $N=2^n$), which deviates substantially from the spectral norm.

The renormalization factor is greatly reduced in~\cite{li2026reducing}, where the optimal value $\|A\|_2$ is achieved using only a single ancilla.  Their method is based on the singular value decomposition $A/\|A\|_2=W_A\Sigma_AV_A^\dag$, where $\Sigma_A=\text{diag}(\cos\theta_1,\cdots,\cos\theta_N)$. We define $\Sigma_{A_1}=\text{diag}(e^{i\theta_1},\cdots,e^{i\theta_N})$,  $\Sigma_{A_2}=\text{diag}(e^{-i\theta_1},\cdots,e^{-i\theta_N})$, and
\begin{align}\label{eq:A12}
A_1=W_A\Sigma_{A_1}V_A^\dag,\quad A_2=W_A\Sigma_{A_2}V_A^\dag.
\end{align}

It can then be verified that $A/\|A\|_2=\frac{A_1+A_2}{2}$, and both $A_1$ and $A_2$ are unitaries. Thus, the following 
\begin{align}\label{eq:UA12}
U_A=(H\otimes I_n)(A_1\oplus A_2)(H\otimes I_n)
\end{align}
is a $(\|A\|_2,1,0)$-block encoding of $A$.  $U_A$ can be constructed by general unitary sysnthesis techniques in Sec.~\ref{sec:unitary}. For a general $2^{n-1}\times2^{n-1}$ dimensional matrix, the leading order of CNOT count is $\frac{11}{48}\times 4^n$.  

The circuit size for general matrices is inevitably exponential in $n$. Below, we consider structured matrices allowing much lower cost for block-encoding.

\subsubsection{Linear combination of unitaries }\label{sec:bl_lcu_matrix}
The Linear combination of unitaries (LCU) framework assumes that $A$ can be decomposed as  $A=\sum_{p=1}^P\alpha_pU_p$, where $\alpha_p>0$ and $U_p$ are  unitaries that typically (but not necessarily) admit simple constructions. Then, we introduce the state preparation unitary $U_{\text{SP}}$ and select unitary $U_{\text{SELECT}}$ satisfying the following
\begin{align}\label{eq:be_lcu}
U_\text{SP}|0\rangle&=\frac{1}{\sqrt{\|\alpha\|_1}}\sum_{p=1}^P\sqrt{\alpha_p}|p\rangle,\\
U_\text{SELECT}&=\sum_{p}|p\rangle\langle p|\otimes U_p,
\end{align} 
with $\|\alpha\|_1=\sum_{p}|\alpha_p|$. It can then be verified that 
\begin{align}\label{eq:be_lcu_2}
U_A=(U_\text{SP}^\dag\otimes I_n) U_\text{SELECT}(U_\text{SP}\otimes I_n)
\end{align}
 is a $(\|\alpha\|_1,\lceil\log_2 P\rceil,0)$-block encoding of $A$. The technique of Eq.~\eqref{eq:be_lcu_2} was first proposed in~\cite{childs2012hamiltonian} and has since become one of the most commonly used approaches for quantum simulation, as it naturally accommodates many-body Hamiltonians. For instance, in spin-$1/2$ models, and in fermionic models via the Jordan-Wigner encoding as in quantum chemistry, the operators $U_p$ are Pauli strings $\{I, \sigma_x, \sigma_y, \sigma_z\}^{\otimes n}$, and each $U_p$ can be realized by a single layer of single-qubit Pauli gates. Provided that $P = O(\text{poly}(n))$, the Hamiltonian $A$ can be block-encoded efficiently.

LCU has circuit size scaling linearly with $P$, and requires one query to each $U_p$.  The practical circuit complexities depend on concrete space-time and $T$-count trade-offs when implementing $U_\text{SP}$ and $U_\text{SELECT}$. For example, when $U_p$ are some Pauli strings, circuit size of LCU is $O(Pn)$, while the circuit depth can achieve $O(\log(Pn))$ with $O(Pn)$ ancillary qubits~\cite{zhang2024circuit}. 
Moreover, specific optimization has been applied to chemistry applications to reduce the $T$ count~\cite{babbush2018encoding}
and to exploit high latency but dense error correcting codes for reducing the total  circuit volume of the SELECT operation~\cite{low2026denser}.

We note that the terminology of LCU has also been studied in~\cite{gui2006general,long2011duality} to describe the so-called duality quantum computer model.

\subsubsection{Structured matrices}
We now introduce block-encoding techniques for some other structured matrices.

\paragraph{Sparse matrices} 
$O_A$ in Eq.~\eqref{eq:bloa} can be constructed by sparse access input model (SAIM) defined in Eq.~\eqref{eq:saim} of Sec.~\ref{sec:sparse_saim}. Thus, if $A$ is $s$-sparse with  $|A_{j,k}|\leqslant1$, the block-encoding of $A$ with renormalization factor $s$ can be constructed with $O(1)$ queries to Eq.~\eqref{eq:saim}~\cite{low2019hamiltonian}. 
 In other words, the efficiency of SAIM directly implies the efficiency of block-encoding. 
 
 Another construction based on $O_A$ and SAIM has also been given in~\cite{gilyen2019quantum},  which shows that when $A$ is $s_r$-row-sparse, $s_c$-column-sparse and $|A_{j,k}|\leqslant1$, the renormalization factor can achieve $\sqrt{s_rs_c}$. A pre-amplification technique is proposed to remove this renormalization factor when $\|A\|_2\leqslant1/2$, at the cost of $\tilde{O}(\sqrt{s_rs_c})$ increased query complexity. A concrete example is also given in Theorem 4.1 of~\cite{camps2024explicit} for $s_r=s_c=s$. Besides, it is worthy to note that the renormalization factor cannot achieve the optimal value $\|A\|_2$ through preamplification~\cite{gilyen2019quantum}, which is different from the construction for universal matrix~\cite{li2026reducing}.  
 
Alternatively,~\citet{yang2025dictionary} introduced a \textit{dictionary data structure} classifying non-zero elements according to their values and indices. The protocol is efficient when data item are polynomial and indexes of non-zero elements are efficiently computable. 

For general sparse matrix without extra assumptions about $O_A$, an alternative method is proposed in~\cite{kuklinski2024s} based on Walsh-Hadamard transformation, which has $2^n\|HAH\|_{\infty}$ renormalization factor, and $O(N\log N)$ circuit size for matrices with $O(N)$ nonzero entries. 

\paragraph{POVM operator} Suppose a quantum circuit $U$ is an implementation scheme for the POVM operator $M$. Then, the block-encoding of $M$ can be realized based on $U$. To be specific, we suppose $$\text{Tr}[\rho M]=\text{Tr}[U (|0\rangle_a\langle 0|\otimes\rho)U^\dag |0\rangle\langle0|\otimes I_{a+n-1} ].$$ In other words, we apply $U$ to input density operator $\rho$, and then the probability of measuring the first ancillary qubit with outcome $0$ is $\text{Tr}[\rho M]$. Then, it can be verified that
\begin{align}
(I_1\otimes U^\dag)(\text{CNOT}\otimes I_{a+s-1})(I_1\otimes U)
\end{align}
is a $(1,1+a,0)$-block-encoding of $M$~\cite{van2018improvements,gilyen2019quantum}.

\paragraph{Density operator}
We suppose unitary $G$ prepares the purification of density matrix $\rho$. Specifically, with ancillary registers $a_1$ and $a_2$, we have $G|0\rangle_{a_1}|0\rangle_{a_2}=\sum_{j}\sqrt{\alpha_j}|j\rangle_{a_1}|\chi_j\rangle_{a_2}$, and $\rho=\sum_{j}\alpha_j|\chi_j\rangle\langle\chi_j|=\text{Tr}[|G\rangle\langle G|]_{a_1}$. We further define $\text{SWAP}_{a_2,s}$ as the a unitary that swaps the ancillary register $a_2$ and system register $s$. 
Then, it has been shown in~\cite{low2019hamiltonian} that 
$G^\dag\text{SWAP}_{a_2,s}\,G$ is a $(1,|a_1|+|a_2|,0)$-block-encoding of density matrix $\rho$, where $|a_1|, |a_2|$ are the number of qubits in registers $a_1, a_2$.

An alternative approach has been suggested in \cite{huang2026eigenstate} for block encoding density matrices as an algorithmic primitive used in eigenstate preparation. The approach exploits the Sum of Squares Spectral Amplification (SOSSA) to further block encode $\sqrt{\rho}$ which has quadratically amplified eigenvalues compared to $\rho$.

\paragraph{Matrices with displacement structures}
A typical matrix with displacement structures is the \textit{Toeplitz matrix} 
\begin{align}\label{eq:prac_Tn}
\bm{T_N}=\begin{pmatrix}t_0&t_{-1}&\cdots&t_{-(N-1)}\\
t_1&t_0&\ddots&\vdots\\
\vdots&\ddots&\ddots&t_{-1}\\
t_{N-1}&\cdots&t_{1}&t_{0}\\
\end{pmatrix}.
\end{align}
In particular, Eq.~\eqref{eq:prac_Tn} becomes \textit{circulant matrix} when $t_{-j}=t_{N-j}$. Moreover, reversing the order of either rows transforms $\bm{T_N}$ to the \textit{Hankel matrix}, and their circuit constructions are largely the same. For example, the block-encoding of $\bm{T_{N}}$ can be realized by the linear combination of permutation unitaries~\cite{zhou2017efficient,daskin2022quantum}, or based on the efficient circuit construction of the query operation $O_A$ in Eq.~\eqref{eq:bloa}~\cite{wan2021block,sunderhauf2024block,camps2024explicit}.

\paragraph{Diagonal unitary for quantum state} Provided quantum state preparation unitary $U_{\rm sp}|0\rangle_n=\sum_{j}\psi_j|j\rangle$,~\citet{rattew2023non} showed that with $O(n)$ circuit depth and $O(1)$ queries to $U_{\rm sp}$, one can construct $(1,n+3,0)$-block encoding of the diagonal matrix $A_{\psi}=\text{diag}(\psi_0,\cdots,\psi_{N-1})$. Combining with QSVT techniques (see Sec.~\ref{sec:qsvt}), one can perform the block-encoding of $A_{P(\psi)}=\text{diag}(P(\psi_0),\cdots,P(\psi_{N-1}))$ for some polynomial $P$. 

Based on this idea, one can further realize the non-linear transformation of quantum state amplitudes. Specifically, our goal is now to prepare the quantum state $\sum_{j}P(\psi_j)|j\rangle_n$, under the restriction $P(0)=0$. One can first define  $h(x)$ such that $P(x)=xh(x)$, and construct the block-encoding for $A_{h(\psi)}$. Applying the block-encoding of $A_{h(\psi)}$ to initial state $\sum_{j}\psi_j|0\cdots0\rangle|j\rangle$ gives the target nonlinear transformation~\cite{rattew2023non}.

\paragraph{Fermionic Hamiltonians}~\citet{liu2025block} considered the block-encoding of Fermionic Hamiltonians in second quantization with optimized $T$ count. 
The oracles are realized with SELECT-SWAP architecture similar to the one in Sec.~\ref{sec:sp_ss}, which quadratically improve the $T$ count from $O(n^4)$ in~\cite{babbush2018encoding} to $\tilde{O}(n^2)$, where $n$ is the number of spin orbitals. The renormalization factor is $O(n^4)$, which can be further reduced to $O(n^2\eta^2)$ if the number of particles is fixed to be $\eta$. Furthermore, block-encodings for so-called first-quantized fermionic Hamiltonians have been developed in a series of works, e.g., \cite{su2021fault}, while~\citet{huang2025fullqubit} additionally block encodes nuclear degrees of freedom for the purposes of fully coherent molecular dynamics simulation.

\paragraph{Low-rank matrices}
We have introduced in Sec.~\ref{sec:bl_general_matrix} a general matrix block-encoding method with optimal renormalization factor and asymptotic circuit size~\cite{li2026reducing}. This approach allows further circuit simplification for low-rank matrices. Specifically, singular vectors corresponding to zero singular values can be neglected during the circuit construction. Thus, the target to be block encoded reduces to a $2^n\times k$ dimensional isometry, instead of a general unitary. For rank-$k$ matrices, the circuit size can be reduced from $O(4^n)$ to $O(k2^n)$.

\subsection{Operations on block-encodings}\label{section:block_encodings:subsection:operations_on_block_encodings}
The block-encoding of simple matrix can be used to produce the block-encoding of more complex matrices. Below, we introduce some typical operations applied on block-encodings. For simplicity, we only introduce error-free results here, while the finite error results can mostly be found in the relevant references. 

\subsubsection{Product operations}
We let $U_A$ and $U_B$ be $(\alpha, a, 0)$- and $(\beta, b, 0)$-block encodings of operators $A$ and $B$, respectively. The product $AB$ admits simple construction
\begin{align}
U_{AB} &= (U_A \otimes I_b)(I_a \otimes U_B),
\end{align}
yielding $(\alpha\beta, a+b, 0)$-block encoding of $AB$. Alternatively, the \textit{compressed gadget} method~\cite{low2018hamiltonian,fang2023time} enables a block-encoding with less ancillary qubits. Generally, let $U_{A_j}$ be an $(\alpha_j, a, 0)$-block encoding of matrix $A_j$, the following circuit, with $\lceil\log_2L\rceil+1$ extra ancillary qubits 

$\\$
\begin{centering}
\Qcircuit @C=.5em @R=1.2em {
&\qw&\gate{{\rm ADD}^L} &\gate{{\rm ADD}^\dag}& \qw   &\gate{{\rm ADD}^\dag}&\push{\;\cdots}\qw  &\qw  &\gate{{\rm ADD}^\dag}&\qw  \\
&{/}\qw &\multigate{1}{U_{A_1}}& \ctrlo{-1}&\multigate{1}{U_{A_2}} & \ctrlo{-1}&\push{\;\cdots\;}\qw &\multigate{1}{U_{A_2}} &\ctrlo{-1}&\qw\\
&{/}\qw &\ghost{U_{A_1}}& \qw&\ghost{U_{A_2}}&\qw&\push{\;\cdots\;}\qw&\ghost{U_{A_L}}&\qw&\qw
}
\end{centering}
$\\$

\noindent is a $(\prod_{j=1}^L\alpha_j, a+\lceil\log_2L\rceil+1, 0)$-block encoding of matrix product $\prod_{j=1}^LA_j$. Here, unitary ADD satisfies ${\rm ADD}|c\rangle=|c+1\mod 2^{\lceil\log_2 L\rceil+1}\rangle$. An alternative method for matrix product with comparable circuit complexity is also provided in~\cite{dong2025products}.  

For Kronecker product,~\cite{camps2020approximate} show that $S_n (U_{A}\otimes U_B) S_n^\dag$ is an $(\alpha\beta,a+b,0)$-block encoding of $A\otimes B$, where $S_n$ is a layer of SWAP gates acting on the main registers of $A$ and $B$.~\cite{dong2025products} further showed that each SWAP can be simplified as two CNOT gates, and discussed the generalization to non-square matrices. 

The element-wise product, denoted as $A \circ B$, is also called Hadamard product. An important observation is that all elements of $A\circ B$ are contained in the tensor product $A\otimes B$. Thus, there exists a permutation unitary $P$, such that $P(U_A\otimes U_B)P^\dag=\begin{pmatrix}\frac{A\circ B}{\alpha\beta}&*\\ *&*\end{pmatrix}$. This implies an $(\alpha\beta,a+b,0)$-block encoding of $A\circ B$. Specifically,~\cite{guo2024quantum} show that the permutation can be realized by $n$ CNOT gates, that is a multi-target
fan-out gate, based on the circuit proposed in~\cite{zhao2021compiling}.  The element-wise product, combined with LCU technique, can be further used to perform element-wise polynomial functions of matrices~\cite{guo2024quantum}. ~\citet{dong2025products} further showed that one can perform the convolution and vectorization of matrices based on element-wise product.

\subsubsection{Quantum singular value transformation}\label{sec:qsvt}

QSVT is a powerful tool performing polynomial transformations of the singular values~\cite{gilyen2019quantum}. Within this framework, seminal quantum algorithms such as Hamiltonian simulation, matrix inversion, Grover search, and amplitude amplification can emerge naturally as specific choices of the target polynomial~\cite{gilyen2019quantum,martyn2021grand}. QSVT achieves this generality by extending the theory of (QSP)~\cite{low2018optimal,low2019hamiltonian}, preliminary for Hermitian matrices, into the non-Hermitian domain.

Let $A=\sum_{k}\sigma_{k}|w_k\rangle\langle v_k|$ be the singular value decomposition of input matrix $A$. The QSVT for function $f(\cdot)$ applied on $A$ can be defined as 
\begin{eqnarray}\label{eq:fsva}
f_{\rm SV}(A)=
 \left\{
\begin{array}{lcl}
\sum_{j}f(\sigma_{j})|w_j\rangle\langle v_j|&    &  d {\;\rm is\; odd}\\
\sum_{j}f(\sigma_{j})|v_j\rangle\langle v_j| &  &  d {\;\rm is\; even}
\end{array} \right.
\end{eqnarray}
The block-encoding of Eq.~\eqref{eq:fsva} can be constructed from the block-encoding of $A$. Let $U_A$ be a $(1, a, 0)$-block encodings of matrix $A$, and define the controlled phase shift 
\begin{align}
R_{\phi}\equiv e^{i2\phi((|0\rangle\langle0|)^{\otimes a}\otimes I_n)}.
\end{align} 
For phase angles $\bm{\phi}=[\phi_1,\cdots,\phi_d]$, the following unitary 
\begin{eqnarray}\label{eq:qsvt_def}
U_{\bm{\phi}}=
 \left\{
\begin{array}{lcl}
R_{\phi_1}U_A\prod_{k=1}^{(d-1)/2}R_{\phi_{2k}}U_A^\dag R_{\phi_{2k+1}}U_A  &  &  d {\;\rm is\; odd}
\\
\prod_{k=1}^{d/2}R_{\phi_{2k-1}}U_A^\dag R_{\phi_{2k}}U_A &    & d {\;\rm is\; even}
\end{array} \right.\notag\\
\end{eqnarray}
is a $(1,a,0)$-block encoding of $P_{SV}(A)$ for some $d$-degree polynomial $P(\cdot)$, which is an odd (even) function if $d$ is odd (even). A more remarkable property is the generality of Eq.~\eqref{eq:qsvt_def}. Namely, the $(1,a,0)$-block encoding of any polynomials admit a construction in Eq.~\eqref{eq:qsvt_def} for some phase angles $\bm{\phi}$, if the following requirements are satisfied
\begin{itemize}
\item $|P(x)|\leqslant1$ for $x\in[-1,1]$
\item $P(x)$ is odd (even) if $d$ is odd (even)
\item The degree of polynomial $P(x)$ is at most $d$
\end{itemize}
We also note that determining the specific phase angles $\bm{\phi}$ for a given target polynomial requires classical preprocessing, and several efficient numerical methods have been developed~\cite{haah2019product,dong2021efficient,yamamoto2024robust}.

Under the QSVT framework, block-encoding serve as the most fundamental level for the \text{end-to-end} quantum computing, which loads classical matrix $A$ to quantum computer as a single unitary operation. At the second level, QSVT enables mapping a block-encoding of $A$ to the block-encoding of its polynomial transformation. At the third level, these mappings can be used to implement subroutines, such as matrix-inversion, Hamiltonian simulation, amplitude amplification, etc. At the highest level, these subroutines are used as building blocks of quantum algorithms for practical tasks, such as for solving partial differential equations, ground state preparation and  for solving linear systems. This hierarchy enables the systematic translation of high-level quantum algorithms into low-level quantum circuits. In Sec.~\ref{sec:exp}, we provide a concrete example of end-to-end quantum algorithm for stimulating the dynamic of molecular system, in which the role of block-encoding is clearly demonstrated.

%% file: secs/practical.tex
\section{Classical optimization for improving data encoding \label{sec:practical}}
Quantum data encoding involves classical preprocessing, so this process can naturally benefit from classical optimization techniques. Here we review several such strategies that enhance its efficiency and accuracy. We begin with classical randomization techniques in Sec.~\ref{sec:cr}, followed by machine learning techniques~\ref{sec:ml}, including supervised~\ref{sec:ml_s} and reinforcement learning~\ref{sec:ml_r}.
\subsection{Classical randomization}\label{sec:cr}
Classical randomization is a simple yet powerful tool in quantum data encoding. The idea is to approximate the target unitary with the incoherent mixture of a set of unitaries $\mathcal{U}$. This incoherent mixture can be much closer to the target channel than any of the unitary in  $\mathcal{U}$.

~\citet{campbell2017shorter,hastings2016turning} considered the problem of randomized unitary synthesis. For example, under certain dense-cover assumption,~\citet{campbell2017shorter} developed an algorithm that for each target unital channel $\mathcal{U}_{\rm targ}$ and coherent accuracy $\varepsilon$, output a mixture of unital channels 
\begin{align}
\mathcal{E}(\rho)=\sum_{j}p_j\mathcal{U}_j(\rho),\label{eq:mix}
\end{align}
where $p_j>0$ and $\sum_{j}p_j=1$.
The improved accuracy measured by diamond norm is
\begin{align}
\|\mathcal{U}_{\rm targ}-\mathcal{E}\|_{\diamond}=O(\varepsilon^2).
\end{align}
Here, each unital channel $\mathcal{U}_j$ satisfies $\|\mathcal{U}_{\rm targ}-\mathcal{U}_j\|_{\diamond}\leqslant\varepsilon$, and can be constructed by gate sequences with a fixed depth depending only on $\varepsilon$.

Taking the fault-tolerant single-qubit rotation as an example, because the circuit size for coherent approximation scales as $O(\log(1/\varepsilon))$, this randomization leads to about a factor of $1/2$ circuit size reduction.~\citet{kliuchnikov2023shorter} then provided particular improvement for single-qubit Z-rotations, with probability $p_j$ efficiently computable.~\citet{akibue2024probabilistic_2} further showed that the optimal probability distribution in Eq.~\eqref{eq:mix} can be computed by semidefinite programming, with classical runtime increasing polynomially with unitary dimension $N$ and the size of $\mathcal{U}$. A similar idea has also been realized for state preparation~\cite{akibue2024probabilistic}. 

For multi-qubit Toffoli gate, the improvement can be even more significant.~\citet{Gosset2025multi_qubit} showed that with probabilistic mixture, the $T$ count can be reduced to $O(\log1/\varepsilon)$ when error $\varepsilon$ is tolerable, which is also proven to be asymptotically optimal. Compared to the $O(n)$ $T$-count for ideal unitary synthesis, randomization provides improvement whenever $n\geqslant\lceil\log(1/\varepsilon)\rceil+3$.

Approximations above are subject to residual errors $\varepsilon$, such that when measuring an observable through circuit sampling, the resulting distribution is biased.
For single-qubit Pauli rotation gates,~\citet{pai2024} introduced an approach for unbiased sampling via quasiprobability decompositions, where Eq.~\eqref{eq:mix} is generalized to the form $\mathcal{U}_{\mathrm{targ}}(\rho) = \sum_j \gamma_j \mathcal{U}_j\rho$ with possibly negative $\gamma_j$. This can be implemented by sampling $U_j$ with probability $p_j \propto |\gamma_j|$, and multiplying measurement outcomes by a proportionality factor and a sign $\operatorname{sign}(\gamma_j)$ in postprocessing.
The sampling variance is slightly larger than $1$, compared to $1$ for the biased methods above. The approach was generalised to unbiased estimators for more general
quantum gates and circuits, such as single-qubit Clifford+T rotations, and beyond in~\cite{koczor2024sparse}.~\citet{bothe2026more} focused on applying the above quasiprobability approach to the specific case of small angle rotations and proved a bound on the number of $T$ gates required to achieve a small angle rotation such that the bound depends on the angle. 

{
Randomisation techniques have also been applied to
quantum simulation~\cite{campbell2018random, Yang.21, zeng2022simple, kiumi2025te, granet2024hamiltonian, zhang2022unbiasedrandomcircuitcompiler, 2cx4-b82c}. For example, given a problem Hamiltonian as a
linear combination
$
H=\sum_{j=1}^{L} c_j\hat{\sigma}_{j}
$
of Pauli strings $\bm{\hat{\sigma}}$
with a vector of prefactors $\bm{c}$,
the qDRIFT approach of~\cite{campbell2018random} simulates time evolution by randomly applying Pauli rotations $e^{-i \Delta \hat{\sigma}_{j}}$ for some rotation angle $\Delta$ according to probabilities $p_j \propto |c_j|$. qDRIFT thereby achieves a circuit depth that is independent of the size of the classical data $L$ (the number of Pauli terms), and scales rather with the L1 $\lVert \bm{c} \rVert_{1}$ norm of the coefficent vector $\bm{c}$ -- while the approach incurs a simulation error $O(\Delta \lVert \bm{c} \rVert_{1} T)$.
}

Classical randomization has also been used for quantum state truncation.
To approximate a target state with many small entries, a straightforward approach is to truncate all small items, thus obtaining a sparse pure state. Instead, randomized truncation~\cite{harrow2026randomized,wang2026efficient} approximates the target state by the mixture of a set of sparse states $\sum_jp_j|\psi_j\rangle\langle\psi_j|$. Due to the simplicity of sparse state preparation (Sec.~\ref{sec:sparse_sp}), this yields greatly reduced circuit complexities. 
The block-encoding techniques detailed in \cref{sec:block_encoding} can also be combined with randomization and thereby reducing resource requirements as shown in, e.g.,~\cite{sun2025randomised, wang2025randomized,paganelli2026randomization}.

It should also be noted that the benefits of classical randomization come at the cost of introducing ``incoherence''. In some scenarios, such as quantum phase estimation, pure input states are assumed and the applicability of randomized algorithms may require careful assessment, e.g., quasiprobability-based techniques may only be combined with sampling-based statistical phase estimation techniques in the present example.

\subsection{Machine learning optimization}\label{sec:ml}
In practical applications, classical data often contain hidden structures that are not directly apparent. Yet, these structures can still help simplifying the quantum data encoding process. Classical machine learning techniques can uncover such structures and the circuit simplifications they enable. Below, we review progress toward this aim, covering both supervised and reinforcement learning approaches.

\subsubsection{Supervised learning}\label{sec:ml_s}

In supervised learning (SL), a model is trained on input–output pairs to capture the underlying mapping. The mappings are typically represented by deep neural networks. After training, the models can predict the output with new input data.  In the context of quantum circuit optimization, the inputs can be target unitaries, states, or noise models, while the outputs can be circuit parameters or gate sequences. 

SL has been applied in quantum data encoding to optimize the circuit cost~\cite{cincio2021machine,weiden2023improving,zhou2022supervised,zhao2024superencoder}. 
 For example,~\cite{cincio2021machine} considered the optimization of quantum circuits to improve the robustness against noise. Using quantum circuit model and noise model as input, the neural network outputs the parameters of the robust quantum circuit. It is shown that the noise-aware circuit learning algorithm works well for W-state preparation and the circuit compilation of quantum Fourier transformation.~\citet{zhao2024superencoder} considered the problem of quantum state preparation. The input and output of the model are target states and the parameters of the quantum circuits, while the loss function is just the state preparation infidelity. Demonstrations on some synthetic dataset and MNIST data set are provided.

\subsubsection{Reinforcement learning}\label{sec:ml_r}
Unlike supervised learning, reinforcement learning (RL)~\cite{sutton1998reinforcement} does not provide the agent with correct actions. Instead, the agent must obtain the optimal strategy through its own exploration. One of the most commonly used approaches is to treat the optimization tasks as a \textit{Markov Decision Process} (MDP), which contains a state space $\mathcal{S}$, action space $\mathcal{A}$, and reward function $r:\mathcal{S}\times\mathcal{A}\rightarrow\mathbb{R}$. At each episode and time step $t$, the agent is in state $s_t\in\mathcal{S}$, and chooses action $a_t\in\mathcal{A}$ according to its policy $\pi(a|s)$. The agent then receives reward $r_t=r(s,a)$ and moves to the next state $s'$. This process continues until current episode terminates, and the agent obtains total discounted reward
$R=\sum_t\gamma^{t-1}r_t$. The discount rate $\gamma\in(0,1)$ ensures that the agent will receive less reward as $t$ increases, so it generally favors a policy with shorter time. The episodes are repeated multiple times, during which the policy $\pi(a|s)$ is trained to optimize the total discounted reward $R$. 

Early attempts have represented the policy with lookup tables and set $\mathcal{S}$ as the discretized quantum state, which works well for single-qubit tasks~\cite{chen2013fidelity,zhang2018automatic,bukov2018reinforcement,zhang2019does}. More advanced treatment is to represent $\pi(a|s)$ by a deep neural network. This treatment, called deep RL, has much stronger representational power, and is able to handle continuous state space $\mathcal{S}$ and even continuous action space $\mathcal{A}$. For example, deep RL has been applied to Hamiltonian-level optimization for multi-qubit state preparation and logic gate constructions~\cite{niu2019universal,zhang2019does,arrazola2019machine,haug2020classifying,baum2021experimental,he2021deep,sivak2022model,wang2025arbitrary,preti2022continuous}. 
Compared to conventional optimization tasks, one of the advantages of RL is its generalizability. During training, the agent can learn the intrinsic structure of problems and their optimal policies. Thus, the well-trained agent can still output good policies even for problem instances that are not in the training set. This property has been observed in e.g.,~\cite{haug2020classifying,he2021deep,baum2021experimental,wang2025arbitrary}.

The gate-level optimization with RL (perhaps any other optimization approaches) is more challenging.  First,  as opposed to the Hamiltonian-level optimization, the strategies are typically discrete, so there is no well-defined \textit{gradient} of the control parameters. Second, the solution landscape is highly complex: circuits achieving near-optimal fidelity may be completely different from those with only slightly lower fidelities. To avoid local minima, extensive exploration across the discrete action space is required, making the learning process more unstable, and the convergence more difficult to guarantee. There are several works that are successful in gate-level optimizations for circuit compiling~\cite{zhang2020topological,moro2021quantum,fosel2021quantum,he2021variational,PhysRevApplied.15.034068,alam2023quantum,rietsch2024unitary,sarra2024discovering,villar2024ai,kremer2025optimizing,olle2025scaling} or state preparation~\cite{haug2020classifying,sivak2022model,kolle2024reinforcement,altmann2024challenges,wang2025arbitrary}. For example,~\cite{zhang2020topological} considered single- and two-qubit gate synthesis by braiding Fibonacci anyon, a task similar to the $\{H,T\}$ decomposition of single-qubit gates. Guided by $A*$ search, the agent achieves comparable performance to brute-force search for errors down to $10^{-3}$ for single-qubit gates.~\citet{fosel2021quantum} considered the compilation of quantum circuits with Z-rotation, Phased-$X$, and CNOT gates. The well-trained agent can improve the mean circuit size from $\approx 159$ to $\approx 98$, which is also better than $\approx105$ obtained by simulated annealing. 
Notably, this agent is further tested in larger quantum circuits with qubit numbers up to $50$, reducing the mean circuit size from $1940$ to $1616$. This result highlights the generalisability of deep RL. 

There are other RL works for circuit compiling that are not under the MDP framework, such as genetic algorithms~\cite{sunkel2023ga4qco} and diffusion models~\cite{furrutter2024quantum}. In particular,~\cite{furrutter2024quantum} applied diffusion models~\cite{rombach2022high} to the problems of entangled state preparation and circuit compilation.

%% file: secs/application.tex
\section{Applications and quantum advantages\label{sec:application}}

\subsection{An illustrative application}\label{sec:exp}

In this section, we illustrate how data-encoding and unitary-synthesis techniques arise in representative quantum applications of practical relevance. In particular, we consider Hamiltonian simulation, detailing how a Hamiltonian matrix $H$ can be block-encoded and how the resulting block-encoding can be transformed into the time-evolution operator $e^{-iHt}$ using quantum singular value transformation (QSVT)~\cite{gilyen2019quantum}.

We consider the toy model of a $\text{He}$-$\text{H}^+$ molecular Hamiltonian at 
a nuclear separation of $90$pm. Using a minimal STO-3G basis, the full configuration interaction 
Hamiltonian is obtained via the Jordan–Wigner transformation as a linear combination
$
H=\sum_{j}c_j\hat{\sigma}_{j}
$
of Pauli strings $\bm{\hat{\sigma}}=[-IX,-IZ,-XI,XX,XZ,-ZI,ZX,ZZ]$
with prefactors $\bm{c}\approx[0.14, 0.66, 0.14,  0.16,  0.14, 0.66,  0.14,  0.15]$~\cite{peruzzo2014variational} (after rounding).
Below, we construct the implementation of the quantum algorithm
by starting at a high level with QSVT and then decomposing further into
lower-level algorithmic and gate compilation primitives. We also
note that while  further optimisation of the construction below is possible,
we aim to present a general approach.

\begin{figure*}[t]
    \centering
          \includegraphics[width=2\columnwidth]{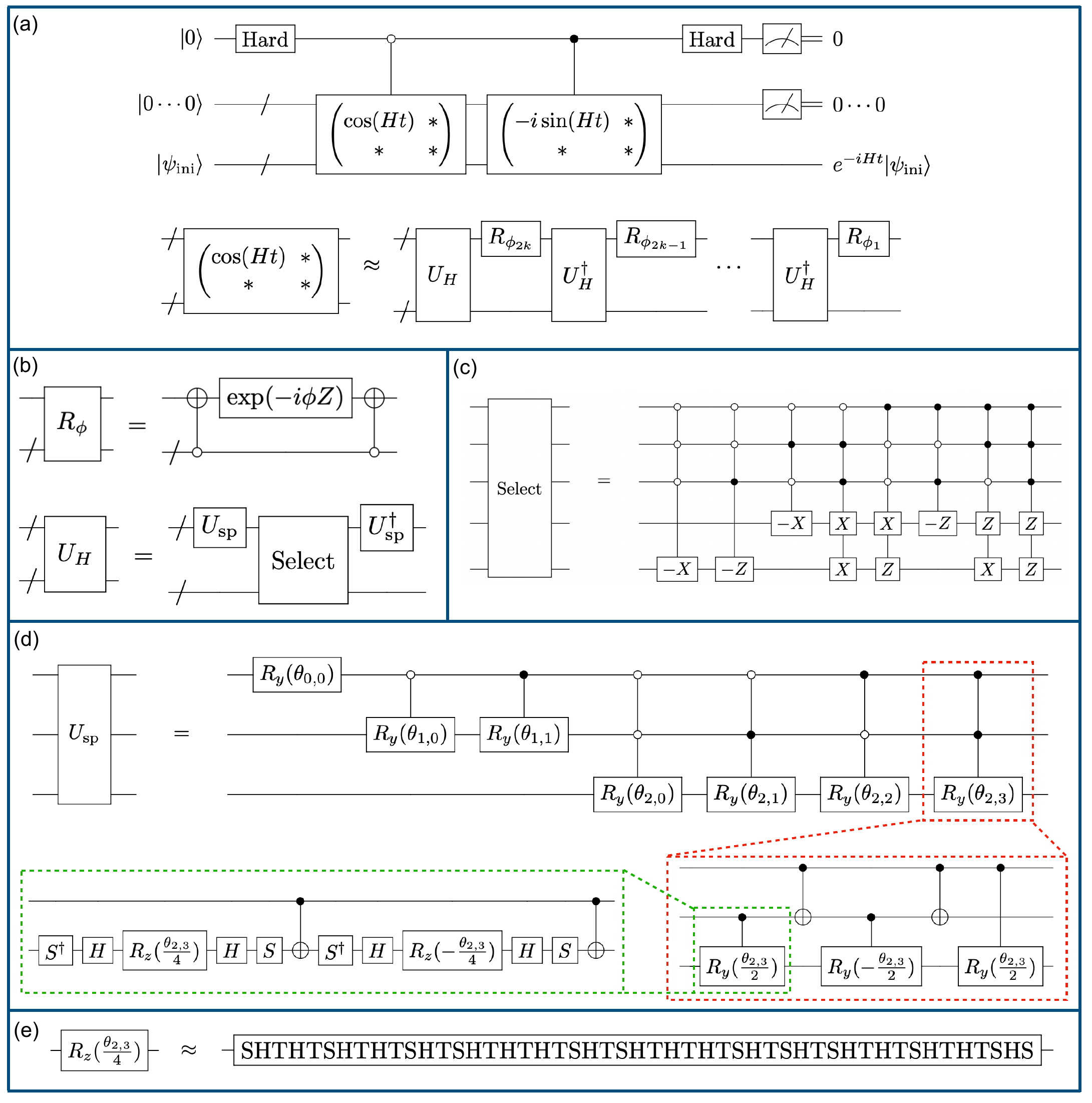}
       \caption{
       	Sketch of the quantum circuits for the illustrative example in Sec.~\ref{sec:exp}. (a) Hamiltonian simulation by QSVT. (b) Circuit for $\Pi_{\phi}$ and block-encoding $U_H$ based on LCU. (c) Select operation realized by UCR. (d) Quantum state preparation by UCR and its subsequent decomposition into CNOT, H, S (and their inverses) and Z-rotation gates. Rotation angles are 
        $\theta_{0,0}=1.57$, $\theta_{1,0}=1.10, \theta_{1,1}=1.08$, $\theta_{2,0}=2.28, \theta_{2,1}=1.64, \theta_{2,2}=2.28, \theta_{2,3}=1.61$
        respectively. (e) Approximation of $R_{z}(\frac{\theta_{2,3}}{4})$ with  $H,T$, and $S=T^2$ with tolerant accuracy $0.01$. The decomposition is based on~\cite{ross2016optimal} and realized by PennyLane~\cite{bergholm2022pennylaneautomaticdifferentiationhybrid}.
} \label{fig:demo}
\end{figure*}

\textbf{QSVT.} The time evolution operator is decomposed into a sum of an odd and an even parity function of the 
Hamiltonian matrix as $e^{-iHt}=\cos(Ht)-i\sin(Ht)$.
Given a unitary $U_{H}$ that block encodes $H$, such that by definition $(\langle0\cdots0|_{\text{anc}}\otimes I)U_H(|0\cdots0\rangle_{\text{anc}}\otimes I)=H/\alpha$
for some normalization factor $\alpha$, implementing
$\cos(Ht)$ and $-i\sin(Ht)$ can be achieved directly by QSVT (Sec.~\ref{sec:qsvt}).
The signal processing operator is applied to the ancillary system as $R_{\phi}\equiv e^{i2\phi((|0\rangle_{\rm anc}\langle0|)\otimes I_n)}.$. Here, anc refers to ancillary systems with qubits initialized to all-zero states.  As illustrated in the lower panel of Fig.~\ref{fig:demo}(a),
QSVT enables the approximation~\cite{gilyen2019quantum} as
\[
\begin{pmatrix}
\cos(Ht)&*\\ *&*\end{pmatrix}
\approx \prod_{k=1}^{d/2}R_{\phi_{2k-1}}U^\dag_{H}R_{\phi_{2k}}U_{H},
\]
for appropriately chosen angles $\phi_{j}$ [see e.g.~\cite{haah2019product,dong2021efficient} for calculating  $\phi_j$].
Constant accuracy in the above approximation is achieved by setting $d=O(\alpha t)$, 
and the approximation of $-i\sin(Ht)$ is implemented via another set of angles $\phi_{j}$.

The even and odd parity contributions are then added together
through the use of one additional ancillary qubit as we illustrate in the upper panel of Fig.~\ref{fig:demo} (a).
The algorithm then terminates by postselecting all ancillary qubits to outcome $|0\rangle$ via, e.g., 
amplitude amplification.

Beyond the block-encoding of $H$, QSVT requires the parametrised gates $R_{\phi}$ which can be implemented as 
a single-qubit $Z$-rotation interleaved between two multi-qubit controlled $X$ gates, 
as illustrated in the upper panel of Fig.~\ref{fig:demo} (b). The multi-qubit controlled $X$ gate is also detailed in Sec.~\ref{sec:cnxg}.
For example, with one ancillary qubit, the $n$-qubit multi-qubit controlled $X$ gate can be realized by
$O(n)$ circuit size and $O(\log n)$ circuit depth~\cite{nie2024quantum} [see also Fig.~\ref{fig:mcu}].

\textbf{Block-encoding.} 
As the present Hamiltonian is a sparse linear combination of Pauli operators, 
its block-encoding $U_H$ is efficiently constructed via the linear combination of unitaries approach (Sec.~\ref{sec:bl_lcu_matrix}).
As illustrated in the lower panel of Fig.~\ref{fig:demo} (b), one can set $U_H=(U_{\text{sp}}^{\dag}\otimes I)U_\text{select}(U_{\text{sp}}\otimes I)$. Here, $U_\text{select}=\sum_{j}|j\rangle_{\text{anc}}\langle j|\otimes \hat{\sigma}_{j}$ can be constructed by uniformly controlled rotations. In the present example, $U_{\text{sp}}$ is a three-qubit state preparation unitary satisfying $U_{\text{sp}}|000\rangle_{\text{anc}}=\frac{1}{\sqrt{\alpha}}\sum_{j}\sqrt{c_j}|j\rangle_{\text{anc}}$. The corresponding normalization factor is $\alpha=\|\bm{c}\|_1=2.19$.

The most straightforward approach for implementing $U_\text{select}$ is applying eight 3-qubit
controlled rotations sequentially (see Fig.~\ref{fig:demo} (c)).  For state preparation, we show in Fig.~\ref{fig:demo} (d) how $U_{\text{sp}}$ can be decomposed into UCRs.  Note that there are many alternative methods for these operations, as introduced in Sec.~\ref{sec:qsp}. Subsequently, each multi-qubit controlled unitary is decomposed into CNOT, $S, S^\dag, H$ and $Z$-rotations.

\textbf{Compilation to hardware native gates.}
We consider the typical example of fault-tolerant quantum
computing in surface codes and assume the elementary gate set
$\{\text{CNOT},H,S, T\}$.
Multi-controlled rotations, required for the signal processing and state preparation stages, can be decomposed into Clifford gates and single-qubit $Z$ rotations, e.g., as illustrated in Fig.~\ref{fig:demo}(d).
Thus, the final step in circuit compilation is to approximate continuous single-qubit $Z$-rotations with $H$, $S$ and $T$ gates,
which can be achieved through the techniques detailed in Sec.~\ref{sec:u_ds}. We illustrate the simple and general approach
of Ref.~\cite{ross2016optimal} that decomposes a single-qubit rotation
into an approximate sequence of $H$, $S$ and $T$ gates in Fig.~\ref{fig:demo}(e). We also note that state-of-the-art compilation typically uses advanced techniques, such as the ones detailed in Sec.~\ref{sec:rm} and in Sec.~\ref{sec:practical}.

\subsection{Relation to quantum advantage}\label{sec:adv}
Quantum data encoding is typically used in an initialization stage and is further processed via quantum algorithms.
However, in several practically important problems, while quantum algorithms may achieve polynomial or exponential speedups,  general end-to-end quantum advantage may be prohibited by costs associated with initialisation. Below we illustrate in three representative tasks
how end-to-end quantum advantage 
 crucially hinges on the efficiency of quantum data encoding.

\paragraph{Ground energy estimation}
One of the most natural applications of quantum computing is the simulation of quantum systems, with the estimation  of their ground energy being a particularly important practical task.
(see also Sec.~\ref{sec:initial_state}).
However, estimating the ground energy of a $k$-local Hamiltonian is in general known to
be hard as it is in the complexity class QMA-complete~\cite{kitaev1997quantum}, which is a quantum analogue
of NP-complete. On the other hand, if a good initial or guiding state can be prepared efficiently,
whose overlap with the true ground state is nontrivial, the problem becomes efficiently
solvable~\cite{kitaev1997quantum} and is known to belong to the class BQP-complete~\cite{gharibian2022dequantizing}.
Thus, quantum advantage can emerge when there is a mapping from a classical description of a
Hamiltonian to a nontrivial initial state. While a broad range of classical heuristic approaches have been developed
for preparing such approximations to ground states of specific Hamiltonian problems, there is no efficient general approach.

\paragraph{Solving linear system}
The above example shows that a quantum advantage can emerge with efficient initial state preparation. Conversely, a quantum advantage may be falsely claimed when quantum data encoding is overlooked~\cite{tang2019quantum,chia2022sampling,huang2025vast}.
As a typical example, we consider the problem of solving linear systems of equations. The argument below also applies to other relevant linear algebra problems. 

\begin{figure}
    \centering
          \includegraphics[width=1\columnwidth]{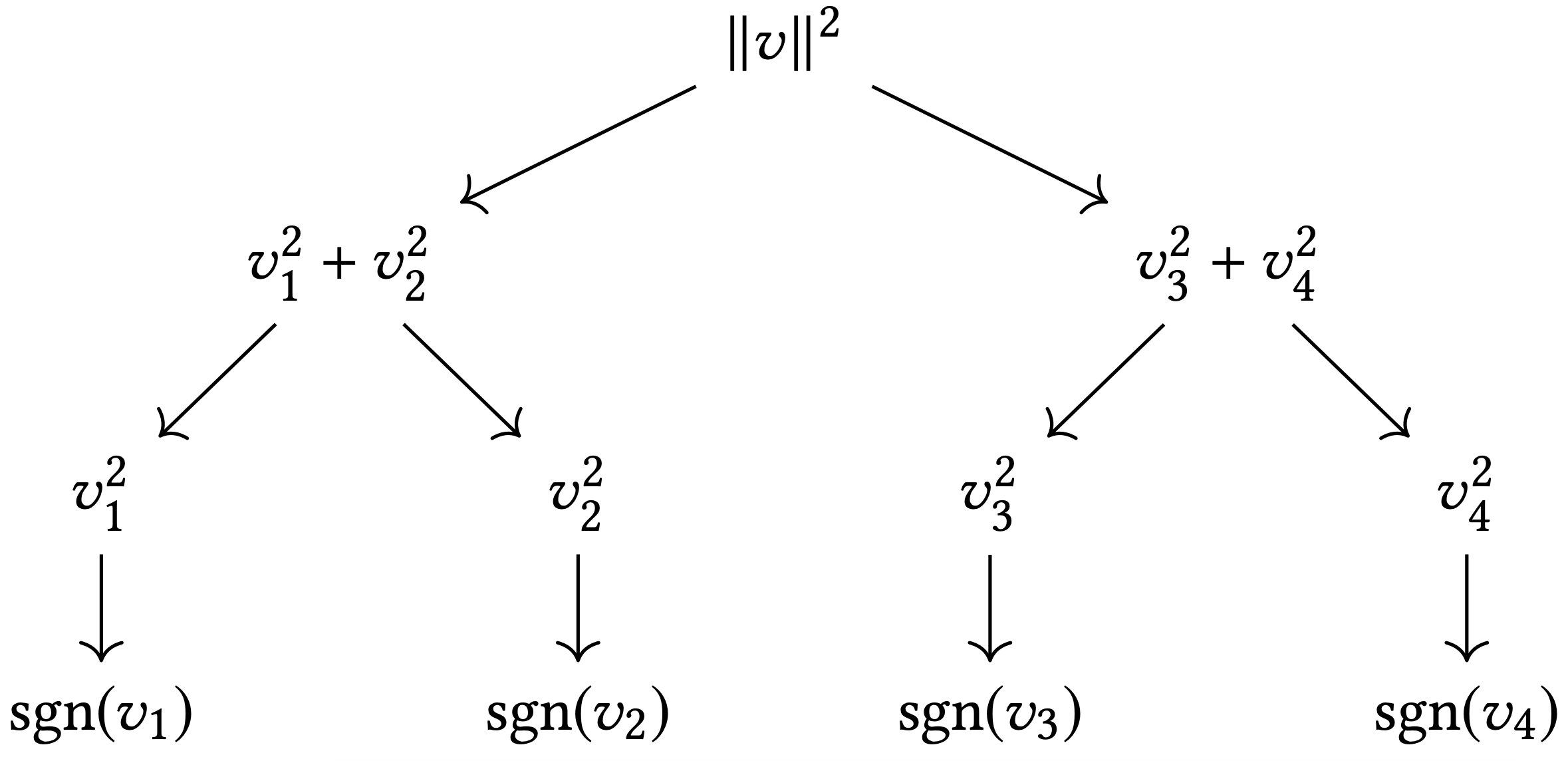}
       \caption{Classical data structure example for $N=4$, which enables logarithmic runtime for query and sampling access.
       	This figure is reproduced from Fig~1 of Ref.~\cite{tang2019quantum}.} \label{fig:cram}
\end{figure}

Given the classical descriptions of a matrix $A\in\mathbb{C}^{2^n\times 2^n}$ and a vector $b\in\mathbb{C}^{2^n}$, the task is to output a quantum state proportional to the solution to $|x\rangle=A^{-1}|b\rangle$.
If the condition number of $A$ is polynomial in $n$, the problem can be solved by a
polynomial number of quantum query accesses to $A$, $b$, and 
using an additional $O(\text{poly}(n))$  quantum gates~\cite{Harrow.09}. Unfortunately, for a general matrix $A$ and vector $b$, the query access has a polynomial circuit size with respect to $N$, meaning that the scaling with respect to $n$ is exponential. Thus, super-polynomial speedup is unlikely to exist for general, unstructured classical data. While the circuit depth can be reduced to $O(\text{poly}(n))$ using additional ancillary qubits, parallelization can similarly be introduced to classical computation. As has been discussed in Sec.~\ref{sec:QRAM}, classical \textit{dequantization} algorithms can further lower classical complexity~\cite{chia2022sampling,tang2022dequantizing}, with a data structure resembling QRAM (see Fig.~\ref{fig:cram} for illustration).

Fortunately, dequantization does not rule out the exponential quantum advantage for \textit{structured} data. \citet{Harrow.09} proved that when query access to $A$ and $b$ is efficient,
solving linear systems is BQP-complete, which is a complexity class that can be considered as a quantum analogue of the class P-complete. This means that there exist instances of linear systems such that a quantum algorithm can achieve an exponential speedup, unless universal quantum circuits can be efficiently simulated on a classical computer.

The argument above implies that there exists structured data, whose query accesses are both quantumly easy and classically hard.  We summarize this as the following open question that is critical for practical quantum advantage.

\begin{challenge}\label{prob:adv}
Find classical representations of structured data, whose quantum query accesses are easy, but classical queries or sampling accesses are hard. 
\end{challenge}

A heuristic example is LCU: When each component of the LCU is efficiently implementable,  their block-encoding can be efficiently realized by methods in Sec.~\ref{sec:bl_lcu_matrix}, with only polylogarithmic space complexity. At the same time, their classical sampling access is believed to be hard. However, to date, concrete data structure instances that simultaneously satisfy both of the above criteria, and also possess practical application value remain scarce.

%% file: secs/discussion.tex
\section{Conclusions and Outlooks\label{sec:discus}}

In this review, we have presented a systematic overview of  quantum data encoding, including quantum state preparation, unitary synthesis, QRAM and block-encoding. We introduce techniques not only for general data, but also for data with structures, e.g. sparsity, Boolean, low entanglement, efficiently computable, that enabling significant reduction of the encoding resources. A unifying perspective is that encoding classical data is not merely a preprocessing step for quantum computing, but also a critical component whose circuit complexity can be deterministic for the overall quantum advantage. 
Looking forward, there are several interesting research directions.

 For general state preparation and unitary synthesis, important circuit complexity gaps remain between upper and lower bounds. This includes Challenge~\ref{prob:para} for the circuit depth of unitary synthesis with ancillas, the unitary synthesis problem stated in Challenge~\ref{prob:usp}. Other open questions include the Clifford+T circuit size of state preparation and unitary synthesis for ancillary-free case, as well as the $T$ count for unitary synthesis. In the practical aspect, it is also important to develop protocols tailored to realistic experimental architectures.  Here, optimization must extend beyond standard circuit complexity to incorporate other factors, such as noise robustness and qubit topology.


For QRAM, most constructions remain active under the total opportunity cost accounting. Existing lower bounds and no-go results apply to restricted settings governed by time-independent local Hamiltonians. An important question is whether the no-go theorem is applied for more general setting. On the constructive side, even if a fully strongly passive QRAM is unlikely to exist (see Challenge.~\ref{prob:qram}), the intermediate regimes remain important and practical. Realizing an end-to-end practically passive QRAM requires not only a query minimal energy and time cost, but also intrinsic noise resilience, and an efficient interface with a fault-tolerant quantum processor. 


In processing classical data, quantum advantage is closely related to the data structure as discussed in Sec.~\ref{sec:adv}. Thus, an important research direction is to identify practically relevant families of structured data, together with concrete computational tasks and explicit input-access models, for which both quantum data encoding and subsequent quantum processing can be performed efficiently. Besides, structure can also benefit classical computing. Any claim of quantum advantage should therefore be supported by evidence of classical hardness. At a minimum, the proposed quantum method should be compared with the best known classical algorithms, including the dequantization techniques, in order to claim the end-to-end quantum advantage.

